\documentclass[aps,prd,preprint,groupedaddress]{revtex4}
\usepackage{hyperref}
\usepackage{amssymb}
\usepackage{amsmath}
\usepackage{slashed}
\usepackage{cases}

\newcommand{\ohf}{\ensuremath{\textstyle \frac{1}{2}}}
\newcommand{\oqt}{\ensuremath{\textstyle \frac{1}{4}}}
\newcommand{\thf}{\ensuremath{\textstyle \frac{3}{2}}}
\newcommand{\shf}{\ensuremath{\textstyle \frac{7}{2}}}
\newcommand{\otd}{\ensuremath{\textstyle \frac{1}{3}}}
\newcommand{\ttd}{\ensuremath{\textstyle \frac{2}{3}}}
\newcommand{\osx}{\ensuremath{\textstyle \frac{1}{6}}}
\newcommand{\fhf}{\ensuremath{\textstyle \frac{5}{2}}}
\newcommand{\oneortwo}{\ensuremath{\textstyle \frac{1}{\sqrt 2}}}
\newcommand{\oneorsix}{\ensuremath{\textstyle \frac{1}{\sqrt 6}}}
\newcommand{\oneortwlv}{\ensuremath{\textstyle \frac{1}{\sqrt{12}}}}
\newcommand{\twoorsix}{\ensuremath{\textstyle \frac{2}{\sqrt 6}}}

\newcommand{\oneorth}{\ensuremath{\textstyle \frac{1}{\sqrt 3}}}
\newcommand{\rtwoorth}{\ensuremath{\textstyle \sqrt{\frac{2}{3}}}}
\newcommand{\sulxsur}[1]{\ensuremath{SU(#1)_L\times SU(#1)_R}}
\newcommand{\gts}{\ensuremath{\mr{GTS}}}

\newcommand{\su}[2]{\ensuremath{SU(#1)_{#2}}}
\newcommand{\sofour}{\ensuremath{SO(4)}}

\newcommand{\rt}{\ensuremath{\Gamma_4\rtimes SO(4)_T}}
\newcommand{\rrt}{\ensuremath{\Gamma_4\rtimes SU(2)_T}}
\newcommand{\swfour}{\ensuremath{SW_{4,\mr{diag}}}}
\newcommand{\gfour}{\ensuremath{\Gamma_4}}
\newcommand{\ls}{\gfour\rtimes\swfour}
\newcommand{\mc}[1]{\ensuremath{\mathcal{#1}}}
\newcommand{\mr}[1]{\ensuremath{\mathrm{#1}}}
\newcommand{\mb}[1]{\ensuremath{\mathbf{#1}}}
\newcommand{\la}{\ensuremath{\mathcal L}}
\newcommand{\ffa}{\ensuremath{^{FF(A)}}}
\newcommand{\ffb}{\ensuremath{^{FF(B)}}}
\newcommand{\order}[1]{\ensuremath{\mathcal{O}(#1)}}
\newcommand{\asq}{\ensuremath{a^2}}
\newcommand{\mq}{\ensuremath{m_q}}

\newcommand{\obs}{MSCO}
\newcommand{\ket}[1]{\ensuremath{|{#1}\rangle}}
\newcommand{\bra}[1]{\ensuremath{\langle{#1}|}}
\newcommand{\bs}[1]{\mbox{\begin{boldmath}\ensuremath{#1}\end{boldmath}}}

\begin{document}

\title{Staggered Heavy Baryon Chiral Perturbation Theory}

\author{Jon A. Bailey}
\email[]{jabailey@wustl.edu}
\affiliation{Washington University in St.\ Louis}

\date{\today}

\begin{abstract}
Although taste violations significantly affect the results of staggered calculations of pseudoscalar and heavy-light mesonic quantities, those entering staggered calculations of baryonic quantities have not been quantified.  Here I develop staggered chiral perturbation theory in the light-quark baryon sector by mapping the Symanzik action into heavy baryon chiral perturbation theory.  For 2+1 dynamical quark flavors, the masses of flavor-symmetric nucleons are calculated to third order in partially quenched and fully dynamical staggered chiral perturbation theory.  To this order the expansion includes the leading chiral logarithms, which come from loops with virtual decuplet-like states, as well as terms of \order{m_\pi^3}, which come from loops with virtual octet-like states.  Taste violations enter through the meson propagators in loops and tree-level terms of \order{\asq}.  The pattern of taste symmetry breaking and the resulting degeneracies and mixings are discussed in detail.  The resulting chiral forms are appropriate to lattice results obtained with operators already in use and could be used to study the restoration of taste symmetry in the continuum limit.  I assume that the fourth root of the fermion determinant can be incorporated in staggered chiral perturbation theory using the replica method.
\end{abstract}


\maketitle

\section{\label{intro}Introduction}
The staggered formulation \cite{Kogut:1974ag} of lattice QCD incorporates the rooting conjecture of~\cite{Marinari:1981qf} to eliminate remnant sea quark doublers.  The past year has seen good progress in our understanding of this conjecture~\cite{Bernard:2006ee,Shamir:2006nj} and the corresponding replica or quark flow method \cite{Damgaard:2000gh,Aubin:2003rg,Aubin:2003mg,Aubin:2003uc} of staggered chiral perturbation theory~\cite{Bernard:2006zw}.  However, the results of \cite{Bernard:2006zw} have yet to be extended to the baryon sector, and many of the assumptions of \cite{Bernard:2006zw,Shamir:2006nj} have not yet been tested.  Lacking proof of these assumptions, successful calculations of experimentally well-known hadronic quantities serve as essential tests of the rooting conjecture and the replica method.  

The advantages \cite{Toussaint:2001zc} of staggered QCD have allowed successful calculations of many hadron masses, decay constants, form factors, and other quantities of phenomenological importance~\cite{Davies:2003ik,Aubin:2004ck,Aubin:2004ej,Aubin:2004fs,Aubin:2004wf,Aubin:2005ar}.  However, to date this success has been most pronounced in the pseudoscalar and heavy-light meson sectors, where the development of staggered chiral perturbation theory (S$\chi$PT)~\cite{Lee:1999zx,Aubin:2003mg,Aubin:2003uc,Aubin:2005aq} facilitated calculations of quark masses, meson masses, and meson decay constants~\cite{Bernard:2001yj,Aubin:2004fs,Aubin:2004ej,Aubin:2005ar}.  Progress calculating baryonic quantities has been impeded by comparatively large statistical errors and two sources of systematic errors:  questions of interpretation specific to the staggered valence sector, and a lack of control over the chiral and continuum extrapolations.  

In contemporary studies employing staggered fermions, three physical quark flavors are explicitly included in the action~\cite{Aubin:2004wf}; the four pseudoflavors per physical quark flavor are called ``tastes.''  The presence of taste quantum numbers implies that the valence sector of staggered QCD contains baryons that have no counterpart in nature.  However, assuming taste symmetry is restored in the continuum limit, the staggered baryons become degenerate within irreps of the taste symmetry group, \su{4}{T}, and baryons composed of valence quarks of the same taste are seen to correspond to physical states~\cite{Bailey:2006zn}.  At nonzero lattice spacing, taste violations lift the \su{4}{T} degeneracies and mix baryons with different \su{4}{T} quantum numbers.  The resulting spectrum contains many sets of nearly degenerate, mixed states.  Since states with the same conserved (lattice) quantum numbers mix, interpolating fields typically overlap multiple nearly degenerate, mixed members of these sets.  

This situation complicates the task of extracting the masses of physical baryons.  To date, staggered calculations of the masses of light-quark baryons have been performed by first taking the continuum limit of lattice data and then using continuum baryon chiral perturbation theory to take the chiral limit~\cite{Aubin:2004wf}.  In principle this approach is perfectly rigorous, but using it to quantify the effects of taste violations, which are \order{\alpha_s^2a^2}, is difficult and has not been done.  However, using S$\chi$PT to study the splittings and mixings due to taste violations is comparatively straightforward.  

Reference~\cite{Bailey:2006zn} considered the questions of interpretation specific to the mass spectrum of light-quark staggered baryons.  It was found that, in calculations of the masses of the nucleon, the lightest decuplet, and certain excited states, one can avoid the complications due to the splittings and mixings introduced by taste violations provided one appropriately chooses the interpolating operators and quark masses.  The work reported here enhances our control over the chiral and continuum limits by developing S$\chi$PT in the baryon sector.  As an example, the masses of certain staggered nucleons are calculated in staggered heavy baryon chiral perturbation theory (SHB$\chi$PT) through leading chiral logarithms; the resulting chiral forms could be used to study the restoration of taste symmetry in the continuum limit and to improve calculations of the nucleon mass.  Calculations of the chiral forms for the masses of other states highlighted in~\cite{Bailey:2006zn} are in progress and will be reported elsewhere~\cite{Bailey:forth}.  

In SHB$\chi$PT, the calculation of baryon masses through leading chiral logarithms requires the inclusion of leading loops, which enter at third order in the staggered chiral expansion.  After discussing the power counting in Sec.~\ref{la}, I write down the Symanzik action and use its symmetries to map the quark-gluon operators into the corresponding interactions of SHB$\chi$PT.  As anticipated in~\cite{Lee:1999zx}, the \order{\asq} Lagrangian breaks the continuum spin-taste symmetry down to the hypercubic group of the lattice.  However, in the rest frame of the heavy baryons, a new remnant spin-taste symmetry eliminates \order{\asq} mixing between spin-\ohf\ and spin-\thf\ baryons.  For calculations to third order, this symmetry allows one to consider the baryon mass matrix in a smaller spin-flavor-taste subspace.  

The nucleon operators currently in use interpolate to baryons that are completely symmetric in flavor~\cite{Aubin:2004wf}.  In Sec.~\ref{id} I recall the symmetry considerations used in~\cite{Bailey:2005ss,Bailey:2006zn} to identify staggered baryons that are degenerate with the nucleon in the continuum limit and to identify the irreducible representations (irreps) of interpolating fields for the flavor-symmetric nucleons.  Sec.~\ref{matrix} contains a discussion of the symmetries of the various terms in the staggered chiral expansion of the masses of the flavor-symmetric nucleons and the implied form of the mass matrix.  Sec.~\ref{mass} contains the staggered chiral forms for the masses of the flavor-symmetric nucleons in partially quenched and fully dynamical SHB$\chi$PT.  In Sec.~\ref{connect}, I show that the pattern of symmetry breaking furnishes the connection between the staggered chiral forms and the irreps of the corresponding interpolating fields; the reasoning is the same as that used to establish the connection between continuum states and lattice irreps~\cite{Bailey:2006zn,Golterman:1984dn}.  Sec.~\ref{sum} summarizes and discusses future work.  

\section{\label{la}The Staggered Heavy Baryon Lagrangian}
Counting $a^2$ as \order{m_q}, where $m_q$ is any of the light quark masses, the staggered chiral power counting is a straightforward generalization of the power counting of the continuum heavy baryon chiral perturbation theory~\cite[pp.~360-362,~476-477]{Scherer:2002tk}.  The staggered heavy baryon Lagrangian is expanded in increasing powers of the quark masses, the squared lattice spacing, and derivatives of the heavy baryons and pseudoscalar mesons.  The baryon masses are calculated in increasing powers of the square roots of the quark masses, the residual off-shell momentum of the heavy baryon, the momenta of the pseudoscalar mesons, the squared lattice spacing, and the average mass splitting between the lightest spin-\ohf\ and spin-\thf\ baryons.  For convenience \[\order{\varepsilon}\equiv\order{m_q^{1/2}}=\order{k}=\order{a}=\order{\Delta},\] where $k$ is the residual heavy baryon momentum or pseudoscalar meson momentum and $\Delta$ is the octet-decuplet mass splitting in the continuum and chiral limits.  

In the continuum theory, the leading order (LO) heavy baryon Lagrangian is of \order{\varepsilon}, while the LO meson Lagrangian is of \order{\varepsilon^2}; the next-to-leading order (NLO) heavy baryon Lagrangian is of \order{\varepsilon^2}, while the NLO meson Lagrangian is of \order{\varepsilon^4}, and so on.  First order contributions to the baryon self-energies do not arise, while tree-level contributions of \order{\varepsilon^3} vanish.  One-loop diagrams first enter the expansion at \order{\varepsilon^3} and can only include vertices from the LO Lagrangian; two-loop diagrams first enter at \order{\varepsilon^5}.  Tree-level contributions at \order{\varepsilon^3} vanish.  Contributions at \order{\varepsilon^2} can contain at most one vertex from the NLO heavy baryon Lagrangian, together with vertices from the LO heavy baryon and meson Lagrangians; in fact, no diagrams with vertices from the LO Lagrangians arise.  

Nontrivially, the situation in SHB$\chi$PT is the same.  The staggered symmetries forbid corrections to the Symanzik action of \order{a} and \order{a^3}~\cite{Sharpe:1993ng,Sharpe:2004is}, so the power counting remains the same through NLO.  Calculating the baryon masses to \order{\varepsilon^3} requires at most the LO heavy baryon and meson Lagrangians for one-loop diagrams, which are \order{\varepsilon^3}, and the NLO heavy baryon Lagrangian for \order{\varepsilon^2} tree-level diagrams.  Tree-level contributions to the baryon masses do not arise at \order{\varepsilon} and \order{\varepsilon^3}.  To derive the LO heavy baryon and meson Lagrangians and the NLO heavy baryon Lagrangian needed for \order{\varepsilon^2} tree-level corrections, we require the Symanzik action through \order{\asq}.  

In partially quenched staggered QCD~\cite{Lee:1999zx,Aubin:2003mg}, the Symanzik action to \order{\asq} is
\begin{eqnarray*}
S_{\mr{eff}}&=&S_4+a^2S_6\\
&=& \int d^4x({\mathcal L}_4+a^2{\mathcal L}_6),\\
\text{where}\quad\quad\la_4&=&\ohf \mr{Tr}(F_{\mu\nu}F_{\mu\nu})+\overline{Q}(\slashed{D}+m)Q\\
\text{and}\quad\quad\la_6&=&\la_6^{\mr{glue}}+\la_6^{\mr{bilin}}+\la_6\ffa+\la_6\ffb.
\end{eqnarray*}
$\la_6^{\mr{glue}}$ contains terms constructed from gluon fields only, $\la_6^{\mr{bilin}}$ contains quark bilinears, and $\la_6^{FF(A,B)}$ contain four-fermion operators constructed from products of quark bilinears.  The terms in $\la_6^{FF(A)}$ respect the remnant taste symmetry \rt~\cite{Lee:1999zx}, while the terms in $\la_6^{FF(B)}$ break the continuum rotation-taste symmetry down to the hypercubic group of the staggered lattice.  

Because the staggered formulation introduces four quark tastes for each physical quark flavor, the chiral symmetry of $\la_4$ is $SU(24|12)_L\times SU(24|12)_R$.  There are twelve valence quarks, twelve sea quarks, and twelve ghost quarks; $Q$ is a 36-element column vector in flavor-taste space.  The quark mass matrix is \[m=\mr{diag}(m_x,m_y,m_z;m_u,m_d,m_s;m_x,m_y,m_z)\otimes I_4,\] where $m_{x,y,z}$ are the masses of the valence quarks, $m_{u,d,s}$, the masses of the sea quarks, and $I_4$ is the identity in taste space.  The masses of the ghost quarks are set equal to the masses of the valence quarks so that valence quarks do not contribute to the fermion determinant.

In the fully dynamical limit, one can drop the graded formalism and consider a Symanzik theory with the chiral symmetry $SU(12)_L\times SU(12)_R$.  For the fully dynamical case with $N$ physical flavors, the symmetries of the various terms in $S_4$ and $S_6$ are given in detail in the Appendix of~\cite{Aubin:2003mg}.  The fourth root of the fermion determinant is implemented in S$\chi$PT calculations by the method of quark flows~\cite{Aubin:2003mg,Labrenz:1996jy} or, equivalently, the replica method~\cite{Damgaard:2000gh,Bernard:2006zw}.  In the continuum limit, the resulting chiral forms are equivalent to the forms that result by considering a partially quenched theory with only three flavors of sea quarks.  

The development of S$\chi$PT in the baryon sector proceeds as in the meson sector~\cite{Aubin:2003mg,Aubin:2003uc}.  In addition to the pseudoscalar mesons, the Lagrangian now includes the lightest spin-\ohf\ and spin-\thf\ baryons.  In the fully dynamical continuum chiral theory~\cite{Jenkins:1990jv}, the lightest spin-\ohf\ baryons transform in the adjoint of \su{3}{F}, so they can be collected in a $3\times 3$ matrix in flavor space:
\begin{eqnarray*}
B_i^{\phantom{i}j}=\left( \begin{array}{ccc}
\oneortwo\Sigma^0+\oneorsix\Lambda & \Sigma^+ & p \\
\Sigma^- & -\oneortwo\Sigma^0+\oneorsix\Lambda & n \\
\Xi^- & \Xi^0 & -\twoorsix\Lambda \end{array} \right).
\end{eqnarray*}
In the quenched and partially quenched continuum chiral theories~\cite{Labrenz:1996jy,Chen:2001yi}, the lightest spin-\ohf\ baryons transform within mixed-symmetry irreps of the appropriate graded symmetry groups; the presence of more than three light quark flavors implies that the baryons no longer transform in the adjoint.  The above matrix parameterization remains valid only in restricted subsectors of the flavor space.  The generalization to an arbitrary number of flavors involves embedding the independent fields in a rank-3 mixed-symmetry tensor of the graded group, $B_{ijk}$.  The embedding respects the defining indicial symmetries of this tensor but is otherwise arbitrary; any convenient basis of the graded irrep suffices.  In the fully dynamical case (alternatively, restricting the indices to fermionic quarks), the standard defining relations of $B$ are
\begin{eqnarray}
B_{ikj}=B_{ijk}\quad\mr{and}\quad B_{ijk}+B_{jki}+B_{kij}=0.\label{Msym}
\end{eqnarray}
Under the chiral symmetry group, the spin-\ohf\ baryon tensor transforms as
\begin{eqnarray*}
B_{ijk}\rightarrow U_i^{\phantom{i}l}U_j^{\phantom{j}m}U_k^{\phantom{k}n}B_{lmn},
\end{eqnarray*}
where $U$ is a local, nonlinear realization of the chiral symmetry defined by the transformation law of the meson fields~\cite{Jenkins:1990jv,Georgi:1985kw,Aubin:2005aq,Scherer:2002tk}
\begin{eqnarray*}
\sigma\equiv {\sqrt \Sigma}\rightarrow \sigma^\prime=L\sigma U^\dag=U\sigma R^\dag.
\end{eqnarray*}

The tastes of S$\chi$PT are effectively additional quark flavors, so the situation in S$\chi$PT is similar to that in the quenched and partially quenched continuum theories:  Even in the fully dynamical case, one makes use of the rank-3 mixed-symmetry tensor in Eq.~(\ref{Msym}).  If the indices range from $1$ to $N$, then the number of independent components is $\frac{1}{3}N(N^2-1)$.  For three quark flavors, $N=12$, so the lightest spin-\ohf\ baryons transform within a $\mb{572_M}$ of \su{12}{f}, the diagonal (vector) subgroup of $SU(12)_L\times SU(12)_R$.

The lightest spin-\thf\ baryons are incorporated in S$\chi$PT similarly.  In the fully dynamical case, the independent fields are embedded in a completely symmetric rank-3 tensor, $T_{ijk}$:
\begin{eqnarray*}
&&T_{ijk}=T_{jki}=T_{kij}=T_{jik}=T_{ikj}=T_{kji},\\
&&T_{ijk}\rightarrow U_i^{\phantom{i}l}U_j^{\phantom{j}m}U_k^{\phantom{k}n}T_{lmn}.
\end{eqnarray*}
The dimension of the symmetric irrep is $\frac{1}{6}N(N+1)(N+2)$, so for three quarks, the lightest spin-\thf\ baryons transform within a $\mb{364_S}$ of \su{12}{f}.  

So far the spacetime indices of the heavy baryon fields have been suppressed.  $B$ is a Dirac spinor, while $T$ is a spin-\thf\ Rarita-Schwinger field and therefore carries a spinor index and a vector index.  The anti-particle components of both fields are projected out in an arbitrary but specific Lorentz frame, and the relativistic chiral Lagrangian is expanded in increasing powers of 1/$m_B$, where $m_B$ is the average octet baryon mass in the chiral limit~\cite{Jenkins:1990jv,Jenkins:1991ne}.  The resulting heavy baryon Lagrangian contains no explicit reference to the Dirac matrices; they are replaced by the classical baryon four-velocity $v_\mu$ and the covariant spin vector $S_\mu\equiv\frac{i}{2}\gamma_5\sigma_{\mu\nu}v^\nu$~\cite{Jenkins:1991ne,Scherer:2002tk}.  For convenience in making contact with the literature on the continuum theory, I use Minkowski space notation throughout; accordingly, the Lorentz indices here and below take values from $\{0,\ 1,\ 2,\ 3\}$.  The Green's functions of SHB$\chi$PT are defined in Euclidean space; a Wick rotation is implicit throughout.

\subsection{\order{a^0} Lagrangian}
In the continuum chiral theory~\cite{Jenkins:1991ne,Chen:2001yi}, the heavy baryon Lagrangian to \order{\varepsilon^2} is constructed of all Hermitian operators containing one derivative or quark mass matrix that respect chiral symmetry, Lorentz invariance, parity, and, before non-relativistic reduction, charge conjugation.  The building blocks are the heavy baryon fields, the covariant derivatives $D_\mu$ of the heavy baryon fields, the mesonic axial vector current $A_\mu\equiv\frac{i}{2}(\sigma\partial_\mu\sigma^{\dag}-\sigma^{\dag}\partial_\mu\sigma)$, the mass matrices $M_\pm\equiv\ohf(\sigma^{\dag}m\sigma^{\dag}\pm\sigma m\sigma)$, the four-vectors $v_\mu$ and $S_\mu$, and the Lorentz-invariant tensors, $g_{\mu\nu}$ and $\epsilon^{\mu\nu\alpha\beta}$.  In the heavy baryon Lagrangian, the vector current $V_\mu\equiv\ohf(\sigma\partial_\mu\sigma^{\dag}+\sigma^{\dag}\partial_\mu\sigma)$ and derivatives of the heavy baryon fields only appear together in the covariant derivative $D_\mu$; the explicit definition of $D_\mu$ is given in~\cite{Chen:2001yi} but will not be required here.  For calculating the baryon masses to \order{\varepsilon^3}, the covariant derivative may be replaced with the partial derivative $\partial_\mu$, $M_-$ does not contribute, and $M_+$ may be replaced with the quark mass matrix, $m$.  The meson fields therefore enter the required operators in the heavy baryon Lagrangian only through $A_\mu$.  

In S$\chi$PT, operators mapped from $S_4$ obey all the symmetries of the continuum theory (after Wick rotation) except for the chiral symmetries, which are simply enlarged to the appropriate flavor-taste group:
\begin{eqnarray*}
\sulxsur{6|3}&\rightarrow&\sulxsur{24|12}\\
&\mr{or}&\\
\sulxsur{3}&\rightarrow&\sulxsur{12}
\end{eqnarray*}
in the partially quenched or fully dynamical cases, respectively.  It follows that the staggered heavy baryon Lagrangian through \order{\varepsilon^2} mapped from $\la_4$ has the same form as the continuum Lagrangian through \order{\varepsilon^2}~\cite{Chen:2001yi}, where the flavor indices in the latter are simply replaced by flavor-taste indices:
\begin{eqnarray*}
\la_4&\rightarrow&\la_{\phi B}^{(1)}+\la_{\phi B}^{(2)}+\dots\\
\la_{\phi B}^{(1)}&\equiv&\overline B iv^{\mu}D_{\mu}B+2\alpha \overline B S^{\mu}BA_{\mu}+2\beta \overline B S^{\mu}A_{\mu}B\\
&-&\overline T^{\nu}(iv^{\mu}D_{\mu}-\Delta)T_{\nu}+\sqrt{\thf}C(\overline T^{\mu}A_{\mu}B+\overline B A_{\mu}T^{\mu})\\
&+&2H\overline T^{\nu}S^{\mu}A_{\mu}T_{\nu}\\
\la_{\phi B}^{(2)}&\equiv&2\alpha_M\overline B BM_++2\beta_M\overline B M_+B+2\sigma_M^{\prime} \overline B B \mathrm{str}(M_+)\\
&+&2\gamma_M\overline T^{\mu}M_+T_{\mu}-2\overline\sigma_M^{\prime}\overline T^{\mu}T_{\mu}\mr{str}(M_+)+\dots\\
\end{eqnarray*}
where the notation is closely based on that of~\cite{Lee:1999zx,Labrenz:1996jy,Chen:2001yi}; $\alpha$, $\beta$, $\alpha_M$, $\beta_M$, $C$, $H$, $\gamma_M$, $\sigma_M^{\prime}$, and $\overline\sigma_M^{\prime}$ are low-energy couplings (LECs), $\Delta$ is the average \mb{572_M}-\mb{364_S} mass splitting in the continuum and chiral limits, and ``\mr{str}'' denotes the supertrace.  The couplings $\sigma_M^{\prime}$ and $\overline\sigma_M^{\prime}$ are normalized so that the supertrace terms reproduce the continuum tree-level contributions to the baryon masses in the continuum limit; $\sigma_M^{\prime}=\frac{1}{4}\sigma_M$ and $\overline\sigma_M^{\prime}=\frac{1}{4}\overline\sigma_M$.  The neglected terms in $\la_{\phi B}^{(2)}$ are higher order in $1/m_B$, contain two derivatives, or are \order{\asq}; through \order{\varepsilon^3}, only the terms of \order{\asq} contribute to the masses of the lightest spin-\ohf\ baryons.  Of the operators listed above, those that contribute are
\begin{eqnarray}
\la_{\phi B}^{(1)\prime}&=&\overline B iv^{\mu}\partial_{\mu}B+\frac{\alpha}{f}\overline B S^{\mu}B\partial_{\mu}\phi+\frac{\beta}{f} \overline B S^{\mu}(\partial_{\mu}\phi) B\label{hblag}\\
&-&\overline T^{\nu}(iv^{\mu}\partial_{\mu}-\Delta)T_{\nu}+\sqrt{\thf}\frac{C}{2f}(\overline T^{\mu}(\partial_{\mu}\phi) B+\overline B (\partial_{\mu}\phi) T^{\mu})\nonumber\\
\la_{\phi B, m}^{(2)\prime}&=&2\alpha_M\overline B Bm+2\beta_M\overline B mB+2\sigma_M^{\prime} \overline B B \mathrm{str}(m)\nonumber
\end{eqnarray}
where the definitions of $M_+$, $A_\mu$, and $\sigma=e^{i\phi/(2f)}=1+\frac{i\phi}{2f}+\order{\phi^2}$~\cite{Aubin:2003mg} were used; $f$ is the pion decay constant in the continuum and chiral limits.  Terms with explicit factors of $\phi$ contribute vertices for the leading loops, and terms with explicit factors of $m$ contribute analytically at \order{\varepsilon^2}.

\subsection{\order{\asq} baryon operators for analytic corrections}
In addition to the chiral operators in $\la_{\phi B}^{(1)}$ and $\la_{\phi B}^{(2)}$ mapped from $\la_4$, one must find the chiral operators in $\la_{\phi B}^{(2)}$ corresponding to $\la_6$.  Here I consider only the subset of these operators needed to calculate irrep-dependent corrections to the masses of the lightest spin-\ohf\ baryons through \order{\varepsilon^3}; through this order, such operators contribute analytic tree-level corrections of \order{\asq}.  I neglect operators contributing only generic \order{\asq} corrections to the LECs in the Lagrangian mapped from $\la_4$ and operators that contribute only to the masses of spin-\thf\ baryons.  

Operators containing the quark mass matrix, the axial vector current, and derivatives of the heavy baryon fields can be ignored because such operators first contribute at $\order{\asq m_q}=\order{\varepsilon^4}$.  Tree-level corrections come only from operators that are bilinear in the heavy baryon fields and have terms that are free of meson fields.  The previous set of building blocks is thus reduced to the heavy baryon fields, the four-vectors $v_\mu$ and $S_\mu$, and the Lorentz-invariant tensors, $g_{\mu\nu}$ and $\epsilon^{\mu\nu\alpha\beta}$.  In addition to these building blocks, one may promote the taste matrices to spurion fields and use the meson field $\sigma$ to construct objects that transform in the adjoint under vector (chiral diagonal) flavor-taste transformations.  Expanding $\sigma=1+\order{\phi}$ at the end of the enumeration, one recovers the Lagrangian that contributes at tree-level.  Finally, the fact that the heavy baryon Lagrangian may be obtained as the non-relativistic limit of the relativistic chiral theory implies that $S_\mu$ occurs at most once in each baryon bilinear, while it suffices for the present purposes to consider operators in which $v_\mu$ occurs at most once for each heavy baryon field.  

Three cases arise.  Chiral operators corresponding to $\la_6^{\mr{glue}}$ and $\la_6^{\mr{bilin}}$ respect all the symmetries of the LO Lagrangian except for spacetime rotation invariance, which is broken to the hypercubic rotations~\cite{Lee:1999zx,Aubin:2003mg}.  
Chiral operators corresponding to $\la_6\ffa$ break chiral flavor-taste to the remnant flavor-taste group; in the fully dynamical case, we have \[U(3)_l\times U(3)_r\times [\rt],\] where $U(3)_l\times U(3)_r$ is the analog of the $U(1)_{\mr{vec}}\times U(1)_A$ of the single-flavor case~\cite{Aubin:2003mg}.  These operators are invariant under all other symmetries of the LO Lagrangian, including spacetime rotations.  Allowing the taste matrices to transform as spurions, these operators also respect chiral flavor-taste transformations.  Finally, chiral operators corresponding to $\la_6\ffb$ break chiral flavor-taste and spacetime rotations to the symmetry group of the lattice theory~\cite{Lee:1999zx,Sharpe:2004is}.  Allowing the taste matrices to transform appropriately, one can work with all the symmetries except spacetime rotations, which are broken to the hypercubic rotations.  We will consider each of these cases in turn.  
\subsubsection{$\la_6^{\mr{glue}}$ and $\la_6^{\mr{bilin}}$}
The chiral operators corresponding to $\la_6^{\mr{glue}}$ and $\la_6^{\mr{bilin}}$ contribute nothing but a generic \order{\asq} correction to the average spin-\ohf\ baryon mass in the chiral limit.  The Symanzik operators contain no taste matrices, so there are no spurions in the corresponding chiral operators.  Displaying flavor-taste indices explicitly, the only possibilities that respect parity are, up to additional factors of $v_{\mu}$,
\begin{eqnarray*}
&&\overline B^{ijk}B_{ijk}v^{\mu}v_{\mu}g_{\mu\mu}\\
&&(\overline B^{ijk}T_{\mu,ijk}+\overline T_{\mu}^{ijk}B_{ijk})v^{\mu}g_{\mu\mu}\\
&&\overline T_{\mu}^{ijk}T_{\mu,ijk}v^{\mu}v_{\mu}\\
&&\overline T_{\mu}^{ijk}T^{\mu}_{ijk}g_{\mu\mu}
\end{eqnarray*}
where the repeated index $\mu$ is summed from $0$ to $3$.  Chiral operators containing $S_\mu$ do not arise because it is an axial vector; bilinears containing it are odd under parity.  The second operator is constructed so that it is invariant under charge conjugation.  One Lorentz index is raised in each operator because I have Wick rotated to Minkowski space.  

The first term is a velocity-dependent correction to the average spin-\ohf\ baryon mass in the chiral and continuum limits.  The underlying lattice theory is not Lorentz invariant, so there is no prohibition against such corrections.  The second term mixes spin-\ohf\ and spin-\thf\ baryons; the continuum spin and taste symmetries are broken by $\la_6$.  The last two terms are generic corrections to the average difference $\Delta$ between the masses of the spin-\ohf\ and spin-\thf\ baryons.  Specializing to the rest frame of the heavy baryon, $v=(1,\mb{0})$, and only the first and last terms survive.  The other two vanish because~\cite{Jenkins:1991ne}
\begin{eqnarray*}
&v_{\mu}T^{\mu}=0,\;\forall\>v\;\Longrightarrow\;T^0=0\;\mr{for}\>v=(1,\mb{0}).&
\end{eqnarray*}
For $\mb{v=0}$, the surviving terms respect all the symmetries of the LO heavy baryon Lagrangian, including invariance under spatial rotations.  The spacetime symmetry is restored for this special case.  Operators with additional factors of $v_\mu$ yield nothing new; such operators either vanish or reduce to the first type.  

The fact that spin-\ohf\ and spin-\thf\ baryons do not mix in the rest frame considerably simplifies, in this frame, the analysis of Sec.~\ref{spec} and calculations of the baryon masses.  Moreover, simulations performed with zero momentum sources must be analyzed in this frame.  Accordingly, the baryon masses given in Sec.~\ref{mass} are specific to this case.  
\subsubsection{$\la_6\ffa$}
There are twelve types of operators in $\la_6\ffa$; following~\cite{Lee:1999zx,Sharpe:2004is}, I label these with the spacetime-taste irreps in which the quark bilinears transform:
\begin{subequations}\label{L_6^ffa}
\begin{eqnarray}
\la_6\ffa&\sim&\bigl([V\times S]+[V\times P]+[V\times T]+(V\to A)\bigr)\label{VA}\\
&+&\bigl([S\times V]+[S\times A]+(S\to P)\bigr)\label{SP}\\
&+&[T\times V]+[T\times A].\label{T}
\end{eqnarray}
\end{subequations}
In these operators, the indices of the spin and taste matrices are contracted separately; e.g., \[[A\times T]\equiv\sum_{\mu<\nu}\sum_\rho\overline Q(\gamma_{\rho5}\otimes \xi_{\mu\nu})Q\>\overline Q(\gamma_{5\rho}\otimes \xi_{\nu\mu})Q.\]
Here $\xi_{\mu\nu}$ is shorthand for $I_9\otimes\xi_{\mu\nu}$, where $I_9$ is the identity matrix in flavor space.  We recall the definitions of the taste matrices:
\[\xi_{\mu\nu}\equiv i\xi_{\mu}\xi_{\nu}\;\text{and}\;\xi_{\rho5}\equiv i\xi_{\rho}\xi_5,\;\text{where}\;\xi_5\equiv \xi_1\xi_2\xi_3\xi_4,\] and similarly for the spin matrices.  In Sec.~\ref{matrix} we will specialize to the Weyl basis for the taste matrices, in which
\begin{equation}\label{Weyl}
\bs{\xi}={\begin{pmatrix}
0 & -i\bs{\sigma} \\
i\bs{\sigma} & 0
\end{pmatrix}}\quad\quad\text{and}\quad\quad
\xi_4={\begin{pmatrix}
0 & I_2 \\
I_2 & 0
\end{pmatrix}}.
\end{equation}
For now, the taste basis is arbitrary; different choices correspond to different interpretations of the underlying staggered degrees of freedom in terms of continuum taste degrees of freedom.

Promoting the taste matrices to spurion fields and displaying the chiral structures of these operators explicitly, one can write each of the operators in $\la_6\ffa$ in one of three distinct forms~\cite{Lee:1999zx,Sharpe:2004is}.  The six operators with vector and axial spin structure~(\ref{VA}) can each be written in the form
\begin{eqnarray}
\mc{O}_F=\pm\sum_\mu\left(\overline Q_R(\gamma_\mu\otimes F_R)Q_R\pm\overline Q_L(\gamma_\mu\otimes F_L)Q_L\right)^2,\label{VAform}
\end{eqnarray}
where $F_{L,R}$ are spurions transforming so that $\mc{O}_F$ is invariant under chiral transformations, Euclidean rotations, parity, and charge conjugation:
\begin{eqnarray}
\sulxsur{N}:&\;&F_L\rightarrow LF_LL^{\dag}\label{VAtrans}\\
&\;&F_R\rightarrow RF_RR^{\dag}\nonumber\\
\sofour_E:&\;&F_L\rightarrow F_L\nonumber\\
&\;&F_R\rightarrow F_R\nonumber\\
\mc{P}:&\;&F_L\leftrightarrow F_R\nonumber\\
\mc{C}:&\;&F_L\leftrightarrow F_R^T.\nonumber
\end{eqnarray}
The four operators with scalar and pseudoscalar spin structure~(\ref{SP}) can be written in the form
\begin{eqnarray}
\mc{O}_F^{\prime}=\left(\overline Q_L(I_4\otimes\widetilde F_L)Q_R\pm\overline Q_R(I_4\otimes\widetilde F_R)Q_L\right)^2,\label{SPform}
\end{eqnarray}
where the spurions $\widetilde F_{L,R}$ transform as follows:
\begin{eqnarray}
\sulxsur{N}:&\;&\widetilde F_L\rightarrow L\widetilde F_LR^{\dag}\label{Ttrans}\\
&\;&\widetilde F_R\rightarrow R\widetilde F_RL^{\dag}\nonumber\\
\sofour_E:&\;&\widetilde F_L\rightarrow \widetilde F_L\nonumber\\
&\;&\widetilde F_R\rightarrow \widetilde F_R\nonumber\\
\mc{P}:&\;&\widetilde F_L\leftrightarrow \widetilde F_R\nonumber\\
\mc{C}:&\;&\widetilde F_L\rightarrow \widetilde F_L^T\nonumber\\
&\;&\widetilde F_R\rightarrow \widetilde F_R^T.\nonumber
\end{eqnarray}
The two tensor operators~(\ref{T}) can be written
\begin{eqnarray}
\mc{O}_F^{\prime\prime}=\sum_{\mu<\nu}\left[\left(\overline Q_L(\gamma_{\mu\nu}\otimes\widetilde F_L)Q_R\right)^2+\left(\overline Q_R(\gamma_{\nu\mu}\otimes\widetilde F_R)Q_L\right)^2\right],\label{Tform}
\end{eqnarray}
where the spurions are the same as in $\mc{O}_F^{\prime}$.  

For convenience in mapping to the heavy baryon theory, I use the meson fields $\sigma$ to construct objects with the spurions that transform in the adjoint of the vector subgroup of the chiral group.  Let
\begin{eqnarray*}
l\equiv\sigma^{\dag}F_L\sigma&\quad\mr{and}\quad&\tilde l\equiv\sigma^{\dag}\widetilde F_L\sigma^{\dag}\\
r\equiv\sigma F_R\sigma^{\dag}&\quad\quad&\tilde r\equiv\sigma\widetilde F_R\sigma.
\end{eqnarray*}
Then
\begin{eqnarray}
\sulxsur{N}:\;l\rightarrow UlU^{\dag}&\quad&\tilde l\rightarrow U\tilde lU^{\dag}\label{trans}\\
r\rightarrow UrU^{\dag}&\quad&\tilde r\rightarrow U\tilde rU^{\dag}\nonumber\\
\sofour_E:\;l\rightarrow l&\quad&\tilde l\rightarrow \tilde l\nonumber\\
r\rightarrow r&\quad&\tilde r\rightarrow \tilde r\nonumber\\
\mc{P}:\;l\leftrightarrow r&\quad&\tilde l\leftrightarrow \tilde r\nonumber\\
\mc{C}:\;l\leftrightarrow r^T&\quad&\tilde l\rightarrow \tilde l^T\nonumber\\
&&\tilde r\rightarrow \tilde r^T.\nonumber
\end{eqnarray}
The spurion-meson objects $l$, $r$, $\tilde l$, and $\tilde r$ can be further combined to construct definite-parity objects that are Hermitian:
\begin{eqnarray}
l\otimes l\pm r\otimes r&\quad\mr{and}\quad& l\otimes r\pm r\otimes l\label{lrs}\\
\tilde l\otimes \tilde l+\tilde r\otimes\tilde r&\quad\quad&\tilde l\otimes \tilde r+\tilde r\otimes\tilde l\nonumber\\
-i(\tilde l\otimes \tilde l-\tilde r\otimes\tilde r)&\quad\quad&-i(\tilde l\otimes \tilde r-\tilde r\otimes\tilde l)\nonumber
\end{eqnarray}
The objects involving $\tilde l$ and $\tilde r$ are effectively Hermitian because one sets the left- and right-handed spurion fields equal to one another at the end of the enumeration, which implies $\tilde l^{\dag}=\tilde r$.  This relation also implies that the above list is complete.  Moreover, in constructing the chiral operators, the only axial vector that is allowed is $S_\mu$, which can occur once.  However, $\la_6\ffa$ is Lorentz invariant, and $v\cdot S=0$ is the only Lorentz invariant containing $S_\mu$ that can be constructed.  Therefore, $S_\mu$ does not appear in the chiral operators corresponding to $\la_6\ffa$, and only the parity-even spurion-meson combinations in~(\ref{lrs}) are allowed.  Lorentz invariance also implies that operators mixing the spin-\thf\ fields $T_\mu$ and the spin-\ohf\ fields do not appear; to be Lorentz invariant, such operators at this order must contain either $v\cdot T$ or $S\cdot T$, but both vanish.

\textit{A priori}, chirally invariant baryon bilinears may contain zero, one, two, or three pairs of contractions between the flavor-taste indices of the baryon fields and the flavor-taste indices of the meson fields.  Explicitly,
\begin{eqnarray}
\overline B^{ijk}WB_{ijk}\quad\quad\overline B^{ijk}X_i^nB_{njk}\quad\quad\overline B^{ijk}Y_{ij}^{nq}B_{nqk}\quad\quad\overline B^{ijk}Z_{ijk}^{nql}B_{nql},\label{cts}
\end{eqnarray}
where $W$, $X$, $Y$, and $Z$ are constructed out of the non-baryonic building blocks and are quadratic in the spurions.  However, it turns out that only operators with two pairs of contractions are relevant here.  The $W$- and $X$-type operators contribute only irrep-independent corrections to the masses, while $Z$-type operators only modify the LECs of the twice-contracted ($Y$-type) operators.  

First consider the $W$-type operators.  Since there are no contractions between the flavor-taste indices of the meson fields in $W$ and the baryon fields, $W$ can be any operator appearing in the remnant-taste symmetric potential $\mc{V}$ of the meson sector~\cite{Lee:1999zx,Aubin:2003mg}.  Setting $\Sigma=1+\order{\phi}$ in these terms, $\mc{V}$ collapses to a single constant.  Therefore, $W$-type operators contribute an \order{\asq}, irrep-independent (generic) correction to the masses of the baryons.

The $X$- and $Z$-type operators reduce similarly.  For example, consider
\begin{eqnarray*}
X_i^n\equiv (lr+rl)_i^{\phantom{i}n}\quad\mr{and}\quad Z_{ijk}^{nql}\equiv (l_i^{\phantom{i}n}r_j^{\phantom{j}q}+r_i^{\phantom{i}n}l_j^{\phantom{j}q})(\Sigma+\Sigma^{\dag})_k^{\phantom{k}l}.
\end{eqnarray*}
The taste matrices all square to the identity, so setting the spurions to their final values and $\sigma=1$ gives $X_i^n\propto\delta_i^n$, and the $X$-type operators reduce to $W$-type operators.  As for the $Z$-type operators, the fact that all operators must be quadratic in the spurions means that only $\Sigma$ can be used to provide a third set of indices in $Z$.  But $\Sigma$ is set to unity at the end of the enumeration, so these operators reduce to the twice-contracted form.  Therefore, all chiral operators leading to distinct spin-taste breaking (irrep-dependent) corrections to the baryon masses have two pairs of contractions between the flavor-taste indices on the baryon fields and those on the meson fields.  

Assuming the indicial symmetries in Eqs.~(\ref{Msym}), there are only two linearly independent, $Y$-type contraction structures:
\begin{eqnarray}
\overline B^{ijk}Y_{ij}^{nq}B_{nqk}&\quad \mr{and}\quad&\overline B^{ijk}Y_{ij}^{nq}B_{knq}.\label{ffconts}
\end{eqnarray}
Although the symmetries in Eqs.~(\ref{Msym}) are not generally valid for the partially quenched case, for which various minus signs must be inserted, only required are those \order{\asq} operators needed for the calculation of tree-level corrections to masses of baryons composed of valence quarks.  In this case, the graded indicial symmetries reduce to those given in Eqs.~(\ref{Msym}), so the latter are correct for the operators considered here, and the contraction structures~(\ref{ffconts}) exhaust the possibilities for the relevant operators.  

For operators with vector or axial spin structure~(\ref{VA},~\ref{VAform}), setting
\begin{eqnarray*}
Y_{ij}^{nq}=l_i^{\phantom{i}n}l_j^{\phantom{j}q}+r_i^{\phantom{i}n}r_j^{\phantom{j}q},\quad l_i^{\phantom{i}n}r_j^{\phantom{j}q}+r_i^{\phantom{i}n}l_j^{\phantom{j}q},
\end{eqnarray*}
inserting the scalar, pseudoscalar, and tensor taste matrices in the spurions, and introducing constants of proportionality gives
\begin{eqnarray*}
[V\times S]\ {\rm and}\ [A\times S]&\rightarrow&(-(\pm b_1+b_2)+2(\pm c_1+c_2)){\overline B}^{ijk}B_{ijk}
\end{eqnarray*}
\begin{eqnarray*}
[V\times P]\ {\rm and}\ [A\times P]&\rightarrow&\pm b_1{\overline B}^{ijk}B_{knq}\left[(\sigma^{\dag}\xi_5\sigma)_i^{\phantom{i}n}(\sigma^{\dag}\xi_5\sigma)_j^{\phantom{j}q}+(\sigma\xi_5\sigma^{\dag})_i^{\phantom{i}n}(\sigma\xi_5\sigma^{\dag})_j^{\phantom{j}q}\right]\\*
&+&b_2{\overline B}^{ijk}B_{knq}\left[(\sigma^{\dag}\xi_5\sigma)_i^{\phantom{i}n}(\sigma\xi_5\sigma^{\dag})_j^{\phantom{j}q}+(\sigma\xi_5\sigma^{\dag})_i^{\phantom{i}n}(\sigma^{\dag}\xi_5\sigma)_j^{\phantom{j}q}\right]\\*
&\pm&c_1{\overline B}^{ijk}B_{nqk}\left[(\sigma^{\dag}\xi_5\sigma)_i^{\phantom{i}n}(\sigma^{\dag}\xi_5\sigma)_j^{\phantom{j}q}+(\sigma\xi_5\sigma^{\dag})_i^{\phantom{i}n}(\sigma\xi_5\sigma^{\dag})_j^{\phantom{j}q}\right]\\*
&+&c_2{\overline B}^{ijk}B_{nqk}\left[(\sigma^{\dag}\xi_5\sigma)_i^{\phantom{i}n}(\sigma\xi_5\sigma^{\dag})_j^{\phantom{j}q}+(\sigma\xi_5\sigma^{\dag})_i^{\phantom{i}n}(\sigma^{\dag}\xi_5\sigma)_j^{\phantom{j}q}\right]
\end{eqnarray*}
\begin{eqnarray*}
[V\times T]\ {\rm and}\ [A\times T]&\rightarrow&\pm b_1{\overline B}^{ijk}B_{knq}\sum _{\mu<\nu}\left[(\sigma^{\dag}\xi_{\mu\nu}\sigma)_i^{\phantom{i}n}(\sigma^{\dag}\xi_{\nu\mu}\sigma)_j^{\phantom{j}q}+(\sigma\xi_{\mu\nu}\sigma^{\dag})_i^{\phantom{i}n}(\sigma\xi_{\nu\mu}\sigma^{\dag})_j^{\phantom{j}q}\right]\\*
&+&b_2{\overline B}^{ijk}B_{knq}\sum _{\mu<\nu}\left[(\sigma^{\dag}\xi_{\mu\nu}\sigma)_i^{\phantom{i}n}(\sigma\xi_{\nu\mu}\sigma^{\dag})_j^{\phantom{j}q}+(\sigma\xi_{\mu\nu}\sigma^{\dag})_i^{\phantom{i}n}(\sigma^{\dag}\xi_{\nu\mu}\sigma)_j^{\phantom{j}q}\right]\\*
&\pm&c_1{\overline B}^{ijk}B_{nqk}\sum _{\mu<\nu}\left[(\sigma^{\dag}\xi_{\mu\nu}\sigma)_i^{\phantom{i}n}(\sigma^{\dag}\xi_{\nu\mu}\sigma)_j^{\phantom{j}q}+(\sigma\xi_{\mu\nu}\sigma^{\dag})_i^{\phantom{i}n}(\sigma\xi_{\nu\mu}\sigma^{\dag})_j^{\phantom{j}q}\right]\\*
&+&c_2{\overline B}^{ijk}B_{nqk}\sum _{\mu<\nu}\left[(\sigma^{\dag}\xi_{\mu\nu}\sigma)_i^{\phantom{i}n}(\sigma\xi_{\nu\mu}\sigma^{\dag})_j^{\phantom{j}q}+(\sigma\xi_{\mu\nu}\sigma^{\dag})_i^{\phantom{i}n}(\sigma^{\dag}\xi_{\nu\mu}\sigma)_j^{\phantom{j}q}\right].
\end{eqnarray*}
Here the taste matrix $\xi_\tau$ is being used as shorthand for $I_3\otimes\xi_\tau$, where $I_3$ is the identity matrix in valence quark flavor space. 

For operators with scalar or pseudoscalar spin structure~(\ref{SP},~\ref{SPform}), setting
\begin{eqnarray*}
Y_{ij}^{nq}=\tilde l_i^{\phantom{i}n}\tilde l_j^{\phantom{j}q}+\tilde r_i^{\phantom{i}n}\tilde r_j^{\phantom{j}q},\quad\tilde l_i^{\phantom{i}n}\tilde r_j^{\phantom{j}q}+\tilde r_i^{\phantom{i}n}\tilde l_j^{\phantom{j}q},
\end{eqnarray*}
inserting the vector and axial taste matrices in the spurions, and introducing constants of proportionality gives
\begin{eqnarray*}
[S\times V]\ {\rm and}\ [P\times V]&\rightarrow&e_1{\overline B}^{ijk}B_{knq}\left[(\sigma^{\dag}\xi_\nu\sigma^{\dag})_i^{\phantom{i}n}(\sigma^{\dag}\xi_\nu\sigma^{\dag})_j^{\phantom{j}q}+(\sigma\xi_\nu\sigma)_i^{\phantom{i}n}(\sigma\xi_\nu\sigma)_j^{\phantom{j}q}\right]\\
&\pm&e_2{\overline B}^{ijk}B_{knq}\left[(\sigma^{\dag}\xi_\nu\sigma^{\dag})_i^{\phantom{i}n}(\sigma\xi_\nu\sigma)_j^{\phantom{j}q}+(\sigma\xi_\nu\sigma)_i^{\phantom{i}n}(\sigma^{\dag}\xi_\nu\sigma^{\dag})_j^{\phantom{j}q}\right]\\*
&+&f_1{\overline B}^{ijk}B_{nqk}\left[(\sigma^{\dag}\xi_\nu\sigma^{\dag})_i^{\phantom{i}n}(\sigma^{\dag}\xi_\nu\sigma^{\dag})_j^{\phantom{j}q}+(\sigma\xi_\nu\sigma)_i^{\phantom{i}n}(\sigma\xi_\nu\sigma)_j^{\phantom{j}q}\right]\\*
&\pm&f_2{\overline B}^{ijk}B_{nqk}\left[(\sigma^{\dag}\xi_\nu\sigma^{\dag})_i^{\phantom{i}n}(\sigma\xi_\nu\sigma)_j^{\phantom{j}q}+(\sigma\xi_\nu\sigma)_i^{\phantom{i}n}(\sigma^{\dag}\xi_\nu\sigma^{\dag})_j^{\phantom{j}q}\right]
\end{eqnarray*}
\begin{eqnarray*}
[S\times A]\ {\rm and}\ [P\times A]&\rightarrow&e_1{\overline B}^{ijk}B_{knq}\left[(\sigma^{\dag}\xi_{\nu5}\sigma^{\dag})_i^{\phantom{i}n}(\sigma^{\dag}\xi_{5\nu}\sigma^{\dag})_j^{\phantom{j}q}+(\sigma\xi_{\nu5}\sigma)_i^{\phantom{i}n}(\sigma\xi_{5\nu}\sigma)_j^{\phantom{j}q}\right]\\*
&\pm&e_2{\overline B}^{ijk}B_{knq}\left[(\sigma^{\dag}\xi_{\nu5}\sigma^{\dag})_i^{\phantom{i}n}(\sigma\xi_{5\nu}\sigma)_j^{\phantom{j}q}+(\sigma\xi_{\nu5}\sigma)_i^{\phantom{i}n}(\sigma^{\dag}\xi_{5\nu}\sigma^{\dag})_j^{\phantom{j}q}\right]\\*
&+&f_1{\overline B}^{ijk}B_{nqk}\left[(\sigma^{\dag}\xi_{\nu5}\sigma^{\dag})_i^{\phantom{i}n}(\sigma^{\dag}\xi_{5\nu}\sigma^{\dag})_j^{\phantom{j}q}+(\sigma\xi_{\nu5}\sigma)_i^{\phantom{i}n}(\sigma\xi_{5\nu}\sigma)_j^{\phantom{j}q}\right]\\*
&\pm&f_2{\overline B}^{ijk}B_{nqk}\left[(\sigma^{\dag}\xi_{\nu5}\sigma^{\dag})_i^{\phantom{i}n}(\sigma\xi_{5\nu}\sigma)_j^{\phantom{j}q}+(\sigma\xi_{\nu5}\sigma)_i^{\phantom{i}n}(\sigma^{\dag}\xi_{5\nu}\sigma^{\dag})_j^{\phantom{j}q}\right].
\end{eqnarray*}

Finally, for operators with tensor spin structure~(\ref{T},~\ref{Tform}), one proceeds as before except that, noting the form of $\mc{O}_F^{\prime\prime}$~(\ref{Tform}), one omits operators corresponding to cross-terms of left- and right-handed spurions.  Setting
\begin{eqnarray*}
Y_{ij}^{nq}=\tilde l_i^{\phantom{i}n}\tilde l_j^{\phantom{j}q}+\tilde r_i^{\phantom{i}n}\tilde r_j^{\phantom{j}q},
\end{eqnarray*}
inserting vector and axial taste matrices, and introducing constants of proportionality gives
\begin{eqnarray*}
[T\times V]&\rightarrow&e_1'{\overline B}^{ijk}B_{knq}\left[(\sigma^{\dag}\xi_\nu\sigma^{\dag})_i^{\phantom{i}n}(\sigma^{\dag}\xi_\nu\sigma^{\dag})_j^{\phantom{j}q}+(\sigma\xi_\nu\sigma)_i^{\phantom{i}n}(\sigma\xi_\nu\sigma)_j^{\phantom{j}q}\right]\\*
&+&f_1'{\overline B}^{ijk}B_{nqk}\left[(\sigma^{\dag}\xi_\nu\sigma^{\dag})_i^{\phantom{i}n}(\sigma^{\dag}\xi_\nu\sigma^{\dag})_j^{\phantom{j}q}+(\sigma\xi_\nu\sigma)_i^{\phantom{i}n}(\sigma\xi_\nu\sigma)_j^{\phantom{j}q}\right]
\end{eqnarray*}
\begin{eqnarray*}
[T\times A]&\rightarrow&e_1'{\overline B}^{ijk}B_{knq}\left[(\sigma^{\dag}\xi_{\nu5}\sigma^{\dag})_i^{\phantom{i}n}(\sigma^{\dag}\xi_{5\nu}\sigma^{\dag})_j^{\phantom{j}q}+(\sigma\xi_{\nu5}\sigma)_i^{\phantom{i}n}(\sigma\xi_{5\nu}\sigma)_j^{\phantom{j}q}\right]\\*
&+&f_1'{\overline B}^{ijk}B_{nqk}\left[(\sigma^{\dag}\xi_{\nu5}\sigma^{\dag})_i^{\phantom{i}n}(\sigma^{\dag}\xi_{5\nu}\sigma^{\dag})_j^{\phantom{j}q}+(\sigma\xi_{\nu5}\sigma)_i^{\phantom{i}n}(\sigma\xi_{5\nu}\sigma)_j^{\phantom{j}q}\right].
\end{eqnarray*}

Although these operators have been deduced within the heavy baryon framework, all the operators listed here do in fact arise when deriving the heavy baryon Lagrangian from the relativistic baryon chiral Lagrangian.  To confirm this assertion, one may map $\la_6\ffa$ to the relativistic chiral theory in Euclidean space, Wick rotate, and then execute the non-relativistic reduction.  For the operators listed above, this analysis has been carried out.  
\subsubsection{$\la_6\ffb$}
The development parallels that for $\la_6\ffa$.  There are four types of operators in $\la_6\ffb$~\cite{Lee:1999zx,Sharpe:2004is}: 
\begin{subequations}\label{L_6^ffb}
\begin{eqnarray}
{\cal L}_6^{FF(B)}&\sim&[V_\mu\times T_\mu]+[A_\mu\times T_\mu]\label{VAmu}\\
&+&[T_\mu\times V_\mu]+[T_\mu\times A_\mu].\label{Tmu}
\end{eqnarray}
\end{subequations}
In these operators, the indices of the spin and taste matrices are contracted together; e.g., \[[V_\mu\times T_\mu]\equiv\sum_\mu\sum_{\nu\neq\mu}\left[\overline Q(\gamma_{\mu}\otimes \xi_{\mu\nu})Q\>\overline Q(\gamma_{\mu}\otimes \xi_{\nu\mu})Q-\overline Q(\gamma_{\mu}\otimes \xi_{\mu\nu5})Q\>\overline Q(\gamma_{\mu}\otimes \xi_{5\nu\mu})Q\right].\]
Up to a taste singlet component that has no influence on the map to the chiral theory~\cite{Sharpe:2004is}, these four operators may each be written in one of two forms.  The two operators with vector and axial spin structure~(\ref{VAmu}) can be written in the form
\begin{eqnarray}
\mc{O}_F\ffb=\pm\sum_\mu\left(\overline Q_R(\gamma_\mu\otimes F_{R\mu})Q_R\pm\overline Q_L(\gamma_\mu\otimes F_{L\mu})Q_L\right)^2,\label{VAmuform}
\end{eqnarray}
where $(F_{L,R})_\mu$ are spurions transforming so that $\mc{O}_F\ffb$ is invariant under chiral transformations, Euclidean hypercubic rotations, parity, and charge conjugation.  Except for the Euclidean spacetime index on the spurions, $\mc{O}_F\ffb$ is the same as $\mc{O}_F$~(\ref{VAform}).  Therefore, excepting Euclidean rotations, the spurions transform as in~(\ref{VAtrans}); under Euclidean rotations, the spurions transform in the fundamental representation (rep):
\begin{eqnarray*}
\sofour_E:&\;&F_{L\mu}\rightarrow \Lambda_{\mu\nu}F_{L\nu}\\
&\;&F_{R\mu}\rightarrow \Lambda_{\mu\nu}F_{R\nu}.
\end{eqnarray*}
The two operators with tensor spin structure~(\ref{Tmu}) can be written in the form
\begin{eqnarray}
\mc{O}_F^{\prime\prime FF(B)}=\sum_\mu\sum_{\nu\not =\mu}\overline Q_L(\gamma_{\mu\nu}\otimes\widetilde F_{L\mu})Q_R\overline Q_R(\gamma_{\nu\mu}\otimes\widetilde F_{R\mu})Q_L,\label{Tmuform}
\end{eqnarray}
where, once again excepting Euclidean rotations, the spurions $(\widetilde F_{L,R})_\mu$ transform as in~(\ref{Ttrans}).  Under Euclidean rotations, the spurions again transform as vectors:
\begin{eqnarray*}
\sofour_E:&\;&\widetilde F_{L\mu}\rightarrow \Lambda_{\mu\nu}\widetilde F_{L\nu}\\
&\;&\widetilde F_{R\mu}\rightarrow \Lambda_{\mu\nu}\widetilde F_{R\nu}.
\end{eqnarray*}
For purposes of mapping to the heavy baryon theory, I again use the meson fields $\sigma$ to construct spurion-meson objects that transform in the adjoint of the vector subgroup of the chiral group.  Let
\begin{eqnarray*}
l_\mu\equiv\sigma^{\dag}F_{L\mu}\sigma&\quad\mr{and}\quad&\tilde l_\mu\equiv\sigma^{\dag}\widetilde F_{L\mu}\sigma^{\dag}\\
r_\mu\equiv\sigma F_{R\mu}\sigma^{\dag}&\quad\quad&\tilde r_\mu\equiv\sigma\widetilde F_{R\mu}\sigma.
\end{eqnarray*}
Then these objects transform as in~(\ref{trans}) but are to be treated as vectors under Euclidean rotations.  The Hermitian, definite-parity objects are constructed as before (cf.~(\ref{lrs})).  However, unlike the operators in $\la_6\ffa$, the operators in $\la_6\ffb$ break Lorentz invariance.  The argument given in the paragraph following~(\ref{lrs}) no longer applies, and the operators can contain both $S_\mu$ and parity-odd spurion-meson objects.  

\textit{A priori}, the chiral operators can have any of the contraction structures given in~(\ref{cts}).  However, the previous arguments against $X$- and $Z$-type operators go through unchanged, while $W$-type operators cannot contribute because they require derivatives to break Lorentz invariance and are therefore necessarily \order{\asq k^2}=\order{\varepsilon^4} or higher, in accord with the central result of~\cite{Lee:1999zx}.  Therefore only the two contraction structures given in~(\ref{ffconts}) enter.

For operators with vector or axial spin structure~(\ref{VAmu},~\ref{VAmuform}), setting
\begin{eqnarray*}
&Y_{ij}^{nq}=v_\mu v^{\mu}((l_\mu)_i^{\phantom{i}n}(l_\mu)_j^{\phantom{j}q}+(r_\mu)_i^{\phantom{i}n}(r_\mu)_j^{\phantom{j}q}),\quad v_\mu v^{\mu}((l_\mu)_i^{\phantom{i}n}(r_\mu)_j^{\phantom{j}q}+(r_\mu)_i^{\phantom{i}n}(l_\mu)_j^{\phantom{j}q}),&\\
&v^{\mu} S_\mu((l_\mu)_i^{\phantom{i}n}(r_\mu)_j^{\phantom{j}q}-(r_\mu)_i^{\phantom{i}n}(l_\mu)_j^{\phantom{j}q}),&
\end{eqnarray*}
inserting the tensor taste matrices in the spurions, introducing constants of proportionality, and summing the remaining tensor index (cf.~\cite{Sharpe:2004is}) gives
\begin{eqnarray*}
[V_\mu\times T_\mu]\ &{\rm and}&\ [A_\mu\times T_\mu]\rightarrow \\&\pm&h_1\sum _{\mu}\sum _{\nu\neq\mu}v_\mu v^{\mu}{\overline B}^{ijk}B_{knq}\left[(\sigma^{\dag}\xi_{\mu\nu}\sigma)_i^{\phantom{i}n}(\sigma^{\dag}\xi_{\nu\mu}\sigma)_j^{\phantom{j}q}+(\sigma\xi_{\mu\nu}\sigma^{\dag})_i^{\phantom{i}n}(\sigma\xi_{\nu\mu}\sigma^{\dag})_j^{\phantom{j}q}\right]\\*
&+&h_2\sum _{\mu}\sum _{\nu\neq\mu}v_\mu v^{\mu}{\overline B}^{ijk}B_{knq}\left[(\sigma^{\dag}\xi_{\mu\nu}\sigma)_i^{\phantom{i}n}(\sigma\xi_{\nu\mu}\sigma^{\dag})_j^{\phantom{j}q}+(\sigma\xi_{\mu\nu}\sigma^{\dag})_i^{\phantom{i}n}(\sigma^{\dag}\xi_{\nu\mu}\sigma)_j^{\phantom{j}q}\right]\\*
&+&h_2'\sum _{\mu}\sum _{\nu\neq\mu}v^{\mu}{\overline B}^{ijk}S_\mu B_{knq}\left[(\sigma^{\dag}\xi_{\mu\nu}\sigma)_i^{\phantom{i}n}(\sigma\xi_{\nu\mu}\sigma^{\dag})_j^{\phantom{j}q}-(\sigma\xi_{\mu\nu}\sigma^{\dag})_i^{\phantom{i}n}(\sigma^{\dag}\xi_{\nu\mu}\sigma)_j^{\phantom{j}q}\right]\\*
&\pm&p_1\sum _{\mu}\sum _{\nu\neq\mu}v_\mu v^{\mu}{\overline B}^{ijk}B_{nqk}\left[(\sigma^{\dag}\xi_{\mu\nu}\sigma)_i^{\phantom{i}n}(\sigma^{\dag}\xi_{\nu\mu}\sigma)_j^{\phantom{j}q}+(\sigma\xi_{\mu\nu}\sigma^{\dag})_i^{\phantom{i}n}(\sigma\xi_{\nu\mu}\sigma^{\dag})_j^{\phantom{j}q}\right]\\*
&+&p_2\sum _{\mu}\sum _{\nu\neq\mu}v_\mu v^{\mu}{\overline B}^{ijk}B_{nqk}\left[(\sigma^{\dag}\xi_{\mu\nu}\sigma)_i^{\phantom{i}n}(\sigma\xi_{\nu\mu}\sigma^{\dag})_j^{\phantom{j}q}+(\sigma\xi_{\mu\nu}\sigma^{\dag})_i^{\phantom{i}n}(\sigma^{\dag}\xi_{\nu\mu}\sigma)_j^{\phantom{j}q}\right]\\*
&+&p_2'\sum _{\mu}\sum _{\nu\neq\mu}v^{\mu}{\overline B}^{ijk}S_\mu B_{nqk}\left[(\sigma^{\dag}\xi_{\mu\nu}\sigma)_i^{\phantom{i}n}(\sigma\xi_{\nu\mu}\sigma^{\dag})_j^{\phantom{j}q}-(\sigma\xi_{\mu\nu}\sigma^{\dag})_i^{\phantom{i}n}(\sigma^{\dag}\xi_{\nu\mu}\sigma)_j^{\phantom{j}q}\right].
\end{eqnarray*}
Two points deserve additional comment.  First, the operator multiplying $h_2'$ is not Hermitian.  The culprit is the coincidence of the second contraction structure in~(\ref{ffconts}) with a spurion-meson object that is anti-symmetric under simultaneous interchange of the upper and lower flavor-taste indices.  This consideration did not arise in mapping $\la_6\ffa$ because parity and Lorentz invariance conspired to allow only symmetric spurion-meson objects.  The remedy is to find a Hermitian linear combination of the contraction structures given in~(\ref{ffconts}) and replace the operator multiplying $h_2'$ with its Hermitian counterpart.  For anti-symmetric spurion-meson objects, the sum of the operators multiplying $h_2'$ and $p_2'$ is anti-Hermitian.  Using the cyclic property in~(\ref{Msym}), one may take
\begin{eqnarray*}
i\overline B^{ijk}Y_{ij}^{nq}B_{qkn}
\end{eqnarray*}
as the second linearly independent contraction structure for anti-symmetric spurion-meson objects.  Then the operator multiplying $h_2'$ above is replaced with the Hermitian operator
\begin{eqnarray*}
\sum _{\mu}\sum _{\nu\neq\mu}v^{\mu} i{\overline B}^{ijk}S_\mu B_{qkn}\left[(\sigma^{\dag}\xi_{\mu\nu}\sigma)_i^{\phantom{i}n}(\sigma\xi_{\nu\mu}\sigma^{\dag})_j^{\phantom{j}q}-(\sigma\xi_{\mu\nu}\sigma^{\dag})_i^{\phantom{i}n}(\sigma^{\dag}\xi_{\nu\mu}\sigma)_j^{\phantom{j}q}\right].
\end{eqnarray*}
Second, at the level of the relativistic baryon chiral Lagrangian, charge conjugation invariance and Hermiticity conspire to forbid operators containing the spurion-meson object
\begin{eqnarray*}
Y_{ij}^{nq}=v^{\mu} S_\mu((l_\mu)_i^{\phantom{i}n}(l_\mu)_j^{\phantom{j}q}-(r_\mu)_i^{\phantom{i}n}(r_\mu)_j^{\phantom{j}q}),
\end{eqnarray*}
and such operators have been eliminated from the above list ($h_1'=p_1'\equiv0$).  For tree-level corrections ($\sigma=\sigma^{\dag}=1$) or quantities calculated in the rest frame ($\mb{v}=0$), this detail does not affect the results.  

For operators with tensor spin structure~(\ref{Tmu},~\ref{Tmuform}), one proceeds as before except that, noting the form of $\mc{O}_F^{\prime\prime FF(B)}$~(\ref{Tmuform}), one omits operators corresponding to direct-terms of left- and right-handed spurions.  Setting
\begin{eqnarray*}
Y_{ij}^{nq}=v_\mu v^{\mu}((\tilde l_\mu)_i^{\phantom{i}n}(\tilde r_\mu)_j^{\phantom{j}q}+(\tilde r_\mu)_i^{\phantom{i}n}(\tilde l_\mu)_j^{\phantom{j}q}),\quad -iv^{\mu} S_\mu((\tilde l_\mu)_i^{\phantom{i}n}(\tilde r_\mu)_j^{\phantom{j}q}-(\tilde r_\mu)_i^{\phantom{i}n}(\tilde l_\mu)_j^{\phantom{j}q}),
\end{eqnarray*}
inserting vector and axial taste matrices, and introducing constants of proportionality gives
\begin{eqnarray*}
[T_\mu\times V_\mu]&\rightarrow&s\sum _{\mu}v_\mu v^{\mu}{\overline B}^{ijk}B_{knq}\left[(\sigma^{\dag}\xi_\mu\sigma^{\dag})_i^{\phantom{i}n}(\sigma\xi_\mu\sigma)_j^{\phantom{j}q}+(\sigma\xi_\mu\sigma)_i^{\phantom{i}n}(\sigma^{\dag}\xi_\mu\sigma^{\dag})_j^{\phantom{j}q}\right]\\*
&+&s'\sum _{\mu}v^{\mu}{\overline B}^{ijk}S_\mu B_{qkn}\left[(\sigma^{\dag}\xi_\mu\sigma^{\dag})_i^{\phantom{i}n}(\sigma\xi_\mu\sigma)_j^{\phantom{j}q}-(\sigma\xi_\mu\sigma)_i^{\phantom{i}n}(\sigma^{\dag}\xi_\mu\sigma^{\dag})_j^{\phantom{j}q}\right]\\*
&+&t\sum _{\mu}v_\mu v^{\mu}{\overline B}^{ijk}B_{nqk}\left[(\sigma^{\dag}\xi_\mu\sigma^{\dag})_i^{\phantom{i}n}(\sigma\xi_\mu\sigma)_j^{\phantom{j}q}+(\sigma\xi_\mu\sigma)_i^{\phantom{i}n}(\sigma^{\dag}\xi_\mu\sigma^{\dag})_j^{\phantom{j}q}\right]\\*
&+&t'\sum _{\mu}-iv^{\mu}{\overline B}^{ijk}S_\mu B_{nqk}\left[(\sigma^{\dag}\xi_\mu\sigma^{\dag})_i^{\phantom{i}n}(\sigma\xi_\mu\sigma)_j^{\phantom{j}q}-(\sigma\xi_\mu\sigma)_i^{\phantom{i}n}(\sigma^{\dag}\xi_\mu\sigma^{\dag})_j^{\phantom{j}q}\right]
\end{eqnarray*}
\begin{eqnarray*}
[T_\mu\times A_\mu]&\rightarrow&s\sum _{\mu}v_\mu v^{\mu}{\overline B}^{ijk}B_{knq}\left[(\sigma^{\dag}\xi_{\mu5}\sigma^{\dag})_i^{\phantom{i}n}(\sigma\xi_{5\mu}\sigma)_j^{\phantom{j}q}+(\sigma\xi_{\mu5}\sigma)_i^{\phantom{i}n}(\sigma^{\dag}\xi_{5\mu}\sigma^{\dag})_j^{\phantom{j}q}\right]\\*
&+&s'\sum _{\mu}v^{\mu}{\overline B}^{ijk}S_\mu B_{qkn}\left[(\sigma^{\dag}\xi_{\mu5}\sigma^{\dag})_i^{\phantom{i}n}(\sigma\xi_{5\mu}\sigma)_j^{\phantom{j}q}-(\sigma\xi_{\mu5}\sigma)_i^{\phantom{i}n}(\sigma^{\dag}\xi_{5\mu}\sigma^{\dag})_j^{\phantom{j}q}\right]\\*
&+&t\sum _{\mu}v_\mu v^{\mu}{\overline B}^{ijk}B_{nqk}\left[(\sigma^{\dag}\xi_{\mu5}\sigma^{\dag})_i^{\phantom{i}n}(\sigma\xi_{5\mu}\sigma)_j^{\phantom{j}q}+(\sigma\xi_{\mu5}\sigma)_i^{\phantom{i}n}(\sigma^{\dag}\xi_{5\mu}\sigma^{\dag})_j^{\phantom{j}q}\right]\\*
&+&t'\sum _{\mu}-iv^{\mu}{\overline B}^{ijk}S_\mu B_{nqk}\left[(\sigma^{\dag}\xi_{\mu5}\sigma^{\dag})_i^{\phantom{i}n}(\sigma\xi_{5\mu}\sigma)_j^{\phantom{j}q}-(\sigma\xi_{\mu5}\sigma)_i^{\phantom{i}n}(\sigma^{\dag}\xi_{5\mu}\sigma^{\dag})_j^{\phantom{j}q}\right],
\end{eqnarray*}
where the procedure followed for the $h_2'$ operator above has already been performed; operators multiplying $s'$ have been replaced with their Hermitian counterparts.

As before, all the operators given here do arise when deriving the heavy baryon Lagrangian from the relativistic baryon chiral Lagrangian; I have mapped $\la_6\ffb$ to the relativistic chiral theory in Euclidean space, performed the Wick rotation, and performed the non-relativistic reduction.  In addition to the operators given here, there are operators that mix spin-\ohf\ and spin-\thf\ fields and operators that contain more powers of the baryon four-velocity.  In the rest frame, the former vanish, while the latter contribute nothing new.  I therefore ignore such operators.  
\subsection{Staggered heavy baryon Lagrangian for \order{\varepsilon^3} baryon octet masses}
In the rest frame of the heavy baryon, $\mb{v}=0$, so the only non-vanishing component of the four-velocity is $v_0=1$.  But $\mb{v}=0$ implies that $S_0=0$, and it follows from the form of the enumerated operators that \order{\asq} operators containing $S_\mu$ do not contribute to the rest frame baryon masses.  Setting $\sigma=\sigma^{\dag}=1$ in the remaining operators and introducing an LEC for each taste-violating term in the Lagrangian that is distinct at tree level,
\begin{eqnarray}
\la_{\phi B,\asq}^{(2)A\prime}&=&A_1{\overline B}^{ijk}B_{knq}(\xi_5)_i^{\phantom{i}n}(\xi_5)_j^{\phantom{j}q}+A_2{\overline B}^{ijk}B_{nqk}(\xi_5)_i^{\phantom{i}n}(\xi_5)_j^{\phantom{j}q}\label{FFA}\\
&+&A_3{\overline B}^{ijk}B_{knq}\sum _{\mu<\nu}(\xi_{\mu\nu})_i^{\phantom{i}n}(\xi_{\mu\nu})_j^{\phantom{j}q}+A_4{\overline B}^{ijk}B_{nqk}\sum _{\mu<\nu}(\xi_{\mu\nu})_i^{\phantom{i}n}(\xi_{\mu\nu})_j^{\phantom{j}q}\nonumber\\
&+&A_5{\overline B}^{ijk}B_{knq}(\xi_\nu)_i^{\phantom{i}n}(\xi_\nu)_j^{\phantom{j}q}+A_6{\overline B}^{ijk}B_{nqk}(\xi_\nu)_i^{\phantom{i}n}(\xi_\nu)_j^{\phantom{j}q}\nonumber\\
&+&A_7{\overline B}^{ijk}B_{knq}(\xi_{\nu5})_i^{\phantom{i}n}(\xi_{\nu5})_j^{\phantom{j}q}+A_8{\overline B}^{ijk}B_{nqk}(\xi_{\nu5})_i^{\phantom{i}n}(\xi_{\nu5})_j^{\phantom{j}q},\nonumber\\
\la_{\phi B,\asq}^{(2)B\prime}&=&B_1{\overline B}^{ijk}B_{knq}\sum _{\nu\neq4}(\xi_{4\nu})_i^{\phantom{i}n}(\xi_{4\nu})_j^{\phantom{j}q}+B_2{\overline B}^{ijk}B_{nqk}\sum _{\nu\neq4}(\xi_{4\nu})_i^{\phantom{i}n}(\xi_{4\nu})_j^{\phantom{j}q}\label{FFB}\\
&+&B_3{\overline B}^{ijk}B_{knq}(\xi_4)_i^{\phantom{i}n}(\xi_4)_j^{\phantom{j}q}+B_4{\overline B}^{ijk}B_{nqk}(\xi_4)_i^{\phantom{i}n}(\xi_4)_j^{\phantom{j}q}\nonumber\\
&+&B_5{\overline B}^{ijk}B_{knq}(\xi_{45})_i^{\phantom{i}n}(\xi_{45})_j^{\phantom{j}q}+B_6{\overline B}^{ijk}B_{nqk}(\xi_{45})_i^{\phantom{i}n}(\xi_{45})_j^{\phantom{j}q}.\nonumber
\end{eqnarray}
In the fully dynamical case, the flavor-taste indices $ijknq$ take the values $1$ to $12$.  Recalling that $\xi_\tau$ for $\tau\in\{I,\ \mu,\ \mu\nu(\mu<\nu),\ \mu5,\ 5\}$ is shorthand for $I_3\otimes\xi_\tau$, we see that the \order{\asq} heavy baryon Lagrangian is invariant under arbitrary flavor transformations, $\su{3}{F}\subset\su{12}{f}$.  However, the taste matrices in $\la_{\phi B,\asq}^{(2)A\prime}$ break taste down to the remnant taste group of~\cite{Lee:1999zx}, \[\rt\subset\su{4}{T},\] while in an arbitrary reference frame, the chiral operators arising from $\la_6\ffb$ (Wick rotated to Euclidean space) break taste to the lattice symmetry group, \[\ls\subset[\rt]\times\sofour_E.\] \swfour\ is the hypercubic group in the diagonal of the taste and spacetime \sofour's; $\Gamma_4$ is the Clifford group generated by shifts by one lattice site~\cite{Lee:1999zx}.  

In contrast, the operators appearing in $\la_{\phi B,\asq}^{(2)B\prime}$ are manifestly invariant under a larger group:  Under the taste subgroup that corresponds to the spatial rotations, which excludes (taste) boosts, the taste matrices $\xi_\nu$ for $\nu=1,\,2,\,3$ transform as components of a three-vector, and the taste matrices $\xi_4$ and $\xi_5$ are invariant.  Moreover, in the rest frame we are free to make arbitrary spatial rotations.  Recalling the taste subgroup corresponding to spatial rotations of spinors, $\su{2}{T}\subset\sofour_T$, one concludes that the \order{\asq} Lagrangian is invariant under
\begin{equation}
U(3)_l\times U(3)_r\times[\rrt]\times SU(2)_E,\label{forms_symmetry}
\end{equation}
where $SU(2)_E$ is the group of spatial rotations.  Table~\ref{Lsyms} lists the symmetries of the various terms of the Lagrangian used in the calculation, \[\la_{\chi}=\la_{\phi B}^{(1)\prime}+\la_{\phi B,m}^{(2)\prime}+\la_\phi^{(1)\prime}-\asq\mc{V}+\asq\la_{\phi B,\asq}^{(2)A\prime}+\asq\la_{\phi B,\asq}^{(2)B\prime},\] where $\la_\phi^{(1)\prime}$ denotes the leading staggered chiral Lagrangian in the meson sector at zero lattice spacing, and \mc{V} is the generalized Lee-Sharpe potential~\cite{Lee:1999zx,Aubin:2003mg}.  
\begin{table}
\begin{ruledtabular}
\begin{tabular}{lcl}
Term in $\mc{L}_{\chi}$ & Case & Flavor-taste-spacetime symmetry\\ \hline
$\la_{\phi B}^{(1)\prime}+\la_{\phi B,m}^{(2)\prime}+\la_\phi^{(1)\prime}$ & isospin limit, $m\neq0$, any \mb{v}, $a$ & $SU(8)_{x,y}\times SU(4)_z\times SO(4)_E$\\
$\mc{V}$ & any $m$, \mb{v}, $a$ & $U(3)_l\times U(3)_r\times [\Gamma_4\rtimes SO(4)_T]\times SO(4)_E$\\
$\la_{\phi B,\asq}^{(2)A\prime}$ & any $m$, $\mb{v}=0$, any $a$ & $U(3)_l\times U(3)_r\times [\Gamma_4\rtimes SO(4)_T]\times SU(2)_E$\\
$\la_{\phi B,\asq}^{(2)B\prime}$ & any $m$, $\mb{v}=0$, any $a$ & $U(3)_l\times U(3)_r\times [\Gamma_4\rtimes SU(2)_T]\times SU(2)_E$\\
\end{tabular}
\end{ruledtabular}
\caption{\label{Lsyms}The valence quark symmetries of terms in the Euclidean staggered heavy baryon Lagrangian that are needed for computing the octet baryon masses to $\order{\varepsilon^3}$.  The taste symmetry in the rest frame of the heavy baryons is not completely broken to the lattice symmetry group.}
\end{table}

\section{\label{spec}The Flavor-Symmetric Nucleons}
\subsection{\label{id}Identifying staggered nucleons and irreps of interpolating fields}
As shown in \cite{Bailey:2006zn}, the \mb{572_M} and \mb{364_S} contain not only baryons that are degenerate in the continuum limit with the octet and decuplet states of nature, but also baryons that have unphysical masses; the latter states are degenerate with certain partially quenched octet and decuplet baryons.  To calculate the masses of baryons degenerate with a given member of the octet or decuplet, one must choose a basis in the flavor-taste space in which degeneracies with the desired states are evident.

Consider the valence sector of the partially quenched theory.  Decomposing the \mb{572_M} and \mb{364_S} of \su{12}{f} into irreps of the flavor-taste subgroup gives
\begin{subequations}
\begin{eqnarray}
\su{12}{f}&\supset&\su{3}{F}\times\su{4}{T}\nonumber\\
\mathbf{572_M}&\rightarrow&\mathbf{(10_S,\ 20_M)\oplus(8_M,\ 20_S)\oplus(8_M,\ 20_M)\oplus(8_M,\ \bar 4_A)\oplus(1_A,\ 20_M)}\label{mix}\\
\mathbf{364_S}&\rightarrow&\mathbf{(10_S,\ 20_S)\oplus(8_M,\ 20_M)\oplus(1_A,\ \bar 4_A)}\label{sym}
\end{eqnarray}
\end{subequations}
Assume that taste is restored in the continuum limit.  Taking the valence quark masses equal so that \su{3}{F} is exact in the valence sector, \su{12}{f} is a good symmetry as well, and all the baryons of the \mb{572_M} are degenerate, as are all the baryons of the \mb{364_S}.  The symmetry of the spin-\ohf\ baryons is the same as the symmetry of the octet, and the symmetry of the spin-\thf\ baryons, the same as that of the decuplet.  Therefore, setting the masses of all three valence quarks and the masses of two sea quarks equal to the average up-down quark mass while setting the mass of the remaining sea quark equal to that of the strange quark mass, the baryons of the \mb{572_M} are degenerate with the nucleon (in the limit of exact isospin), and those of the \mb{364_S}, with the $\Delta$.  

Increasing the mass of the strange valence quark to its physical value does not change the masses of baryons that do not contain a strange valence quark; such baryons remain degenerate with the nucleon or the $\Delta$.  In particular, the isospin-\thf\ members of the \mb{(10_S,\ 20_M)} remain degenerate with the nucleon.  These states are flavor-symmetric; interpolating fields for the single-flavor members of the \mb{(10_S,\ 20_M)} were constructed in \cite{Golterman:1984dn}.  In the continuum limit, the $\su{8}{x,y}\times\su{4}{z}$ symmetry of the valence sector allows one to rotate these flavor-symmetric nucleons into those with physical flavor and simple taste structure, i.e., baryons that are manifestly degenerate with the nucleon~\cite{Bailey:2006zn}.  

The interpolating fields for the flavor-symmetric nucleons fall into irreps of the geometrical time-slice group (\mr{GTS})~\cite{Golterman:1984dn}.  Decomposing the continuum irrep of the flavor-symmetric nucleons into irreps of \mr{GTS} gives~\cite{Golterman:1984dn}
\begin{eqnarray*}
\mb{(\ohf,\ 20_M)}\rightarrow3\mb{(8)}\oplus\mb{16},
\end{eqnarray*}
where a direct product with continuum parity is suppressed on both sides of the decomposition.  At nonzero lattice spacing, discretization effects lift the degeneracy of the \mr{GTS} irreps and introduce mixing among corresponding members of the \mb{8}'s.  Interpolating fields transforming in the \mb{8} of \mr{GTS} overlap corresponding members of each of the three \mb{8}'s, while interpolating fields transforming in the \mb{16} overlap corresponding members of the \mb{16}.  

\subsection{\label{matrix}Taste symmetry and the mass matrix}
Consider one of the single-flavor, isospin-\thf\ members of the \mb{10_S} of the \mb{(10_S,\ 20_M)}.  At nonzero lattice spacing, taste violations partially lift the degeneracy of the members of the \mb{20_M}.  The baryons are degenerate within irreps of the remnant taste symmetries respected by the relevant propagators and vertices at a given order in the staggered chiral expansion.  To \order{\varepsilon^3} in the staggered chiral expansion, the masses receive analytic contributions of \order{m_q} and \order{\asq} at tree level, non-analytic contributions of \order{m_q^{3/2}} from loops with virtual spin-\ohf\ baryons, and non-analytic contributions of \order{\Delta m_q\ln m_q} from loops with virtual spin-\thf\ baryons.  

The analytic contributions proportional to $m_q$ come from $\la_{\phi B,m}^{(2)\prime}$; referring to Table~\ref{Lsyms}, we see that they respect $SU(8)_{x,y}\times SU(4)_z\times SO(4)_E$.  The vertices and heavy baryon propagators in the loops come from $\la_{\phi B}^{(1)\prime}$, while the meson propagators come from $\la_\phi^{(1)\prime}$ and $\mc{V}$.  Therefore the loops respect $SU(2)_{x,y}\times U(1)_z\times [\Gamma_4\rtimes SO(4)_T]\times SO(4)_E$; the loops break taste from \su{4}{T} to the remnant taste symmetry \rt.  Under \rt, the \mb{20_M} of \su{4}{T} decomposes into one 12-dimensional irrep and two 4-dimensional irreps:
\begin{eqnarray*}
\su{4}{T}&\supset&\rt \\
\mb{20_M}&\rightarrow&\mb{12}\oplus2\mb{(4)}.
\end{eqnarray*}
Finally, analytic contributions proportional to \asq\ come from $\la_{\phi B,\asq}^{(2)A\prime}$ and $\la_{\phi B,\asq}^{(2)B\prime}$.  The former do not break taste further than the loops; the latter break taste to \rrt.  Under \rrt, the \mb{12} of \rt\ decomposes further into an \mb{8} and a \mb{4}, while the two \mb{4}'s of \rt\ are irreducible under \rrt:  
\begin{eqnarray*}
\rt&\supset&\rrt\\
\mb{12}&\rightarrow&\mb{8}\oplus\mb{4}\\
\mb{4}&\rightarrow&\mb{4}.
\end{eqnarray*}
Taste violations in the loops and in the \order{\asq} terms from $\la_{\phi B,\asq}^{(2)A\prime}$ lift the continuum degeneracy between baryons in the \mb{12} and those in the two \mb{4}'s of \rt\ and introduce mixing between corresponding states in the two \mb{4}'s; such states have the same \rt\ quantum numbers.  Taste violations in the \order{\asq} terms from $\la_{\phi B,\asq}^{(2)B\prime}$ lift the remaining degeneracy between baryons in the \mb{8} and those in the \mb{4} of \rrt\ and mix corresponding states in the three \mb{4}'s of \rrt; in this case, corresponding states have the same \rrt\ quantum numbers.

The quantum numbers of these remnant taste symmetries are eigenvalues of maximal sets of commuting observables (\obs) corresponding to each symmetry.  To construct such sets, first consider the decomposition of the \rt\ irreps under the $\sofour_T$ subgroup:\begin{subequations}\label{so4inrt}
\begin{eqnarray}
\rt&\supset&\sofour_T\nonumber\\
\mb{12}&\rightarrow&{\textstyle \bigl(1,\frac{1}{2}\bigr)\oplus\bigl(\frac{1}{2},1\bigr)}\\
\mb{4}&\rightarrow&{\textstyle \bigl(0,\frac{1}{2}\bigr)\oplus\bigl(\frac{1}{2},0\bigr)}\label{so4inrtfun}
\end{eqnarray}
\end{subequations}
where the isomorphism $SO(4)\simeq SU(2)\times SU(2)/Z_2$ has been used to label the $SO(4)_T$ irreps; the conserved observables are simply the spins corresponding to the $SU(2)$'s.  The decomposition (\ref{so4inrtfun}) is explicit in the Weyl representation of the taste matrices, so it is convenient to work in this representation.  In the Weyl representation, the only diagonal member of $\Gamma_4$ that commutes with but is not redundant with the spins is $\xi_5$.  Therefore, one \obs\ of \rt\ is $\{\xi_5,\>J_L^2,\>J_R^2,\>J_{Lz},\>J_{Rz}\}$, where \mb{J_L} and \mb{J_R} denote the spins.  

Tree-level corrections of \order{\asq} from $\la_{\phi B,\asq}^{(2)B\prime}$ break \rt\ to \rrt\ by breaking $SO(4)_T$ to \su{2}{T}, where \su{2}{T} is the subgroup of $SO(4)_T$ obtained by rotating left-handed and right-handed Weyl spinors together; decomposing the $SO(4)_T$ irreps under \su{2}{T} is an elementary exercise in addition of angular momentum:
\begin{subequations}
\label{su2inso4}
\begin{eqnarray}
\sofour_T&\supset&\su{2}{T}\nonumber\\
\bigl(1,\ohf\bigr)\;\;\text{and}\;\;\bigl(\ohf,1\bigr)&\rightarrow&\thf\oplus\ohf\\
\bigl(0,\ohf\bigr)\;\;\text{and}\;\;\bigl(\ohf,0\bigr)&\rightarrow&\ohf
\end{eqnarray}
\end{subequations}
The sum of left- and right-handed spins is respected by \rrt, and the corresponding \obs\ is $\{\xi_5,\>(\mathbf{J_L+J_R})^2,\> J_{Lz}+J_{Rz}\}$.  Table~\ref{Qnums} lists the \obs's of the remnant taste symmetries and notation for the corresponding quantum numbers.

It is not difficult to see that $\Gamma_4$ contains a taste parity operation:  $\xi_4$ interchanges the left- and right-handed spin quantum numbers in the product irreps of~(\ref{so4inrt}); moreover, under \su{2}{T} in \rrt,
\begin{subequations}
\label{su2inrrt}
\begin{eqnarray}
\rrt&\supset&\su{2}{T}\nonumber\\
\mb{8}&\rightarrow&\thf\oplus\thf\\
\mb{4}&\rightarrow&\ohf\oplus\ohf
\end{eqnarray}
\end{subequations}
where taste parity interchanges the two \su{2}{T} irreps in each of the decompositions~(\ref{su2inrrt}).  Taste parity allows one to identify irreps of the remnant taste symmetries from the spin irreps appearing in the decompositions (\ref{so4inrt}) and (\ref{su2inrrt}).  
\begin{table}
\begin{ruledtabular}
\begin{tabular}{lll}
Remnant taste symmetry & \obs\ & Quantum numbers\\ \hline
\rt\ & $\xi_5$, $J_L^2$, $J_R^2$, $J_{Lz}$, $J_{Rz}$ & $\xi_5$, $j_L$, $j_R$, $m_L$, $m_R$ \\
\rrt\ & $\xi_5$, $(\mathbf{J_L+J_R})^2$, $J_{Lz}+J_{Rz}$ & $\xi_5$, $j$, $m$
\end{tabular}
\end{ruledtabular}
\caption{\label{Qnums}Maximal sets of commuting observables (\obs's) for each remnant taste symmetry.  The quantum numbers of each \obs\ are used to distinguish generically nonzero mixing elements in the mass matrix from off-diagonal elements that must vanish.}
\end{table}

The explicit form of the mass matrix depends on the basis chosen for the \mb{20_M}.  Taking tensor products of fundamental representations of \su{4}{T}, projecting onto the \mb{20_M}, and demanding that the states also be eigenvectors of the \obs\ of \rt\ leads to
\begin{subequations}\label{basis}
\begin{eqnarray}
|N_{aab}\rangle&\equiv&\oneorsix\left(|aab\rangle+|aba\rangle-2|baa\rangle\right),\quad a\neq b,\quad a,\;b\in\{1,\>2,\>3,\>4\}\\
|\Sigma_{abc}\rangle&\equiv&\oneortwlv\left(|abc\rangle+|bac\rangle-2|cab\rangle+|acb\rangle+|bca\rangle-2|cba\rangle\right)\label{siggy}\\
|\Lambda_{abc}\rangle&\equiv&\ohf\left(|abc\rangle+|acb\rangle-|bac\rangle-|bca\rangle\right)\label{lammy}
\end{eqnarray}
\end{subequations}
where in (\ref{siggy}) and (\ref{lammy}), $abc\in \{123,\>124,\>341,\>342\}$.  The names for the states are motivated by analogy with the states of the mixed irrep of \su{3}{F}; of the 20 vectors spanning the $\mathbf{20_M}$, 12 have a taste structure analogous to the flavor structure of the nucleon, 4 have a taste structure analogous to the flavor structure of the neutral $\Sigma$, and 4, taste structure analogous to the flavor structure of the $\Lambda$.  

The convenience of this basis is evident when one considers taste violations from loops and tree-level \order{\asq} contributions respecting \rt.  Although they are both linear combinations of states with the same taste labels, $\Sigma_{abc}$ and $\Lambda_{abc}$ do not mix because they are eigenstates of $J_L^2$ and $J_R^2$ with different eigenvalues; the $\Sigma$'s are symmetric under interchange of the first two indices, and so are triplets, $j_{L(R)}=1$, while the $\Lambda$'s are antisymmetric under interchange of the first two indices, and so are singlets, $j_{L(R)}=0$.  The \rt\ quantum numbers of each of the states in the basis (\ref{basis}) are listed in Table~\ref{RTeigen1}.
\begin{table}
\begin{ruledtabular}
\begin{tabular}{lrcrr}
State of $\mathbf{20_M}$ & $\xi_5$ & $\bigl(j_L,j_R\bigr)$ & $m_L$ & $m_R$\\ \hline
$N_{112}$ & $+1$ & ${ \bigl(0,\ohf \bigr)}$ & $0$ & $+{ \ohf }$\\
$N_{221}$ & $+1$ & ${ \bigl(0,\ohf \bigr)}$ & $0$ & $-{ \ohf }$ \\
$N_{334}$ & $-1$ & ${ \bigl(\ohf ,0\bigr)}$ & $+{ \ohf }$ & $0$  \\
$N_{443}$ & $-1$ & ${ \bigl(\ohf ,0\bigr)}$ & $-{ \ohf }$ & $0$  \\ \hline

$\Lambda_{341}$ & $+1$ & ${ \bigl(0,\ohf \bigr)}$ & $0$ & $+{ \ohf }$\\
$\Lambda_{342}$ & $+1$ & ${ \bigl(0,\ohf \bigr)}$ & $0$ & $-{ \ohf }$\\
$\Lambda_{123}$ & $-1$ & ${ \bigl(\ohf ,0\bigr)}$ & $+{ \ohf }$& $0$  \\
$\Lambda_{124}$ & $-1$ & ${ \bigl(\ohf ,0\bigr)}$ & $-{ \ohf }$& $0$  \\\hline

$N_{331}$ & $+1$ & ${ \bigl(1,\ohf \bigr)}$ & $+1$ & $+{ \ohf }$ \\
$N_{332}$ & $+1$ & ${ \bigl(1,\ohf \bigr)}$ & $+1$ & $-{ \ohf }$ \\
$N_{441}$ & $+1$ & ${ \bigl(1,\ohf \bigr)}$ & $-1$ & $+{ \ohf }$ \\
$N_{442}$ & $+1$ & ${ \bigl(1,\ohf \bigr)}$ & $-1$ & $-{ \ohf }$ \\
$\Sigma_{341}$ & $+1$ & ${ \bigl(1,\ohf \bigr)}$ & $0$ & $+{\ohf }$ \\
$\Sigma_{342}$ & $+1$ & ${\bigl(1,\ohf \bigr)}$ & $0$ & $-{\ohf }$\\
$N_{113}$ & $-1$ & ${\bigl(\ohf ,1\bigr)}$ & $+{\ohf }$& $+1$  \\
$N_{114}$ & $-1$ & ${\bigl(\ohf ,1\bigr)}$ & $-{\ohf }$& $+1$  \\
$N_{223}$ & $-1$ & ${\bigl(\ohf ,1\bigr)}$ & $+{\ohf }$& $-1$  \\
$N_{224}$ & $-1$ & ${\bigl(\ohf ,1\bigr)}$ & $-{\ohf }$& $-1$  \\
$\Sigma_{123}$ & $-1$ & ${\bigl(\ohf ,1\bigr)}$ & $+{\ohf }$& $0$  \\
$\Sigma_{124}$ & $-1$ & ${\bigl(\ohf ,1\bigr)}$ & $-{\ohf }$& $0$
\end{tabular}
\end{ruledtabular}
\caption{\label{RTeigen1}Quantum numbers of \rt\ for each of the states in the \mb{20_M}.  States having the same set of \rt\ eigenvalues are mixed by taste violations.  States with distinct eigenvalues are not mixed by interactions that respect \rt.}
\end{table}
As a set, the \rt\ quantum numbers of members of the \mb{12} differ from the \rt\ quantum numbers of the two \mb{4}'s; interactions respecting \rt\ do not mix members of the \mb{12} with members of the two \mb{4}'s.  In the same way, the \rt\ quantum numbers serve to completely distinguish members within each of the \rt\ irreps; within each irrep, the mass submatrix of contributions that respect \rt\ is diagonal.  This observation is consistent with the fact that members of a given irrep are not mixed by interactions respecting the corresponding symmetry group.  However, each member of each \mb{4} has the same quantum numbers as one member of the other \mb{4}.  The taste violations mix $N_{112}$ with $\Lambda_{341}$, $N_{221}$ with $\Lambda_{342}$, and so on.  An 8-dimensional submatrix corresponding to the two \mb{4}'s contains generically nonzero off-diagonal elements due to contributions from the loops and tree-level \order{\asq} corrections that respect \rt.  Note that states having the same \rt\ quantum numbers are nonetheless distinct states; they have different \su{4}{T} quantum numbers.  

The form of the contributions to the mass matrix that respect \rt\ follows from the symmetry of the quark flows and the \rt\ symmetry; in the basis given in Table~\ref{RTeigen1}, the 8-dimensional submatrix corresponding to the two \mb{4}'s (for any given member of the \mb{10_S}) has the form
\begin{equation}\label{rtmat}
\left( \begin{array}{cccc|cccc}
 c_1 & 0 & 0 & 0 & c_3 & 0 & 0 & 0\cr
 0 & c_1 & 0 & 0 & 0 & -c_3& 0 & 0\cr
 0 & 0 & c_1 & 0 & 0 & 0 & c_3 & 0\cr
 0 & 0 & 0 & c_1 & 0 & 0 & 0 & -c_3\cr 
 \hline
 c_3 & 0 & 0 & 0 & c_2 & 0 & 0 & 0\cr
 0 & -c_3& 0 & 0 & 0 & c_2 & 0 & 0\cr
 0 & 0 & c_3 & 0 & 0 & 0 & c_2 & 0\cr
 0 & 0 & 0 & -c_3& 0 & 0 & 0 & c_2\end{array} \right)
\end{equation}
To \order{\varepsilon^3} in the staggered chiral expansion, the contributions parameterized here are given in Sec.~\ref{mass}.  The results have been checked to verify that they conform to the pattern of degeneracies and mixings given here.  The minus signs in the off-diagonal elements arise because some of the states in Table~\ref{RTeigen1} are unconventionally normalized with respect to the ladder operators.  Specifically,
\begin{eqnarray*}
J_{R-}\ket{N_{112}}&=&-\ket{N_{221}}\\
J_{L-}\ket{N_{334}}&=&-\ket{N_{443}}\\
&\text{\mr{while}}&\\
J_{R-}\ket{\Lambda_{341}}&=&\ket{\Lambda_{342}}\\
J_{L-}\ket{\Lambda_{123}}&=&\ket{\Lambda_{124}}.
\end{eqnarray*}

In the continuum limit, the restoration of \su{4}{T} implies that $c_3$ must vanish and $c_1$ must equal $c_2$.  The tree-level taste violations vanish trivially; a straightforward exercise with the loop contributions completes the consistency check (cf. Appendix~\ref{xirep}).  

Tree-level \order{\asq} contributions from $\la_{\phi B,\asq}^{(2)B\prime}$ may be parameterized in much the same way.  The taste violations in such contributions split the baryons in the \rt\ \mb{12} between the \mb{8} and a \mb{4} of \rrt\ and introduce mixing between corresponding states in the resulting three \mb{4}'s of \rrt.  The two \mb{4}'s of Table~\ref{RTeigen1} are also irreducible under \rrt; however, because the \obs\ of \rrt\ includes the total angular momentum $\mb{(J_L+J_R)^2}$, the components $J_{Lz}$ and $J_{Rz}$ are no longer separately conserved, and contributions from $\la_{\phi B,\asq}^{(2)B\prime}$ are not diagonal in the basis given in Table~\ref{RTeigen1} for the \mb{12}.  Constructing eigenstates of the new \obs\ gives the bases for the \mb{8} and \mb{4} of \rrt\ shown in Table~\ref{RTeigen2}.  These states are related to those in Table~\ref{RTeigen1} by the Clebsch-Gordan coefficients for adding spin $1$ and spin \ohf.  

From the eigenvalues listed in Table~\ref{RTeigen2}, we see that the submatrix containing generically nonzero off-diagonal elements is 12-dimensional and corresponds to mixing among corresponding members of the three \mb{4}'s of \rrt.  In the basis of Table~\ref{RTeigen2}, the form of the mixing matrix implied by the symmetry of the quark flows and \rrt\ is
\begin{equation}\label{rrtmat}\addtocounter{MaxMatrixCols}{2}
\left( \begin{array}{cccc|cccc|cccc}
 c_4 & 0 & 0 & 0 & c_7 & 0 & 0 & 0 & c_8 & 0 & 0 & 0 \cr
 0 & c_4 & 0 & 0 & 0 & -c_7& 0 & 0 & 0 & -c_8& 0 & 0 \cr
 0 & 0 & c_4 & 0 & 0 & 0 & c_7 & 0 & 0 & 0 & c_8 & 0 \cr
 0 & 0 & 0 & c_4 & 0 & 0 & 0 & -c_7& 0 & 0 & 0 & -c_8\cr
 \hline
 c_7 & 0 & 0 & 0 & c_5 & 0 & 0 & 0 & c_9 & 0 & 0 & 0 \cr
 0 & -c_7& 0 & 0 & 0 & c_5 & 0 & 0 & 0 & c_9 & 0 & 0 \cr
 0 & 0 & c_7 & 0 & 0 & 0 & c_5 & 0 & 0 & 0 & c_9 & 0 \cr
 0 & 0 & 0 & -c_7& 0 & 0 & 0 & c_5 & 0 & 0 & 0 & c_9 \cr
 \hline
 c_8 & 0 & 0 & 0 & c_9 & 0 & 0 & 0 & c_6 & 0 & 0 & 0 \cr
 0 & -c_8& 0 & 0 & 0 & c_9 & 0 & 0 & 0 & c_6 & 0 & 0 \cr
 0 & 0 & c_8 & 0 & 0 & 0 & c_9 & 0 & 0 & 0 & c_6 & 0 \cr
 0 & 0 & 0 & -c_8& 0 & 0 & 0 & c_9 & 0 & 0 & 0 & c_6 \end{array} \right)
\end{equation}
The contributions parameterized here are presented in Sec.~\ref{mass}; a straightforward exercise shows that they are consistent with this parameterization (cf.~\ref{xirep}).  The minus signs again arise from the phase convention for $\ket{N_{221}}$ and $\ket{N_{443}}$.
\begin{table}
\begin{ruledtabular}
\begin{tabular}{lrcr}
State of $\mathbf{20_M}$ & $\xi_5$ & $j$ & $m$ \\ \hline
$N_{112}$ & $+1$ & \ohf & $+{\ohf }$\\
$N_{221}$ & $+1$ & \ohf & $-{\ohf }$ \\
$N_{334}$ & $-1$ & \ohf & $+{\ohf }$  \\
$N_{443}$ & $-1$ & \ohf & $-{\ohf }$  \\ \hline

$\Lambda_{341}$ & $+1$ & \ohf & $+{\ohf }$\\
$\Lambda_{342}$ & $+1$ & \ohf & $-{\ohf }$\\
$\Lambda_{123}$ & $-1$ & \ohf & $+{\ohf }$\\
$\Lambda_{124}$ & $-1$ & \ohf & $-{\ohf }$\\\hline

$+{\textstyle \sqrt \frac{2}{3}}\,N_{332}-{\textstyle \frac{1}{\sqrt 3}}\,\Sigma_{341}$ &
$+1$ & \ohf & $+{\ohf }$\\
$-{\textstyle \sqrt \frac{2}{3}}\,N_{441}+{\textstyle \frac{1}{\sqrt 3}}\,\Sigma_{342}$ &
$+1$ & \ohf & $-{\ohf }$\\
$+{\textstyle \sqrt \frac{2}{3}}\,N_{114}-{\textstyle \frac{1}{\sqrt 3}}\,\Sigma_{123}$ &
$-1$ & \ohf & $+{\ohf }$\\
$-{\textstyle \sqrt \frac{2}{3}}\,N_{223}+{\textstyle \frac{1}{\sqrt 3}}\,\Sigma_{124}$ &
$-1$ & \ohf & $-{\ohf }$\\\hline

$N_{331}$ & $+1$ & ${\textstyle \frac{3}{2}}$ & $+{\textstyle \frac{3}{2}}$ \\
${\textstyle \frac{1}{\sqrt 3}}\,N_{332}+{\textstyle \sqrt \frac{2}{3}}\,\Sigma_{341}$ &
$+1$ & ${\textstyle \frac{3}{2}}$ & $+{\ohf }$\\
${\textstyle \frac{1}{\sqrt 3}}\,N_{441}+{\textstyle \sqrt \frac{2}{3}}\,\Sigma_{342}$ &
$+1$ & ${\textstyle \frac{3}{2}}$ & $-{\ohf }$\\
$N_{442}$ & $+1$ & ${\textstyle \frac{3}{2}}$ & $-{\textstyle \frac{3}{2}}$ \\
$N_{113}$ & $-1$ & ${\textstyle \frac{3}{2}}$ & $+{\textstyle \frac{3}{2}}$ \\
${\textstyle \frac{1}{\sqrt 3}}\,N_{114}+{\textstyle \sqrt \frac{2}{3}}\,\Sigma_{123}$ &
$-1$ & ${\textstyle \frac{3}{2}}$ & $+{\ohf }$\\
${\textstyle \frac{1}{\sqrt 3}}\,N_{223}+{\textstyle \sqrt \frac{2}{3}}\,\Sigma_{124}$ &
$-1$ & ${\textstyle \frac{3}{2}}$ & $-{\ohf }$\\
$N_{224}$ & $-1$ & ${\textstyle \frac{3}{2}}$ & $-{\textstyle \frac{3}{2}}$
\end{tabular}
\end{ruledtabular}
\caption{\label{RTeigen2}Eigenvalues of the \obs\ of \rrt\ for each of the eigenstates in the \mb{20_M}.  States with the same eigenvalues are mixed by contributions respecting \rrt.}
\end{table}

\subsection{\label{mass}Masses of the flavor-symmetric nucleons}
We now consider the staggered chiral forms for the masses of the flavor-symmetric nucleons having degenerate valence quarks of mass $m_x$.  The forms are given through \order{\varepsilon^3} in fully dynamical and partially quenched SHB$\chi$PT.  Let the tree-level contributions proportional to $m_q$ be denoted by $\mr{Tree}(\mq)$, the tree-level contributions proportional to $a^2$, by $\mr{Tree}(\asq)$, the loops with virtual spin-\ohf\ baryons, by $\mr{Loop}(\ohf)$, and the loops with virtual spin-\thf\ baryons, by $\mr{Loop}(\thf)$.  These contributions are of \order{\mq}, \order{\asq}, \order{m_q^{3/2}}, and \order{\Delta\mq\ln \mq}, respectively.  The mass matrix is given by
\begin{eqnarray}
M&=&M_0+\Sigma(0),\\
\Sigma(0)&=&\mr{Tree}(\mq)+\mr{Tree}(\asq)+\mr{Loop}(\ohf)+\mr{Loop}(\thf)
\end{eqnarray}
where $M_0$ is the average mass of the \mb{572_M} in the continuum and chiral limits, and $\Sigma(0)$ is the heavy baryon self-energy evaluated at $v\cdot r=0$; here $r$ is the residual 4-momentum of the heavy baryon.

In accord with the discussion of Sec.~\ref{matrix}, $M_0$ and $\mr{Tree}(\mq)$ are diagonal matrices in the baryon-taste subspace corresponding to the \mb{20_M} of \su{4}{T}; explicitly, \[M_0+\mr{Tree}(\mq)=[M_0-2(\alpha_M+\beta_M)m_x-2\sigma_M(m_u+m_d+m_s)]\,I_{20},\] where $I_{20}$ is the identity matrix.  $\mr{Loop}(\ohf)$, $\mr{Loop}(\thf)$, and the contributions of $\la_{\phi B,\asq}^{(2)A\prime}$ to $\mr{Tree}(\asq)$ are diagonal in the \mb{12} of \rt\ but have the form given in~(\ref{rtmat}) in the subspace corresponding to the two \mb{4}'s of \rt.  The contributions of $\la_{\phi B,\asq}^{(2)B\prime}$ to $\mr{Tree}(\asq)$ are diagonal in the \mb{8} of \rrt\ but have the form given in~(\ref{rrtmat}) in the three \mb{4}'s of \rrt.  

For convenience in writing the results, let
\begin{eqnarray*}
\mr{Loop}({\ohf})&=&\frac{1}{6f^2}\left[-\frac{1}{32\pi}\sigma_{1/2}^{\mr{conn}}+\sigma_{1/2}^{\mr{disc}}\right],\\
\mr{Loop}({\thf})&=&\left(\frac{C}{4f}\right)^2\left[\frac{1}{(2\pi)^2}\sigma_{3/2}^{\mr{conn}}+\sigma_{3/2}^{\mr{disc}}\right],\\
\mr{Tree}(\asq)&=&-\asq\left[\sigma^{FF(A)}+\sigma^{FF(B)}\right],
\end{eqnarray*}
where $\sigma^{\mr{conn}}$ is the sum of loops with connected meson propagators, $\sigma^{\mr{disc}}$ is the sum of loops with disconnected (hairpin) propagators, and $\sigma^{FF(A,B)}$ are tree-level contributions from $\la_{\phi B,\asq}^{(2)A,B\prime}$.  At tree-level, the pseudoscalar meson masses are degenerate within irreps of \rt, so we label them with an index $t\in\{I,\ V,\ T,\ A,\ P\}$ denoting the $SO(4)_T$ irrep:  \[(m_{ij}^t)^2=\lambda(m_i+m_j)+\asq\Delta^t,\] where $i,\ j\in\{u,\ d,\ s,\ x,\ y,\ z\}$, $\lambda$ is proportional to the chiral condensate, and $\Delta^t$ is the meson taste-splitting~\cite{Aubin:2003mg}.  Let $\tau\in\{I,\ \mu,\ \mu\nu(\mu<\nu),\ \mu5,\ 5\}$ be the meson taste index, and $s_t$, the set of meson tastes in irrep $t$:  $s_I=\{I\}$, $s_V=\{\mu\}$, $s_T=\{\mu\nu(\mu<\nu)\}$, $s_A=\{\mu5\}$, and $s_P=\{5\}$.  Finally, let $n_t$ denote the number of meson tastes corresponding to irrep $t$.  Then 
\[n_t=\sum_{\tau\in s_t} 1,\quad\text{and}\quad\sum_\tau f(\tau)=\sum_t \sum_{\tau\in s_t} f(\tau)\] for any function $f(\tau)$.  If $f$ is a function of $t$ only, then \[\sum_\tau f(\tau)=\sum_t \sum_{\tau\in s_t} f(t)=\sum_t n_t f(t).\]  

For the contributions to $\mr{Loop}(\ohf)$, we have
\begin{eqnarray}
\sigma_{1/2}^{\mr{conn}}&=&\sum _t \left[c_{1/2,t}^{\mr{sea}}\sum_q(m_{xq}^t)^3+c_{1/2,t}^{\mr{val}}(m_{xx}^t)^3\right]\label{loopohf}\\
\sigma_{1/2}^{\mr{disc}}&=&6(2(\alpha+\beta))^2iD_{XX}^I+(4c_{1/2,V}^{\mr{sea}}+c_{1/2,V}^{\mr{val}})iD_{XX}^V+(V\rightarrow A)\nonumber
\end{eqnarray}
where $q\in\{u,\ d,\ s\}$, the $D_{XX}^t$ are hairpin loop integrals, and $c_{1/2,t}^{\mr{sea}}$ and $c_{1/2,t}^{\mr{val}}$ are symmetric matrices depending on the irrep $t$ and containing the LECs.  In dimensional regularization,
\begin{eqnarray*}
D_{XX}^I&\equiv &i{\mu}^{4-n}\int\frac{d^nk}{(2\pi)^n}\;\frac{(S\cdot k)^2}{v\cdot k+i\epsilon}\;\frac{i}{3}\;\frac{\mc{X}_I}{(k^2-(m_{xx}^{I})^2+i\epsilon)^2}\\
D_{XX}^{V,\,A}&\equiv &a^2\delta_{V,\,A}^{\prime}{\mu}^{4-n}\int\frac{d^nk}{(2\pi)^n}\;\frac{(S\cdot k)^2}{v\cdot k+i\epsilon} \;\frac{\mc{X}_{V,\,A}}{(k^2-(m_{xx}^{V,\,A})^2+i\epsilon)^2\;(k^2-m_{\eta^{\prime}_{V,\,A}}^2+i\epsilon)}
\end{eqnarray*}
where \[\mc{X}_t\equiv\frac{(k^2-(m_{uu}^{t})^2+i\epsilon)\;(k^2-(m_{dd}^{t})^2+i\epsilon)\;(k^2-(m_{ss}^{t})^2+i\epsilon)}{(k^2-m_{\pi_{t}}^2+i\epsilon)\;(k^2-m_{\eta_{t}}^2+i\epsilon)}.\]
Contributions from quark flows that contain a sea quark circulating in the loop are represented by $c_{1/2,t}^{\mr{sea}}$; an overall factor of $\frac{1}{4}$ included in $c_{1/2,t}^{\mr{sea}}$ accounts for taking the fourth root of the fermion determinant in simulations.  Contributions from quark flows containing only valence quarks in the meson propagator are represented by $c_{1/2,t}^{\mr{val}}$.  In the basis (\ref{basis}) of Table~\ref{RTeigen1}, the diagonal elements of $c_{1/2,t}^{\mr{sea}}$ are
\begin{equation}
\langle c_{1/2,t}^{\mr{sea}}\rangle=n_t(\textstyle{\frac{5}{8}}\alpha^2+\beta^2+\textstyle{\frac{1}{2}}\alpha\beta),\nonumber
\end{equation}
and the off-diagonal elements of $c_{1/2,t}^{\mr{sea}}$ all vanish.  Unlike $c_{1/2,t}^{\mr{sea}}$, $c_{1/2,t}^{\mr{val}}$ depends on the representation used for the taste matrices.  Employing the Weyl representation, the distinct, nontrivial matrix elements of $c_{1/2,t}^{\mr{val}}$ are listed in Table~\ref{Ctohf}.  The results for $c_{1/2,t}^{\mr{sea,val}}$ are consistent with the degeneracies and mixings parameterized in (\ref{rtmat}).  Appendix~\ref{xirep} gives the general forms of the matrix elements of $c_{1/2,t}^{\mr{val}}$ in terms of the elements of the taste matrices.  These forms are independent of the representation used for the taste matrices and can therefore be used to show that the chiral forms of continuum HB$\chi$PT are recovered in any representation.  
\begin{table}
\begin{ruledtabular}
\begin{tabular}{lccccc}
State(s) & $\langle c_{1/2,I}^{\mr{val}}\rangle$ & $\langle c_{1/2,V}^{\mr{val}}\rangle$ & $\langle c_{1/2,T}^{\mr{val}}\rangle$ & $\langle c_{1/2,A}^{\mr{val}}\rangle$ & $\langle c_{1/2,P}^{\mr{val}}\rangle$ \\ \hline
      $N_{112}$ & $(\shf,\ 2,\ 10)$ & $(0,\ 0,\ 0)$ & $(1,\ -20,\ -28)$ & $(0,\ 0,\ 0)$ & $(\shf,\ 2,\ 10)$ \\
 $N_{331}$ & $(\shf,\ 2,\ 10)$ & $(\otd,\ -\textstyle{\frac{20}{3}},\ -\textstyle{\frac{28}{3}})$ & $(\textstyle{\frac{11}{3}},\ -\textstyle{\frac{4}{3}},\ \textstyle{\frac{16}{3}})$ & $(\otd,\ -\textstyle{\frac{20}{3}},\ -\textstyle{\frac{28}{3}})$ & $(\textstyle{\frac{1}{6}},\ -\textstyle{\frac{10}{3}},\ -\textstyle{\frac{14}{3}})$ \\
$\Lambda_{123}$ & $(\shf,\ 2,\ 10)$ & $(5,\ -4,\ 4)$ & $(-3,\ -12,\ -24)$ & $(5,\ -4,\ 4)$ & $(-\textstyle{\frac{5}{2}},\ 2,\ -2)$\\
$N_{112}|\Lambda_{341}$ & $(0,\ 0,\ 0)$ & $({\sqrt 6},\ 4{\sqrt 6},\ 8{\sqrt 6})$ & $(0,\ 0,\ 0)$ & $(-{\sqrt 6},\ -4{\sqrt 6},\ -8{\sqrt 6})$ & $(0,\ 0,\ 0)$
\end{tabular}
\end{ruledtabular}
\caption{\label{Ctohf}Coefficients of $(\alpha^2,\ \beta^2,\ \alpha\beta)$ for the distinct, nontrivial matrix elements of $c_{1/2,t}^{\mr{val}}$.}
\end{table}

For the contributions to $\mr{Loop}(\thf)$, we have
\begin{eqnarray}
\sigma_{3/2}^{\mr{conn}}&=&\sum _t \left[c_{3/2,t}^{\mr{sea}}\sum_q\mc{F}(m_{xq}^t)+c_{3/2,t}^{\mr{val}}\mc{F}(m_{xx}^t)\right]\label{loopthf}\\
\sigma_{3/2}^{\mr{disc}}&=&(4c_{3/2,V}^{\mr{sea}}+c_{3/2,V}^{\mr{val}})iE_{XX}^V+(V\rightarrow A)\nonumber
\end{eqnarray}
where
\begin{equation*}
\mc{F}(m)\equiv \textstyle{\frac{\Delta}{6}}\left[(3m^2-2\Delta^2)\ln (m^2/\mu^2)-4m^2+\textstyle{\frac{10}{3}}\Delta^2\right]-\textstyle{\frac{2}{3}}m^3g(\Delta/m),
\end{equation*}
\begin{numcases}{g(x)=}
(1-x^2)^{3/2}\arccos x & $0\leq x\leq 1$\nonumber\\
(x^2-1)^{3/2}\ln(x+\sqrt{x^2-1}) & $x>1$\nonumber
\end{numcases}
and
\begin{eqnarray*}
E_{XX}^{V,\,A}\equiv a^2\delta_{V,\,A}^{\prime}{\mu}^{4-n}\int\frac{d^nk}{(2\pi)^n}\;\frac{k_{\nu}k_{\lambda}P^{\nu\lambda}}{v\cdot k-\Delta+i\epsilon}\;\frac{\mc{X}_{V,\,A}}{(k^2-(m_{xx}^{V,\,A})^2+i\epsilon)^2\;(k^2-m_{\eta^{\prime}_{V,\,A}}^2+i\epsilon)};
\end{eqnarray*}
the projection operator $P^{\nu\lambda}$, in $n$ spacetime dimensions, is
\begin{eqnarray*}
P^{\nu\lambda}=v^\nu v^\lambda-g^{\nu\lambda}-4\left(\frac{n-3}{n-1}\right)S^\nu S^\lambda.
\end{eqnarray*}
As before, the $c_{3/2,t}^{\mr{sea}}$ and $c_{3/2,t}^{\mr{val}}$ are symmetric matrices in the baryon taste space.  However, the coefficients for spin-\thf\ contributions are simpler than for the spin-\ohf\ contributions.  The diagonal elements of $c_{3/2,t}^{\mr{sea}}$ are
\begin{eqnarray*}
\langle c_{3/2,t}^{\mr{sea}}\rangle=\ohf n_t,
\end{eqnarray*}
while the off-diagonal elements vanish.  In the Weyl representation, the distinct, nontrivial matrix elements of $c_{3/2,t}^{\mr{val}}$ are listed in Table~\ref{Ctthf}.  As for the spin-\ohf\ contributions, the results for $c_{3/2,t}^{\mr{sea,val}}$ are consistent with the degeneracies and mixings parameterized in (\ref{rtmat}).  The general forms of the matrix elements of $c_{3/2,t}^{\mr{val}}$, in terms of the elements of the taste matrices, are given in Appendix~\ref{xirep}.  
\begin{table}
\begin{ruledtabular}
\begin{tabular}{lccccc}
State(s) & $\langle c_{3/2,I}^{\mr{val}}\rangle$ & $\langle c_{3/2,V}^{\mr{val}}\rangle$ & $\langle c_{3/2,T}^{\mr{val}}\rangle$ & $\langle c_{3/2,A}^{\mr{val}}\rangle$ & $\langle c_{3/2,P}^{\mr{val}}\rangle$ \\ \hline
      $N_{112}$ & $-2$ & $0$ & $20$ & $0$ & $-2$ \\
 $N_{331}$ & $-2$ & $\textstyle{\frac{20}{3}}$ & $\textstyle{\frac{4}{3}}$ & $\textstyle{\frac{20}{3}}$ & $\textstyle{\frac{10}{3}}$ \\
$\Lambda_{123}$ & $-2$ & $4$ & $12$ & $4$ & $-2$\\
$N_{112}|\Lambda_{341}$ & $0$ & $-4{\sqrt 6}$ & $0$ & $4{\sqrt 6}$ & $0$
\end{tabular}
\end{ruledtabular}
\caption{\label{Ctthf}The distinct, nontrivial matrix elements of $c_{3/2,t}^{\mr{val}}$.}
\end{table}

Like the loops, $\sigma^{FF(A)}$ is a symmetric matrix.  In the Weyl representation, the distinct, nontrivial matrix elements of $\sigma^{FF(A)}$ are listed in Table~\ref{sigFFA}; they are consistent with the degeneracies and mixings parameterized in~(\ref{rtmat}).  The LECs $A_i$, $i=1,\dots,8$ are defined in Eq.~(\ref{FFA}).  Appendix~\ref{FF} contains the general forms of the matrix elements of $\sigma^{FF(A)}$ in terms of the taste matrices appearing in (\ref{FFA}).
\begin{table}
\begin{ruledtabular}
\begin{tabular}{lcccccccc}
State(s) & $A_1$ & $A_3$ & $A_5$ & $A_7$ & $A_2$ & $A_4$ & $A_6$ & $A_8$ \\ \hline
      $N_{112}$ & $-\ohf$ & $-1$ & $0$ & $0$ & $1$ & $-4$ & $0$ & $0$ \\
 $N_{331}$ & $-\textstyle{\frac{1}{6}}$ & $-\textstyle{\frac{2}{3}}$ & $-\otd$ & $-\otd$ & $-\textstyle{\frac{2}{3}}$ & $\otd$ & $-\textstyle{\frac{4}{3}}$ &  $-\textstyle{\frac{4}{3}}$\\
$\Lambda_{123}$ & $\ohf$ & $0$ & $-1$ & $-1$ & $0$ & $-3$ & $0$ & $0$ \\
$N_{112}|\Lambda_{341}$ & $0$ & $0$ & $0$ & $0$ & $0$ & $0$ & ${\sqrt 6}$ & $-{\sqrt 6}$
\end{tabular}
\end{ruledtabular}
\caption{\label{sigFFA}Coefficients of $A_i$, $i=1,\dots,8$, for the matrix elements of $\sigma^{FF(A)}$.}
\end{table}

For the contributions to the symmetric matrix $\sigma^{FF(B)}$, we consider the basis of Table~\ref{RTeigen2}.  In the Weyl representation, the distinct, nontrivial matrix elements are listed in Table~\ref{sigFFB}; they are consistent with the degeneracies and mixings parameterized in~(\ref{rrtmat}).  The LECs $B_i$, $i=1,\dots,6$ are defined in Eq.~(\ref{FFB}).  Appendix~\ref{FF} contains the general forms of the matrix elements of $\sigma^{FF(B)}$ in terms of the taste matrices appearing in (\ref{FFB}).
\begin{table}
\begin{ruledtabular}
\begin{tabular}{lcccccc}
State(s) & $B_1$ & $B_3$ & $B_5$ & $B_2$ & $B_4$ & $B_6$ \\ \hline
      $N_{112}$ & $-\ohf$ & $0$ & $0$ & $-2$ & $0$ & $0$ \\
      $N_{331}$ & $-\osx$ & $-\osx$ & $-\osx$ & $-\ttd$ & $-\ttd$ & $-\ttd$ \\
$\Lambda_{123}$ & $0$ & $-\oqt$ & $-\oqt$ & $-\thf$ & $0$ & $0$ \\
${\textstyle \sqrt \frac{2}{3}}\,N_{332}-{\textstyle \frac{1}{\sqrt 3}}\,\Sigma_{341}$ & $-\ttd$ & $\textstyle{\frac{1}{12}}$ & $\textstyle{\frac{1}{12}}$ & $\textstyle{\frac{11}{6}}$ & $\otd$ & $\otd$ \\
$N_{112}|\Lambda_{341}$ & $0$ & $0$ & $0$ & $0$ & $\textstyle{\sqrt{\frac{3}{8}}}$ & $-\textstyle{\sqrt{\frac{3}{8}}}$\\
$N_{112}|{\textstyle \sqrt \frac{2}{3}}\,N_{332}-{\textstyle \frac{1}{\sqrt 3}}\,\Sigma_{341}$ & $0$ & $-\textstyle{\frac{1}{\sqrt 6}}$ & $\textstyle{\frac{1}{\sqrt 6}}$ & $0$ & $\textstyle{\frac{1}{2{\sqrt 6}}}$ & $-\textstyle{\frac{1}{2{\sqrt 6}}}$ \\
$\Lambda_{341}|{\textstyle \sqrt \frac{2}{3}}\,N_{332}-{\textstyle \frac{1}{\sqrt 3}}\,\Sigma_{341}$ & $\otd$ & $-\osx$ & $-\osx$ & $\otd$ & $\otd$ & $\otd$
\end{tabular}
\end{ruledtabular}
\caption{\label{sigFFB}Coefficients of $B_i$, $i=1,\dots,6$, for the matrix elements of $\sigma^{FF(B)}$.}
\end{table}

Taking the continuum and isospin limits, the staggered chiral forms reduce to that of continuum HB$\chi$PT~\cite{Chen:2001yi,Walker-Loud:2004hf} for the partially quenched nucleon.  For the fully dynamical 2+1 flavor case, the hairpin loop integrals simplify significantly; we have
\begin{eqnarray*}
D_{UU}^I&=&i{\mu}^{4-n}\int\frac{d^nk}{(2\pi)^n}\;\frac{(S\cdot k)^2}{v\cdot k+i\epsilon}\;\frac{i}{3}\mc{X^{\prime}}_I\\
D_{UU}^{V,\,A}&=&a^2\delta_{V,\,A}^{\prime}{\mu}^{4-n}\int\frac{d^nk}{(2\pi)^n}\;\frac{(S\cdot k)^2}{v\cdot k+i\epsilon}\; \frac{\mc{X^{\prime}}_{V,\,A}}{(k^2-m_{\eta^{\prime}_{V,\,A}}^2+i\epsilon)}\\
E_{UU}^{V,\,A}&=& a^2\delta_{V,\,A}^{\prime}{\mu}^{4-n}\int\frac{d^nk}{(2\pi)^n}\;\frac{k_{\nu}k_{\lambda}P^{\nu\lambda}}{v\cdot k-\Delta+i\epsilon}\;\frac{\mc{X^{\prime}}_{V,\,A}}{(k^2-m_{\eta^{\prime}_{V,\,A}}^2+i\epsilon)},
\end{eqnarray*}
where \[\mc{X^{\prime}}_t\equiv\frac{\mc{X}_t}{(k^2-(m_{uu}^{t})^2+i\epsilon)^2}=\frac{(k^2-(m_{ss}^{t})^2+i\epsilon)}{(k^2-m_{\pi_{t}}^2+i\epsilon)(k^2-m_{\eta_{t}}^2+i\epsilon)}.\]
Recalling the residue relations of~\cite{Aubin:2003uc}, evaluating the integrals, and renormalizing the tree-level LECs gives
\begin{eqnarray*}
D_{UU}^I&\rightarrow&\frac{i}{192\pi}(m_{\eta_I}^3-3m_{\pi_I}^3),\\
D_{UU}^{V,\,A}&\rightarrow&-\frac{ia^2\delta_{V,\,A}^{\prime}}{32\pi}\left(R_{\pi_{V,\,A}}^{[3,1]}m_{\pi_{V,\,A}}^3+R_{\eta_{V,\,A}}^{[3,1]}m_{\eta_{V,\,A}}^3+R_{\eta^{\prime}_{V,\,A}}^{[3,1]}m_{\eta^{\prime}_{V,\,A}}^3\right),\\
E_{UU}^{V,\,A}&\rightarrow&\frac{ia^2\delta_{V,\,A}^{\prime}}{(2\pi)^2}\left(R_{\pi_{V,\,A}}^{[3,1]}\mc{F}(m_{\pi_{V,\,A}})+R_{\eta_{V,\,A}}^{[3,1]}\mc{F}(m_{\eta_{V,\,A}})+R_{\eta^{\prime}_{V,\,A}}^{[3,1]}\mc{F}(m_{\eta^{\prime}_{V,\,A}})\right),
\end{eqnarray*}
where $R_{l_{V,\,A}}^{[3,1]}$ is shorthand for the residue $R_{l_{V,\,A}}^{[3,1]}(\{m_{\pi_{V,\,A}},\,m_{\eta_{V,\,A}},\,m_{\eta^{\prime}_{V,\,A}}\};\{m_{ss}^{V,\,A}\})$.  

Combining these results for the hairpin integrals with the coefficients $c^{\mr{sea}}$ and those listed in Tables~\ref{Ctohf}, \ref{Ctthf}, \ref{sigFFA}, and \ref{sigFFB}, the parameters of (\ref{rtmat}) and (\ref{rrtmat}) are, in the fully dynamical 2+1 flavor case,
\begin{eqnarray}
c_1&=&M_0-2(\alpha_M+\beta_M+2\sigma_M)m_u-2\sigma_Mm_s\label{c1}\\
&-&\frac{1}{192\pi f^2}\biggl[(\textstyle{\frac{5}{8}}\alpha^2+\beta^2+\textstyle{\frac{1}{2}}\alpha\beta)\sum_t n_t(2m_{\pi_t}^3+m_{K_t}^3)\nonumber\\
&&\phantom{\frac{1}{192\pi f^2}}+(\shf\alpha^2+2\beta^2+10\alpha\beta)(m_{\pi_I}^3+m_{\pi_P}^3)+(\alpha^2-20\beta^2-28\alpha\beta)m_{\pi_T}^3\nonumber\\
&&\phantom{\frac{1}{192\pi f^2}}+4(\alpha+\beta)^2(m_{\eta_I}^3-3m_{\pi_I}^3)-16a^2(\textstyle{\frac{5}{8}}\alpha^2+\beta^2+\textstyle{\frac{1}{2}}\alpha\beta)\times\nonumber\\
&&\phantom{\frac{1}{192\pi f^2}}\times\left(\delta_V^{\prime}(R_{\pi_{V}}^{[3,1]}m_{\pi_{V}}^3+R_{\eta_{V}}^{[3,1]}m_{\eta_{V}}^3+R_{\eta^{\prime}_{V}}^{[3,1]}m_{\eta^{\prime}_{V}}^3)+(V\rightarrow A)\right)\biggr]\nonumber\\
&+&\left(\frac{C}{8\pi f}\right)^2\biggl[\ohf\sum_t n_t[2\mc{F}(m_{\pi_t})+\mc{F}(m_{K_t})]\nonumber\\
&&\phantom{\left(\frac{C}{8\pi f}\right)^2}-2[\mc{F}(m_{\pi_I})+\mc{F}(m_{\pi_P})]+20\mc{F}(m_{\pi_T})\nonumber\\
&&\phantom{\left(\frac{C}{8\pi f}\right)^2}-8a^2\left(\delta_V^{\prime}(R_{\pi_{V}}^{[3,1]}\mc{F}(m_{\pi_{V}})+R_{\eta_{V}}^{[3,1]}\mc{F}(m_{\eta_{V}})+R_{\eta^{\prime}_{V}}^{[3,1]}\mc{F}(m_{\eta^{\prime}_{V}}))+(V\rightarrow A)\right)\biggr]\nonumber\\
&-&a^2\left[-\ohf A_1-A_3+A_2-4A_4\right],\nonumber
\end{eqnarray}
\begin{eqnarray}
c_2&=&M_0-2(\alpha_M+\beta_M+2\sigma_M)m_u-2\sigma_Mm_s\label{c2}\\
&-&\frac{1}{192\pi f^2}\biggl[(\textstyle{\frac{5}{8}}\alpha^2+\beta^2+\textstyle{\frac{1}{2}}\alpha\beta)\sum_t n_t(2m_{\pi_t}^3+m_{K_t}^3)\nonumber\\
&&\phantom{\frac{1}{192\pi f^2}}+(\shf\alpha^2+2\beta^2+10\alpha\beta)m_{\pi_I}^3+(5\alpha^2-4\beta^2+4\alpha\beta)(m_{\pi_V}^3+m_{\pi_A}^3)\nonumber\\
&&\phantom{\frac{1}{192\pi f^2}}+(-3\alpha^2-12\beta^2-24\alpha\beta)m_{\pi_T}^3+(-\fhf\alpha^2+2\beta^2-2\alpha\beta)m_{\pi_P}^3\nonumber\\
&&\phantom{\frac{1}{192\pi f^2}}+4(\alpha+\beta)^2(m_{\eta_I}^3-3m_{\pi_I}^3)\nonumber\\
&&\phantom{\frac{1}{192\pi f^2}}-a^2\bigl(16(\textstyle{\frac{5}{8}}\alpha^2+\beta^2+\textstyle{\frac{1}{2}}\alpha\beta)+(5\alpha^2-4\beta^2+4\alpha\beta)\bigr)\times\nonumber\\
&&\phantom{\frac{1}{192\pi f^2}}\times\left(\delta_V^{\prime}(R_{\pi_{V}}^{[3,1]}m_{\pi_{V}}^3+R_{\eta_{V}}^{[3,1]}m_{\eta_{V}}^3+R_{\eta^{\prime}_{V}}^{[3,1]}m_{\eta^{\prime}_{V}}^3)+(V\rightarrow A)\right)\biggr]\nonumber\\
&+&\left(\frac{C}{8\pi f}\right)^2\biggl[\ohf\sum_t n_t[2\mc{F}(m_{\pi_t})+\mc{F}(m_{K_t})]\nonumber\\
&&\phantom{\left(\frac{C}{8\pi f}\right)^2}-2[\mc{F}(m_{\pi_I})+\mc{F}(m_{\pi_P})]+4[\mc{F}(m_{\pi_V})+\mc{F}(m_{\pi_A})]+12\mc{F}(m_{\pi_T})\nonumber\\
&&\phantom{\left(\frac{C}{8\pi f}\right)^2}-12a^2\left(\delta_V^{\prime}(R_{\pi_{V}}^{[3,1]}\mc{F}(m_{\pi_{V}})+R_{\eta_{V}}^{[3,1]}\mc{F}(m_{\eta_{V}})+R_{\eta^{\prime}_{V}}^{[3,1]}\mc{F}(m_{\eta^{\prime}_{V}}))+(V\rightarrow A)\right)\biggr]\nonumber\\
&-&a^2\left[\ohf A_1-A_5-A_7-3A_4\right],\nonumber
\end{eqnarray}
\begin{eqnarray}
c_3&=&-\frac{1}{192\pi f^2}\biggl[{\sqrt 6}(\alpha^2+4\beta^2+8\alpha\beta)(m_{\pi_V}^3-m_{\pi_A}^3)-a^2{\sqrt 6}(\alpha^2+4\beta^2+8\alpha\beta)\times\label{c3}\\
&&\phantom{\frac{1}{192\pi f^2}}\times\left(\delta_V^{\prime}(R_{\pi_{V}}^{[3,1]}m_{\pi_{V}}^3+R_{\eta_{V}}^{[3,1]}m_{\eta_{V}}^3+R_{\eta^{\prime}_{V}}^{[3,1]}m_{\eta^{\prime}_{V}}^3)-(V\rightarrow A)\right)\biggr]\nonumber\\
&+&\left(\frac{C}{8\pi f}\right)^2\biggl[-4{\sqrt 6}[\mc{F}(m_{\pi_V})-\mc{F}(m_{\pi_A})]\nonumber\\
&&\phantom{\left(\frac{C}{8\pi f}\right)^2}+4a^2{\sqrt 6}\left(\delta_V^{\prime}(R_{\pi_{V}}^{[3,1]}\mc{F}(m_{\pi_{V}})+R_{\eta_{V}}^{[3,1]}\mc{F}(m_{\eta_{V}})+R_{\eta^{\prime}_{V}}^{[3,1]}\mc{F}(m_{\eta^{\prime}_{V}}))-(V\rightarrow A)\right)\biggr]\nonumber\\
&-&a^2{\sqrt 6}\left[A_6-A_8\right],\nonumber
\end{eqnarray}
\begin{eqnarray}
c_4&=&-a^2\left[-\ohf B_1-2B_2\right],\label{c4}\\
c_5&=&-a^2\left[-\oqt B_3-\oqt B_5-\thf B_2\right],\label{c5}\\
c_6&=&-a^2\left[-\ttd B_1+\textstyle{\frac{1}{12}}B_3+\textstyle{\frac{1}{12}}B_5+\textstyle{\frac{11}{6}}B_2+\otd B_4+\otd B_6\right],\label{c6}\\
c_7&=&-a^2\left[\textstyle{\sqrt{\frac{3}{8}}}B_4-\textstyle{\sqrt{\frac{3}{8}}}B_6\right],\label{c7}\\
c_8&=&-a^2\left[-\textstyle{\frac{1}{\sqrt 6}}B_3+\textstyle{\frac{1}{\sqrt 6}}B_5+\textstyle{\frac{1}{2{\sqrt 6}}}B_4-\textstyle{\frac{1}{2{\sqrt 6}}}B_6\right],\label{c8}\\
c_9&=&-a^2\left[\otd B_1-\osx B_3-\osx B_5+\otd B_2+\otd B_4+\otd B_6\right].\label{c9}  
\end{eqnarray}
In addition to the parameters $c_i$, $i=1,\dots,9$, which correspond to members of \rrt\ \mb{4}'s, define the parameters $d_1$ and $d_2$ to respectively correspond to the \rt\ \mb{12} and \rrt\ \mb{8}.  Then
\begin{eqnarray}
d_1&=&M_0-2(\alpha_M+\beta_M+2\sigma_M)m_u-2\sigma_Mm_s\label{d1}\\
&-&\frac{1}{192\pi f^2}\biggl[(\textstyle{\frac{5}{8}}\alpha^2+\beta^2+\textstyle{\frac{1}{2}}\alpha\beta)\sum_t n_t(2m_{\pi_t}^3+m_{K_t}^3)\nonumber\\
&&\phantom{\frac{1}{192\pi f^2}}+(\shf\alpha^2+2\beta^2+10\alpha\beta)m_{\pi_I}^3+\otd(\alpha^2-20\beta^2-28\alpha\beta)(m_{\pi_V}^3+m_{\pi_A}^3)\nonumber\\
&&\phantom{\frac{1}{192\pi f^2}}+\otd(11\alpha^2-4\beta^2+16\alpha\beta)m_{\pi_T}^3+\otd(\ohf\alpha^2-10\beta^2-14\alpha\beta)m_{\pi_P}^3\nonumber\\
&&\phantom{\frac{1}{192\pi f^2}}+4(\alpha+\beta)^2(m_{\eta_I}^3-3m_{\pi_I}^3)\nonumber\\
&&\phantom{\frac{1}{192\pi f^2}}-a^2\bigl(16(\textstyle{\frac{5}{8}}\alpha^2+\beta^2+\textstyle{\frac{1}{2}}\alpha\beta)+\otd(\alpha^2-20\beta^2-28\alpha\beta)\bigr)\times\nonumber\\
&&\phantom{\frac{1}{192\pi f^2}}\times\left(\delta_V^{\prime}(R_{\pi_{V}}^{[3,1]}m_{\pi_{V}}^3+R_{\eta_{V}}^{[3,1]}m_{\eta_{V}}^3+R_{\eta^{\prime}_{V}}^{[3,1]}m_{\eta^{\prime}_{V}}^3)+(V\rightarrow A)\right)\biggr]\nonumber\\
&+&\left(\frac{C}{8\pi f}\right)^2\biggl[\ohf\sum_t n_t[2\mc{F}(m_{\pi_t})+\mc{F}(m_{K_t})]\nonumber\\
&&\phantom{\left(\frac{C}{8\pi f}\right)^2}-2\mc{F}(m_{\pi_I})+\textstyle{\frac{20}{3}}[\mc{F}(m_{\pi_V})+\mc{F}(m_{\pi_A})]+\textstyle{\frac{4}{3}}\mc{F}(m_{\pi_T})+\textstyle{\frac{10}{3}}\mc{F}(m_{\pi_P})\nonumber\\
&&\phantom{\left(\frac{C}{8\pi f}\right)^2}-\textstyle{\frac{44}{3}}a^2\Bigl(\delta_V^{\prime}(R_{\pi_{V}}^{[3,1]}\mc{F}(m_{\pi_{V}})+R_{\eta_{V}}^{[3,1]}\mc{F}(m_{\eta_{V}})+R_{\eta^{\prime}_{V}}^{[3,1]}\mc{F}(m_{\eta^{\prime}_{V}}))+(V\rightarrow A)\Bigr)\biggr]\nonumber\\
&+&a^2\left(\otd\right)\bigl[\ohf A_1+2A_3+A_5+A_7+2A_2-A_4+4A_6+4A_8],\nonumber
\end{eqnarray}
\begin{eqnarray}
d_2&=&a^2\left(\otd\right)[\ohf B_1+\ohf B_3+\ohf B_5+2B_2+2B_4+2B_6\bigr].\label{d2}
\end{eqnarray}
The mass submatrices equal to $d_1\,I_{12}$ and $d_2\,I_8$ were suppressed in (\ref{rtmat}) and (\ref{rrtmat}), respectively.  

In the 2+1 flavor partially quenched case, double poles in the integrands of the hairpin loop integrals complicate the results for $c_1$, $c_2$, $c_3$, and $d_1$.  Evaluating the integrals and renormalizing the tree-level LECs gives
\begin{eqnarray*}
D_{XX}^I&\rightarrow&-\frac{i}{96\pi}\left(\thf R_{X_I}^{[2,2]}\cdot(m_{xx}^I)+D_{X_I,\,X_I}^{[2,2]}(m_{xx}^I)^3+D_{\eta_I,\,X_I}^{[2,2]}(m_{\eta_I})^3\right),\\
D_{XX}^{V,\,A}&\rightarrow&\frac{ia^2\delta_{V,\,A}^{\prime}}{32\pi}\Bigl(\thf R_{X_{V,\,A}}^{[3,2]}\cdot(m_{xx}^{V,\,A})+D_{X_{V,\,A},\,X_{V,\,A}}^{[3,2]}(m_{xx}^{V,\,A})^3\\
&&\phantom{\frac{ia^2\delta_{V,\,A}^{\prime}}{32\pi}}+D_{\eta_{V,\,A},\,X_{V,\,A}}^{[3,2]}(m_{\eta_{V,\,A}})^3+D_{\eta_{V,\,A}^{\prime},\,X_{V,\,A}}^{[3,2]}(m_{\eta_{V,\,A}^{\prime}})^3\Bigr),\\
E_{XX}^{V,\,A}&\rightarrow&-\frac{ia^2\delta_{V,\,A}^{\prime}}{(2\pi)^2}\biggl(R_{X_{V,\,A}}^{[3,2]}\frac{\mc{F}^{\prime}(m_{xx}^{V,\,A})}{2(m_{xx}^{V,\,A})}+D_{X_{V,\,A},\,X_{V,\,A}}^{[3,2]}\mc{F}(m_{xx}^{V,\,A})\\
&&\phantom{-\frac{ia^2\delta_{V,\,A}^{\prime}}{(2\pi)^2}}+D_{\eta_{V,\,A},\,X_{V,\,A}}^{[3,2]}\mc{F}(m_{\eta_{V,\,A}})+D_{\eta_{V,\,A}^{\prime},\,X_{V,\,A}}^{[3,2]}\mc{F}(m_{\eta_{V,\,A}^{\prime}})\biggr),
\end{eqnarray*}
where $X_{I,\,V,\,A}$ denote the states with masses $m_{xx}^{I,\,V,\,A}$, and, adapting the notation of \cite{Aubin:2003uc}, $R_{l_{V,\,A}}^{[3,2]}$ is shorthand for $R_{l_{V,\,A}}^{[3,2]}(\{m_{xx}^{V,\,A},\,m_{\eta_{V,\,A}},\,m_{\eta^{\prime}_{V,\,A}}\};\{m_{uu}^{V,\,A},\,m_{ss}^{V,\,A}\})$, \[D_{l_{V,\,A},\,X_{V,\,A}}^{[3,2]}\equiv+\frac{d}{d[(m_{xx}^{V,\,A})^2]}R_{l_{V,\,A}}^{[3,2]},\] $R_{l_I}^{[2,2]}$ is shorthand for $R_{l_I}^{[2,2]}(\{m_{xx}^I,\,m_{\eta_I}\};\{m_{uu}^I,\,m_{ss}^I\})|_{m^2\rightarrow-m^2}$, and \[D_{l_I,\,X_I}^{[2,2]}\equiv+\frac{d}{d[(m_{xx}^I)^2]}R_{l_I}^{[2,2]}.\]  Working in Minkowski space leads to the plus signs before the derivatives and the necessity of evaluating $R_{l_I}^{[2,2]}$ with all masses $m$ replaced, relative to the conventions of \cite{Aubin:2003uc}, by $im$, where $i^2=-1$; $R_{l_{V,\,A}}^{[3,2]}$ is invariant under this substitution.  The prime on $\mc{F}^{\prime}(m)$ represents the derivative with respect to $m$.  

Combining the results for the hairpin integrals with the coefficients $c^{\mr{sea}}$, the coefficients listed in Tables~\ref{Ctohf} and \ref{Ctthf}, and the results (\ref{loopohf}) and (\ref{loopthf}), the parameters $c_1$, $c_2$, $c_3$, and $d_1$ are, in the 2+1 flavor partially quenched case,
\begin{eqnarray}
c_1&=&M_0-2(\alpha_M+\beta_M)m_x-2\sigma_M(2m_u+m_s)\label{pqc1}\\
&-&\frac{1}{192\pi f^2}\biggl[(\textstyle{\frac{5}{8}}\alpha^2+\beta^2+\textstyle{\frac{1}{2}}\alpha\beta)\sum_t n_t\left(2(m_{xu}^t)^3+(m_{xs}^t)^3\right)\nonumber\\
&&\phantom{\frac{1}{192\pi f^2}}+(\shf\alpha^2+2\beta^2+10\alpha\beta)\left((m_{xx}^I)^3+(m_{xx}^P)^3\right)+(\alpha^2-20\beta^2-28\alpha\beta)(m_{xx}^T)^3\nonumber\\
&&\phantom{\frac{1}{192\pi f^2}}-8(\alpha+\beta)^2\left(\thf R_{X_I}^{[2,2]}\cdot(m_{xx}^I)+D_{X_I,\,X_I}^{[2,2]}(m_{xx}^I)^3+D_{\eta_I,\,X_I}^{[2,2]}(m_{\eta_I})^3\right)\nonumber\\
&&\phantom{\frac{1}{192\pi f^2}}+16a^2(\textstyle{\frac{5}{8}}\alpha^2+\beta^2+\textstyle{\frac{1}{2}}\alpha\beta)\Bigl[\delta_V^{\prime}\Bigl(\thf R_{X_{V}}^{[3,2]}\cdot(m_{xx}^{V})+D_{X_{V},\,X_{V}}^{[3,2]}(m_{xx}^{V})^3\nonumber\\
&&\phantom{\frac{1}{192\pi f^2}+16a^2(\textstyle{\frac{5}{8}}\alpha^2+\beta^2+\textstyle{\frac{1}{2}}\alpha\beta)}+D_{\eta_{V},\,X_{V}}^{[3,2]}(m_{\eta_{V}})^3+D_{\eta_{V}^{\prime},\,X_{V}}^{[3,2]}(m_{\eta_{V}^{\prime}})^3\Bigr)+(V\rightarrow A)\Bigr]\biggr]\nonumber\\
&+&\left(\frac{C}{8\pi f}\right)^2\biggl[\ohf\sum_t n_t[2\mc{F}(m_{xu}^t)+\mc{F}(m_{xs}^t)]\nonumber\\
&&\phantom{\left(\frac{C}{8\pi f}\right)^2}-2[\mc{F}(m_{xx}^I)+\mc{F}(m_{xx}^P)]+20\mc{F}(m_{xx}^T)\nonumber\\
&&\phantom{\left(\frac{C}{8\pi f}\right)^2}+8a^2\Bigl[\delta_V^{\prime}\Bigl(R_{X_{V}}^{[3,2]}\frac{\mc{F}^{\prime}(m_{xx}^{V})}{2(m_{xx}^{V})}+D_{X_{V},\,X_{V}}^{[3,2]}\mc{F}(m_{xx}^{V})\nonumber\\
&&\phantom{\left(\frac{C}{8\pi f}\right)^2+8a^2}+D_{\eta_{V},\,X_{V}}^{[3,2]}\mc{F}(m_{\eta_{V}})+D_{\eta_{V}^{\prime},\,X_{V}}^{[3,2]}\mc{F}(m_{\eta_{V}^{\prime}})\Bigr)+(V\rightarrow A)\Bigr]\biggr]\nonumber\\
&-&a^2\left[-\ohf A_1-A_3+A_2-4A_4\right],\nonumber
\end{eqnarray}
\begin{eqnarray}
c_2&=&M_0-2(\alpha_M+\beta_M)m_x-2\sigma_M(2m_u+m_s)\label{pqc2}\\
&-&\frac{1}{192\pi f^2}\biggl[(\textstyle{\frac{5}{8}}\alpha^2+\beta^2+\textstyle{\frac{1}{2}}\alpha\beta)\sum_t n_t\left(2(m_{xu}^t)^3+(m_{xs}^t)^3\right)\nonumber\\
&&\phantom{\frac{1}{192\pi f^2}}+(\shf\alpha^2+2\beta^2+10\alpha\beta)(m_{xx}^I)^3+(5\alpha^2-4\beta^2+4\alpha\beta)((m_{xx}^V)^3+(m_{xx}^A)^3)\nonumber\\
&&\phantom{\frac{1}{192\pi f^2}}+(-3\alpha^2-12\beta^2-24\alpha\beta)(m_{xx}^T)^3+(-\fhf\alpha^2+2\beta^2-2\alpha\beta)(m_{xx}^P)^3\nonumber\\
&&\phantom{\frac{1}{192\pi f^2}}-8(\alpha+\beta)^2\left(\thf R_{X_I}^{[2,2]}\cdot(m_{xx}^I)+D_{X_I,\,X_I}^{[2,2]}(m_{xx}^I)^3+D_{\eta_I,\,X_I}^{[2,2]}(m_{\eta_I})^3\right)\nonumber\\
&&\phantom{\frac{1}{192\pi f^2}}+a^2\bigl(16(\textstyle{\frac{5}{8}}\alpha^2+\beta^2+\textstyle{\frac{1}{2}}\alpha\beta)+(5\alpha^2-4\beta^2+4\alpha\beta)\bigr)\times\nonumber\\
&&\phantom{\frac{1}{192\pi f^2}}\times\Bigl[\delta_V^{\prime}\Bigl(\thf R_{X_{V}}^{[3,2]}\cdot(m_{xx}^{V})+D_{X_{V},\,X_{V}}^{[3,2]}(m_{xx}^{V})^3\nonumber\\
&&\phantom{\frac{1}{192\pi f^2}\times\ }+D_{\eta_{V},\,X_{V}}^{[3,2]}(m_{\eta_{V}})^3+D_{\eta_{V}^{\prime},\,X_{V}}^{[3,2]}(m_{\eta_{V}^{\prime}})^3\Bigr)+(V\rightarrow A)\Bigr]\biggr]\nonumber\\
&+&\left(\frac{C}{8\pi f}\right)^2\biggl[\ohf\sum_t n_t[2\mc{F}(m_{xu}^t)+\mc{F}(m_{xs}^t)]\nonumber\\
&&\phantom{\left(\frac{C}{8\pi f}\right)^2}-2[\mc{F}(m_{xx}^I)+\mc{F}(m_{xx}^P)]+4[\mc{F}(m_{xx}^V)+\mc{F}(m_{xx}^A)]+12\mc{F}(m_{xx}^T)\nonumber\\
&&\phantom{\left(\frac{C}{8\pi f}\right)^2}+12a^2\Bigl[\delta_V^{\prime}\Bigl(R_{X_{V}}^{[3,2]}\frac{\mc{F}^{\prime}(m_{xx}^{V})}{2(m_{xx}^{V})}+D_{X_{V},\,X_{V}}^{[3,2]}\mc{F}(m_{xx}^{V})\nonumber\\
&&\phantom{\left(\frac{C}{8\pi f}\right)^2+12a^2}+D_{\eta_{V},\,X_{V}}^{[3,2]}\mc{F}(m_{\eta_{V}})+D_{\eta_{V}^{\prime},\,X_{V}}^{[3,2]}\mc{F}(m_{\eta_{V}^{\prime}})\Bigr)+(V\rightarrow A)\Bigr]\biggr]\nonumber\\
&-&a^2\left[\ohf A_1-A_5-A_7-3A_4\right],\nonumber
\end{eqnarray}
\begin{eqnarray}
c_3&=&-\frac{1}{192\pi f^2}\biggl[{\sqrt 6}(\alpha^2+4\beta^2+8\alpha\beta)((m_{xx}^V)^3-(m_{xx}^A)^3)+a^2{\sqrt 6}(\alpha^2+4\beta^2+8\alpha\beta)\times\label{pqc3}\\
&&\phantom{\frac{1}{192\pi f^2}}\times\Bigl[\delta_V^{\prime}\Bigl(\thf R_{X_{V}}^{[3,2]}\cdot(m_{xx}^{V})+D_{X_{V},\,X_{V}}^{[3,2]}(m_{xx}^{V})^3\nonumber\\
&&\phantom{\frac{1}{192\pi f^2}\times\ }+D_{\eta_{V},\,X_{V}}^{[3,2]}(m_{\eta_{V}})^3+D_{\eta_{V}^{\prime},\,X_{V}}^{[3,2]}(m_{\eta_{V}^{\prime}})^3\Bigr)+(V\rightarrow A)\Bigr]\biggr]\nonumber\\
&+&\left(\frac{C}{8\pi f}\right)^2\biggl[-4{\sqrt 6}[\mc{F}(m_{xx}^V)-\mc{F}(m_{xx}^A)]\nonumber\\
&&\phantom{\left(\frac{C}{8\pi f}\right)^2}-4a^2{\sqrt 6}\Bigl[\delta_V^{\prime}\Bigl(R_{X_{V}}^{[3,2]}\frac{\mc{F}^{\prime}(m_{xx}^{V})}{2(m_{xx}^{V})}+D_{X_{V},\,X_{V}}^{[3,2]}\mc{F}(m_{xx}^{V})\nonumber\\
&&\phantom{\left(\frac{C}{8\pi f}\right)^2+12a^2}+D_{\eta_{V},\,X_{V}}^{[3,2]}\mc{F}(m_{\eta_{V}})+D_{\eta_{V}^{\prime},\,X_{V}}^{[3,2]}\mc{F}(m_{\eta_{V}^{\prime}})\Bigr)+(V\rightarrow A)\Bigr]\biggr]\nonumber\\
&-&a^2{\sqrt 6}\left[A_6-A_8\right],\nonumber
\end{eqnarray}
\begin{eqnarray}
d_1&=&M_0-2(\alpha_M+\beta_M)m_x-2\sigma_M(2m_u+m_s)\label{pqd1}\\
&-&\frac{1}{192\pi f^2}\biggl[(\textstyle{\frac{5}{8}}\alpha^2+\beta^2+\textstyle{\frac{1}{2}}\alpha\beta)\sum_t n_t\left(2(m_{xu}^t)^3+(m_{xs}^t)^3\right)\nonumber\\
&&\phantom{\frac{1}{192\pi f^2}}+(\shf\alpha^2+2\beta^2+10\alpha\beta)(m_{xx}^I)^3+\otd(\alpha^2-20\beta^2-28\alpha\beta)((m_{xx}^V)^3+(m_{xx}^A)^3)\nonumber\\
&&\phantom{\frac{1}{192\pi f^2}}+\otd(11\alpha^2-4\beta^2+16\alpha\beta)(m_{xx}^T)^3+\otd(\ohf\alpha^2-10\beta^2-14\alpha\beta)(m_{xx}^P)^3\nonumber\\
&&\phantom{\frac{1}{192\pi f^2}}-8(\alpha+\beta)^2\left(\thf R_{X_I}^{[2,2]}\cdot(m_{xx}^I)+D_{X_I,\,X_I}^{[2,2]}(m_{xx}^I)^3+D_{\eta_I,\,X_I}^{[2,2]}(m_{\eta_I})^3\right)\nonumber\\
&&\phantom{\frac{1}{192\pi f^2}}+a^2\bigl(16(\textstyle{\frac{5}{8}}\alpha^2+\beta^2+\textstyle{\frac{1}{2}}\alpha\beta)+\otd(\alpha^2-20\beta^2-28\alpha\beta)\bigr)\times\nonumber\\
&&\phantom{\frac{1}{192\pi f^2}}\times\Bigl[\delta_V^{\prime}\Bigl(\thf R_{X_{V}}^{[3,2]}\cdot(m_{xx}^{V})+D_{X_{V},\,X_{V}}^{[3,2]}(m_{xx}^{V})^3\nonumber\\
&&\phantom{\frac{1}{192\pi f^2}\times\ }+D_{\eta_{V},\,X_{V}}^{[3,2]}(m_{\eta_{V}})^3+D_{\eta_{V}^{\prime},\,X_{V}}^{[3,2]}(m_{\eta_{V}^{\prime}})^3\Bigr)+(V\rightarrow A)\Bigr]\biggr]\nonumber\\
&+&\left(\frac{C}{8\pi f}\right)^2\biggl[\ohf\sum_t n_t[2\mc{F}(m_{xu}^t)+\mc{F}(m_{xs}^t)]\nonumber\\
&&\phantom{\left(\frac{C}{8\pi f}\right)^2}-2\mc{F}(m_{xx}^I)+\textstyle{\frac{20}{3}}[\mc{F}(m_{xx}^V)+\mc{F}(m_{xx}^A)]+\textstyle{\frac{4}{3}}\mc{F}(m_{xx}^T)+\textstyle{\frac{10}{3}}\mc{F}(m_{xx}^P)\nonumber\\
&&\phantom{\left(\frac{C}{8\pi f}\right)^2}+{\textstyle\frac{44}{3}}a^2\Bigl[\delta_V^{\prime}\Bigl(R_{X_{V}}^{[3,2]}\frac{\mc{F}^{\prime}(m_{xx}^{V})}{2(m_{xx}^{V})}+D_{X_{V},\,X_{V}}^{[3,2]}\mc{F}(m_{xx}^{V})\nonumber\\
&&\phantom{\left(\frac{C}{8\pi f}\right)^2+12a^2}+D_{\eta_{V},\,X_{V}}^{[3,2]}\mc{F}(m_{\eta_{V}})+D_{\eta_{V}^{\prime},\,X_{V}}^{[3,2]}\mc{F}(m_{\eta_{V}^{\prime}})\Bigr)+(V\rightarrow A)\Bigr]\biggr]\nonumber\\
&+&a^2\left(\otd\right)\bigl[\ohf A_1+2A_3+A_5+A_7+2A_2-A_4+4A_6+4A_8].\nonumber
\end{eqnarray}

The distinct, nontrivial elements of the mass matrix of the single-flavor nucleons are simply related to the parameters $c_i$ and $d_j$, $j=1,\,2$:
\begin{subequations}\label{finalforms}
\begin{eqnarray}
\bra{N_{112}}M\ket{N_{112}}&=&c_1+c_4\\
\bra{\Lambda_{341}}M\ket{\Lambda_{341}}&=&c_2+c_5\\
\bra{{\textstyle \sqrt \frac{2}{3}}\,N_{332}-{\textstyle \frac{1}{\sqrt 3}}\,\Sigma_{341}}M\ket{{\textstyle \sqrt \frac{2}{3}}\,N_{332}-{\textstyle \frac{1}{\sqrt 3}}\,\Sigma_{341}}&=&d_1+c_6\\
\bra{N_{331}}M\ket{N_{331}}&=&d_1+d_2,\\
\nonumber\\
\bra{N_{112}}M\ket{\Lambda_{341}}&=&c_3+c_7\\
\bra{N_{112}}M\ket{{\textstyle \sqrt \frac{2}{3}}\,N_{332}-{\textstyle \frac{1}{\sqrt 3}}\,\Sigma_{341}}&=&c_8\\
\bra{\Lambda_{341}}M\ket{{\textstyle \sqrt \frac{2}{3}}\,N_{332}-{\textstyle \frac{1}{\sqrt 3}}\,\Sigma_{341}}&=&c_9.
\end{eqnarray}
\end{subequations}
The \order{\asq} tree-level terms can be roughly estimated by noting that $\mr{Tree}(\mq)-a^2\sigma^{FF(A)}$ absorbs the renormalization-scale dependence of the loop diagrams with virtual spin-\thf\ baryons~\cite{Manohar:1983md}.  At a given scale, one expects the counterterms to be at least as large as the change in the loop diagrams under a change in the renormalization scale by an amount of order unity.  If at some scale the counterterms were much smaller, then changing the scale would make the magnitude of the counterterms comparable to the change in the loops.  For quark masses and lattice spacings for which the staggered chiral power counting is meaningful, the tree-level corrections of \order{\mq} should be roughly equal to the tree-level corrections of \order{\asq}; one expects the expansion in the tree-level masses of the staggered mesons to be meaningful at tree-level.  
Varying the renormalization scale in the loops and associating the resulting \order{\asq} terms with valence and sea contributions of \order{\mq} gives
\begin{eqnarray*}
\mr{Tree}(\asq)\gtrsim -2(\alpha_M+\beta_M)\frac{\asq}{56\lambda}\sum_t \Delta^{t}({\textstyle \frac{3}{4}} n_t+c_{3/2,t}^{\mr{val}})-2\sigma_M\frac{3\asq}{32\lambda}\sum_t \Delta^{t},
\end{eqnarray*}
where the term proportional to $(\alpha_M+\beta_M)$ comes from the valence sector, and the term proportional to $\sigma_M$, from the sea.  Given an estimate of the LECs $\alpha_M$, $\beta_M$, $\sigma_M$, and $\lambda$ from the continuum and the measured values of the meson mass splittings $\Delta^{t}$~\cite{Aubin:2004fs}, one estimates the splittings to be roughly 10-40 MeV for lattice spacings of current interest.  This splitting in the flavor-symmetric nucleon masses may have been observed already with unimproved staggered quarks~\cite{Fukugita:1993}.  Different nucleon operators transforming in the \mb{8} of \gts\ would generically have differing overlaps with the nondegenerate states in the \mb{20_M}, which could generate the observed operator dependence of the central mass values.
\subsection{\label{connect}Correspondence between chiral forms and interpolating fields}
The connection between the interpolating operators and the staggered chiral forms may be obtained by decomposing the irreps of the forms into irreps of the operators.  To \order{\varepsilon^3}, the staggered chiral forms respect the parity--spin--remnant-taste symmetry $\mc{P}\times SU(2)_E\times[\rrt]$, while the operators transform within irreps of \gts.  For the flavor-symmetric nucleons, the relevant decompositions are
\begin{eqnarray*}
SU(2)_E\times[\rrt]&\supset&\gts\\
\mb{(\ohf,\ 8)}&\rightarrow&\mb{16}\\
\mb{(\ohf,\ 4)}&\rightarrow&\mb{8},
\end{eqnarray*}
where the direct product with continuum parity has been suppressed on both sides of the decompositions.  Operators transforming in the \mb{8} of \gts\ overlap members of $SU(2)_E\times[\rrt]$ $\mb{(\ohf,\ 4)}$'s; operators transforming in the \mb{16} overlap members of $\mb{(\ohf,\ 8)}$'s.  To \order{\varepsilon^3} in S$\chi$PT, members of a given $SU(2)_E\times[\rrt]$ irrep are degenerate, so the exact linear combination of states within the $SU(2)_E\times[\rrt]$ multiplet created by a given operator is unimportant; the staggered chiral forms are degenerate.  

To \order{\varepsilon^3}, the flavor-symmetric nucleons transform in one $\mb{(\ohf,\ 8)}$ and three $\mb{(\ohf,\ 4)}$'s.  Operators transforming in the \gts\ \mb{8} overlap states in the three \mb{4}'s of \rrt:  Representative members of these irreps are the $N_{112}$, $\Lambda_{341}$, and ${\textstyle \sqrt \frac{2}{3}}\,N_{332}-{\textstyle \frac{1}{\sqrt 3}}\,\Sigma_{341}$.  Operators transforming in the \mb{16} overlap states in the \mb{8} of \rrt:  for example, the $N_{331}$.  For the case of 2+1 fully dynamical flavors, the staggered chiral forms for the masses of these states are given in Eq.~(\ref{finalforms}) with Eqs.~(\ref{c1}) through (\ref{d2}); for the 2+1 flavor partially quenched case, the corresponding results for $c_{1,2,3}$ and $d_1$ are given in Eqs.~(\ref{pqc1}), (\ref{pqc2}), (\ref{pqc3}), and (\ref{pqd1}).

\section{\label{sum}Summary and Further Directions}
Staggered, partially quenched chiral perturbation theory has been formulated in the light-quark baryon sector by introducing taste degrees of freedom in heavy baryon chiral perturbation theory and breaking the taste symmetry by mapping the operators of the \order{\asq} Symanzik action to the heavy baryon Lagrangian.  Including operators of \order{\asq} in the Symanzik action allows one to calculate octet and decuplet baryon masses to \order{\varepsilon^3}=\order{m_q^{3/2}} in the staggered chiral expansion.  As an example, the masses of the single-flavor nucleons have been calculated; the result for the mass matrix is consistent with the pattern of degeneracies and mixings implied by the remnant taste symmetries, \rt\ and \rrt.  In the rest frame of the heavy baryon, the symmetry \rrt\ emerges as the taste symmetry of the chiral Lagrangian mapped from type $B$ four-fermion operators in the Symanzik action~\cite{Lee:1999zx}; the resulting spin-taste $SU(2)_E\times[\rrt]$ protects against mixing between states of different spin.  In the continuum limit, taste restoration forces all off-diagonal elements of the mass matrix to vanish, while the diagonal elements reduce to the result obtained for the partially quenched nucleon in continuum HB$\chi$PT~\cite{Chen:2001yi,Walker-Loud:2004hf}.  

The splittings in the nucleon mass must vanish as taste is restored, so they could be used to test taste restoration.  The staggered chiral forms given here are those needed to quantify taste violations in simulation results obtained from the local corner-wall operators of~\cite{Aubin:2004wf}.  At small quark masses, the lattice spacing and quark mass dependences of the data qualitatively agree with the staggered chiral forms given here.  At larger quark masses, the \order{\varepsilon^3} chiral forms decrease as $-m_\phi^3$, while the data continues to increase.  Fits to continuum $\chi$PT that include analytic terms of \order{m_q^2} seem to accurately describe the general trend of the data at large quark mass.  At sufficiently large quark mass and fixed lattice spacing, continuum \order{\varepsilon^4}=\order{m_q^2} terms will dominate over terms that are formally \order{\varepsilon^4}=\order{m_q\asq}=\order{a^4}, so it might be possible to describe the data by supplementing the results given here with continuum terms of \order{m_q^2}.  Alternatively, some authors argue that dimensional regularization in B$\chi$PT incorporates spurious high-energy physics~\cite{Donoghue:1998bs}.  If true, then it is possible that the data at larger quark masses would be better described by using a cut-off regulator.  Such an approach amounts to resumming the perturbative expansion and might not require terms of \order{\varepsilon^4}.  Minimally, departing from dimensional regularization would require recalculating the loop integrals at \order{\varepsilon^3}.  

Operators transforming in the \mb{8} of \gts\ interpolate to three nucleon states and two $\Delta$ states.  Operators transforming in the \mb{16} of \gts\ interpolate to only one nucleon, but to three $\Delta$'s~\cite{Bailey:2006zn}.  For extracting the mass spectrum, one would prefer operators that do not interpolate to states in \su{4}{T}-degenerate multiplets, which are split and mixed by discretization effects.  By these criteria, operators that would be ideal for extracting the masses of the nucleon and the lightest decuplet were identified and constructed in~\cite{Bailey:2006zn}.  The associated chiral forms for the nucleon, the $\Delta$, and the $\Omega^-$ have been calculated to \order{\varepsilon^3}; these and the chiral forms for the $\Sigma^*$ and $\Xi^*$ will be reported in a future publication.  There are plans to use these operators and chiral forms in the near future~\cite{Bailey:coming}.

\begin{acknowledgments}
The guidance and assistance of C.~Bernard were essential at nearly every stage of this project.  Funding was supplied in part by the U.S. Department of Energy under grant DE-FG02-91ER40628.
\end{acknowledgments}

\appendix
\section{\label{xirep}Quark-flow coefficients and taste matrices}
Here I write down the matrix elements of the quark-flow coefficients $c_{1/2,t}^{\mr{sea,val}}$ and $c_{3/2,t}^{\mr{sea,val}}$ in terms of the elements of the taste matrices $\xi_\tau$, $\tau\in\{I,\ \mu,\ \mu\nu(\mu<\nu),\ \mu5,\ 5\}$.  These explicit forms can be used to verify that, in the Weyl representation of the taste matrices, the loop contributions are consistent with the parameterization~(\ref{rtmat}).  Because they are independent of the representation used for the taste matrices, they can also be used to verify that the continuum limits of the \order{\varepsilon^3} staggered chiral forms equal, in any representation, the corresponding \order{\varepsilon^3} chiral forms of continuum HB$\chi$PT.  The result is immediate if one uses the completeness relation of the taste matrices:
\begin{equation}
\sum_\tau \xi_{ab}^\tau\xi_{cd}^\tau=4\delta_{ad}\delta_{bc},\nonumber
\end{equation}
where $\xi^\tau\equiv\xi_\tau$ and $\xi^\tau_{ab}\equiv\bra{a}\xi^\tau\ket{b}$, the $(a,\,b)$-element of $\xi^\tau$ in the fundamental representation of $SU(4)_T$.  

In the basis (\ref{basis}) of Table~\ref{RTeigen1}, the diagonal elements of $c_{1/2,t}^{\mr{sea}}$ and $c_{1/2,t}^{\mr{val}}$ are
\begin{eqnarray*}
\bra{N_{aab}}c_{1/2,t}^{\mr{sea}}\ket{N_{aab}}&=&n_t(\textstyle{\frac{5}{8}}\alpha^2+\beta^2+\textstyle{\frac{1}{2}}\alpha\beta)\\
\bra{N_{aab}}c_{1/2,t}^{\mr{val}}\ket{N_{aab}}&=&\sum_{\tau\in s_t}\bigl\{\alpha^2\left[\textstyle{\frac{1}{6}}(\xi_{ab}^{\tau}\xi_{ba}^{\tau}+10\xi_{aa}^{\tau}\xi_{bb}^{\tau}+11(\xi_{aa}^{\tau})^2)\right]\\
&+&\beta^2\left[\textstyle{\frac{2}{3}}(-(\xi_{aa}^{\tau})^2-5\xi_{ab}^{\tau}\xi_{ba}^{\tau}+4\xi_{aa}^{\tau}\xi_{bb}^{\tau})\right]\\
&+&2\alpha\beta\left[\textstyle{\frac{1}{3}}(-7\xi_{ab}^{\tau}\xi_{ba}^{\tau}+11\xi_{aa}^{\tau}\xi_{bb}^{\tau}+4(\xi_{aa}^{\tau})^2)\right]\bigr\}
\end{eqnarray*}
\begin{eqnarray*}
\bra{\Sigma_{abc}}c_{1/2,t}^{\mr{sea}}\ket{\Sigma_{abc}}&=&n_t(\textstyle{\frac{5}{8}}\alpha^2+\beta^2+\textstyle{\frac{1}{2}}\alpha\beta)\\
\bra{\Sigma_{abc}}c_{1/2,t}^{\mr{val}}\ket{\Sigma_{abc}}&=&\sum_{\tau\in s_t}\bigl\{\alpha^2\left[\textstyle{\frac{1}{6}}(11(\xi_{aa}^{\tau}\xi_{bb}^{\tau}+\xi_{ab}^{\tau}\xi_{ba}^{\tau})+\textstyle{\frac{1}{2}}(\xi_{ac}^{\tau}\xi_{ca}^{\tau}+\xi_{bc}^{\tau}\xi_{cb}^{\tau})+5(\xi_{aa}^{\tau}\xi_{cc}^{\tau}+\xi_{bb}^{\tau}\xi_{cc}^{\tau}))\right]\\
&+&\beta^2\left[-\textstyle{\frac{1}{3}}(2(\xi_{aa}^{\tau}\xi_{bb}^{\tau}+\xi_{ab}^{\tau}\xi_{ba}^{\tau})+5(\xi_{ac}^{\tau}\xi_{ca}^{\tau}+\xi_{bc}^{\tau}\xi_{cb}^{\tau})-4(\xi_{aa}^{\tau}\xi_{cc}^{\tau}+\xi_{bb}^{\tau}\xi_{cc}^{\tau}))\right]\\
&+&2\alpha\beta\left[\textstyle{\frac{1}{3}}(4(\xi_{aa}^{\tau}\xi_{bb}^{\tau}+\xi_{ab}^{\tau}\xi_{ba}^{\tau})-\textstyle{\frac{7}{2}}(\xi_{ac}^{\tau}\xi_{ca}^{\tau}+\xi_{bc}^{\tau}\xi_{cb}^{\tau})+\textstyle{\frac{11}{2}}(\xi_{aa}^{\tau}\xi_{cc}^{\tau}+\xi_{bb}^{\tau}\xi_{cc}^{\tau}))\right]\bigr\}
\end{eqnarray*}
\begin{eqnarray*}
\bra{\Lambda_{abc}}c_{1/2,t}^{\mr{sea}}\ket{\Lambda_{abc}}&=&n_t(\textstyle{\frac{5}{8}}\alpha^2+\beta^2+\textstyle{\frac{1}{2}}\alpha\beta)\\
\bra{\Lambda_{abc}}c_{1/2,t}^{\mr{val}}\ket{\Lambda_{abc}}&=&\sum_{\tau\in s_t}\bigl\{\alpha^2\left[\textstyle{\frac{1}{2}}(\xi_{aa}^{\tau}\xi_{bb}^{\tau}-\xi_{ab}^{\tau}\xi_{ba}^{\tau}+\textstyle{\frac{5}{2}}(\xi_{ac}^{\tau}\xi_{ca}^{\tau}+\xi_{bc}^{\tau}\xi_{cb}^{\tau})+3(\xi_{aa}^{\tau}\xi_{cc}^{\tau}+\xi_{bb}^{\tau}\xi_{cc}^{\tau}))\right]\\
&+&\beta^2\left[2(\xi_{aa}^{\tau}\xi_{bb}^{\tau}-\xi_{ab}^{\tau}\xi_{ba}^{\tau})-(\xi_{ac}^{\tau}\xi_{ca}^{\tau}+\xi_{bc}^{\tau}\xi_{cb}^{\tau})\right]\\
&+&2\alpha\beta\left[\textstyle{\frac{1}{2}}(4(\xi_{aa}^{\tau}\xi_{bb}^{\tau}-\xi_{ab}^{\tau}\xi_{ba}^{\tau})+(\xi_{ac}^{\tau}\xi_{ca}^{\tau}+\xi_{bc}^{\tau}\xi_{cb}^{\tau})+3(\xi_{aa}^{\tau}\xi_{cc}^{\tau}+\xi_{bb}^{\tau}\xi_{cc}^{\tau}))\right]\bigr\}.
\end{eqnarray*}
The remnant taste symmetry \rt\ forces many of the off-diagonal elements to vanish.  The distinct off-diagonal elements that do not vanish by this symmetry are
\begin{eqnarray*}
\bra{N_{aab}}c_{1/2,t}^{\mr{sea}}\ket{\Lambda_{dfg}}&=&0 \\
\bra{N_{aab}}c_{1/2,t}^{\mr{val}}\ket{\Lambda_{dfg}}&=&\sum_{\tau\in s_t}\bigl\{\alpha^2\bigl[\textstyle{\frac{1}{2{\sqrt6}}}(7\xi_{ag}^{\tau}(\delta_{ad}\xi_{bf}^{\tau}-\delta_{af}\xi_{bd}^{\tau})+4\xi_{bg}^{\tau}(\delta_{ad}\xi_{af}^{\tau}-\delta_{af}\xi_{ad}^{\tau})\\
&+&3\delta_{ag}(\xi_{ad}^{\tau}\xi_{bf}^{\tau}-\xi_{af}^{\tau}\xi_{bd}^{\tau})+11\xi_{ag}^{\tau}(\delta_{bf}\xi_{ad}^{\tau}-\delta_{bd}\xi_{af}^{\tau}))\bigr] \\
&+&\beta^2\bigl[-\textstyle{\frac{2}{\sqrt 6}}(3\delta_{ag}(\xi_{af}^{\tau}\xi_{bd}^{\tau}-\xi_{ad}^{\tau}\xi_{bf}^{\tau})+\delta_{bf}\xi_{ag}^{\tau}(\xi_{ad}^{\tau}-\xi_{af}^{\tau})\\&+&\xi_{ag}^{\tau}(\delta_{af}\xi_{bd}^{\tau}-\delta_{ad}\xi_{bf}^{\tau})+2\xi_{bg}^{\tau}(\delta_{ad}\xi_{af}^{\tau}-\delta_{af}\xi_{ad}^{\tau}))\bigr] \\
&+&2\alpha\beta\bigl[\textstyle{\frac{1}{\sqrt 6}}(\xi_{bg}^{\tau}(\delta_{af}\xi_{ad}^{\tau}-\delta_{ad}\xi_{af}^{\tau})+5\xi_{ag}^{\tau}(\delta_{ad}\xi_{bf}^{\tau}-\delta_{af}\xi_{bd}^{\tau}) \\
&+&6\delta_{ag}(\xi_{ad}^{\tau}\xi_{bf}^{\tau}-\xi_{af}^{\tau}\xi_{bd}^{\tau})+4\xi_{ag}^{\tau}(\delta_{bf}\xi_{ad}^{\tau}-\delta_{bd}\xi_{af}^{\tau}))\bigr]\bigr\}
\end{eqnarray*}
The matrix element $\bra{N_{aab}}c_{1/2,t}^{\mr{sea}}\ket{\Lambda_{dfg}}$ vanishes accidentally; calculation shows it is proportional to $\delta_{ag}(\delta_{bf}\delta_{ad}-\delta_{bd}\delta_{af})$, and no two of the indices $dfg$ are ever equal for the states $\Lambda_{dfg}$.  In the Weyl representation, these matrix elements of $c_{1/2,t}^{\mr{sea,val}}$ are consistent with the degeneracies and mixings parameterized in (\ref{rtmat}).  In the Weyl representation, the distinct, nontrivial matrix elements of $c_{1/2,t}^{\mr{val}}$ are listed in Table~\ref{Ctohf}.  

The diagonal elements of $c_{3/2,t}^{\mr{sea}}$ and $c_{3/2,t}^{\mr{val}}$ are
\begin{eqnarray*}
\bra{N_{aab}}c_{3/2,t}^{\mr{sea}}\ket{N_{aab}}&=&\ohf n_t\\
\bra{N_{aab}}c_{3/2,t}^{\mr{val}}\ket{N_{aab}}&=&\sum_{\tau\in s_t}\textstyle{\frac{2}{3}}\left[(\xi_{aa}^{\tau})^2+5\xi_{ab}^{\tau}\xi_{ba}^{\tau}-4\xi_{aa}^{\tau}\xi_{bb}^{\tau}\right]\\
\bra{\Sigma_{abc}}c_{3/2,t}^{\mr{sea}}\ket{\Sigma_{abc}}&=&\ohf n_t\\
\bra{\Sigma_{abc}}c_{3/2,t}^{\mr{val}}\ket{\Sigma_{abc}}&=&\sum_{\tau\in s_t}\textstyle{\frac{1}{3}}\left[2(\xi_{aa}^{\tau}\xi_{bb}^{\tau}+\xi_{ab}^{\tau}\xi_{ba}^{\tau})+5(\xi_{ac}^{\tau}\xi_{ca}^{\tau}+\xi_{bc}^{\tau}\xi_{cb}^{\tau})-4(\xi_{aa}^{\tau}\xi_{cc}^{\tau}+\xi_{bb}^{\tau}\xi_{cc}^{\tau})\right]\\
\bra{\Lambda_{abc}}c_{3/2,t}^{\mr{sea}}\ket{\Lambda_{abc}}&=&\ohf n_t\\
\bra{\Lambda_{abc}}c_{3/2,t}^{\mr{val}}\ket{\Lambda_{abc}}&=&\sum_{\tau\in s_t}\left[-2(\xi_{aa}^{\tau}\xi_{bb}^{\tau}-\xi_{ab}^{\tau}\xi_{ba}^{\tau})+\xi_{ac}^{\tau}\xi_{ca}^{\tau}+\xi_{bc}^{\tau}\xi_{cb}^{\tau}\right],
\end{eqnarray*}
while the off-diagonal elements that are not required to vanish by symmetry are
\begin{eqnarray*}
\bra{N_{aab}}c_{3/2,t}^{\mr{sea}}\ket{\Lambda_{dfg}}&=&0\\
\bra{N_{aab}}c_{3/2,t}^{\mr{val}}\ket{\Lambda_{dfg}}&=&\sum_{\tau\in s_t}\bigl[\textstyle{\frac{2}{\sqrt 6}}(3\delta_{ag}(\xi_{af}^{\tau}\xi_{bd}^{\tau}-\xi_{ad}^{\tau}\xi_{bf}^{\tau})+\delta_{bf}\xi_{ag}^{\tau}(\xi_{ad}^{\tau}-\xi_{af}^{\tau})\\&+&\xi_{ag}^{\tau}(\delta_{af}\xi_{bd}^{\tau}-\delta_{ad}\xi_{bf}^{\tau})+2\xi_{bg}^{\tau}(\delta_{ad}\xi_{af}^{\tau}-\delta_{af}\xi_{ad}^{\tau}))\bigr].
\end{eqnarray*}
In the Weyl representation, these matrix elements of $c_{3/2,t}^{\mr{sea,val}}$ are consistent with the degeneracies and mixings parameterized in (\ref{rtmat}).  In this representation, the distinct, nontrivial matrix elements of $c_{3/2,t}^{\mr{val}}$ are listed in Table~\ref{Ctthf}.  

\section{\label{FF}Four-fermion contributions and taste matrices}
Here I write down the matrix elements of $\sigma^{FF(A)}$ and $\sigma^{FF(B)}$ in terms of the elements of the taste matrices appearing in (\ref{FFA}) and (\ref{FFB}).  These results can be used to show that, in the Weyl representation, $\sigma^{FF(A)}$ and $\sigma^{FF(B)}$ are consistent with the forms (\ref{rtmat}) and (\ref{rrtmat}), respectively.  

For the diagonal $\sigma^{FF(A)}$ contributions, we have
\begin{eqnarray*}
\bra{N_{aab}}\sigma^{FF(A)}\ket{N_{aab}}&=&A_1\left[-\textstyle{\frac{1}{6}}(2(\xi_{aa}^5)^2+\xi_{ab}^5\xi_{ba}^5+\xi_{aa}^5\xi_{bb}^5)\right]\\
&+&A_3\left[-\textstyle{\frac{1}{6}}(2\xi_{aa}^{\mu\nu}\xi_{aa}^{\mu\nu}+\xi_{ab}^{\mu\nu}\xi_{ba}^{\mu\nu}+\xi_{aa}^{\mu\nu}\xi_{bb}^{\mu\nu})\right]\\
&+&A_5\left[-\textstyle{\frac{1}{6}}(2\xi_{aa}^{\nu}\xi_{aa}^{\nu}+\xi_{ab}^{\nu}\xi_{ba}^{\nu}+\xi_{aa}^{\nu}\xi_{bb}^{\nu})\right]\\
&+&A_7\left[-\textstyle{\frac{1}{6}}(2\xi_{aa}^{\nu5}\xi_{aa}^{\nu5}+\xi_{ab}^{\nu5}\xi_{ba}^{\nu5}+\xi_{aa}^{\nu5}\xi_{bb}^{\nu5})\right]\\
&+&A_2\left[\textstyle{\frac{1}{6}}((\xi_{aa}^5)^2-4\xi_{ab}^5\xi_{ba}^5+5\xi_{aa}^5\xi_{bb}^5)\right]\\
&+&A_4\left[\textstyle{\frac{1}{6}}(\xi_{aa}^{\mu\nu}\xi_{aa}^{\mu\nu}-4\xi_{ab}^{\mu\nu}\xi_{ba}^{\mu\nu}+5\xi_{aa}^{\mu\nu}\xi_{bb}^{\mu\nu})\right]\\
&+&A_6\left[\textstyle{\frac{1}{6}}(\xi_{aa}^{\nu}\xi_{aa}^{\nu}-4\xi_{ab}^{\nu}\xi_{ba}^{\nu}+5\xi_{aa}^{\nu}\xi_{bb}^{\nu})\right]\\
&+&A_8\left[\textstyle{\frac{1}{6}}(\xi_{aa}^{\nu5}\xi_{aa}^{\nu5}-4\xi_{ab}^{\nu5}\xi_{ba}^{\nu5}+5\xi_{aa}^{\nu5}\xi_{bb}^{\nu5})\right],
\end{eqnarray*}
\begin{eqnarray*}
\bra{\Sigma_{abc}}\sigma^{FF(A)}\ket{\Sigma_{abc}}&=&A_1\left[-\textstyle{\frac{1}{12}}(4(\xi_{ab}^5\xi_{ba}^5+\xi_{aa}^5\xi_{bb}^5)+\xi_{ac}^5\xi_{ca}^5+\xi_{aa}^5\xi_{cc}^5+\xi_{bc}^5\xi_{cb}^5+\xi_{bb}^5\xi_{cc}^5)\right]\\
&+&A_3\left[-\textstyle{\frac{1}{12}}(4(\xi_{ab}^{\mu\nu}\xi_{ba}^{\mu\nu}+\xi_{aa}^{\mu\nu}\xi_{bb}^{\mu\nu})+\xi_{ac}^{\mu\nu}\xi_{ca}^{\mu\nu}+\xi_{aa}^{\mu\nu}\xi_{cc}^{\mu\nu}+\xi_{bc}^{\mu\nu}\xi_{cb}^{\mu\nu}+\xi_{bb}^{\mu\nu}\xi_{cc}^{\mu\nu})\right]\\
&+&A_5\left[-\textstyle{\frac{1}{12}}(4(\xi_{ab}^{\nu}\xi_{ba}^{\nu}+\xi_{aa}^{\nu}\xi_{bb}^{\nu})+\xi_{ac}^{\nu}\xi_{ca}^{\nu}+\xi_{aa}^{\nu}\xi_{cc}^{\nu}+\xi_{bc}^{\nu}\xi_{cb}^{\nu}+\xi_{bb}^{\nu}\xi_{cc}^{\nu})\right]\\
&+&A_7\left[-\textstyle{\frac{1}{12}}(4(\xi_{ab}^{\nu5}\xi_{ba}^{\nu5}+\xi_{aa}^{\nu5}\xi_{bb}^{\nu5})+\xi_{ac}^{\nu5}\xi_{ca}^{\nu5}+\xi_{aa}^{\nu5}\xi_{cc}^{\nu5}+\xi_{bc}^{\nu5}\xi_{cb}^{\nu5}+\xi_{bb}^{\nu5}\xi_{cc}^{\nu5})\right]\\
&+&A_2\left[\textstyle{\frac{1}{12}}(2(\xi_{ab}^5\xi_{ba}^5+\xi_{aa}^5\xi_{bb}^5)-4\xi_{ac}^5\xi_{ca}^5+5\xi_{aa}^5\xi_{cc}^5-4\xi_{bc}^5\xi_{cb}^5+5\xi_{bb}^5\xi_{cc}^5)\right]\\
&+&A_4\left[\textstyle{\frac{1}{12}}(2(\xi_{ab}^{\mu\nu}\xi_{ba}^{\mu\nu}+\xi_{aa}^{\mu\nu}\xi_{bb}^{\mu\nu})-4\xi_{ac}^{\mu\nu}\xi_{ca}^{\mu\nu}+5\xi_{aa}^{\mu\nu}\xi_{cc}^{\mu\nu}-4\xi_{bc}^{\mu\nu}\xi_{cb}^{\mu\nu}+5\xi_{bb}^{\mu\nu}\xi_{cc}^{\mu\nu})\right]\\
&+&A_6\left[\textstyle{\frac{1}{12}}(2(\xi_{ab}^{\nu}\xi_{ba}^{\nu}+\xi_{aa}^{\nu}\xi_{bb}^{\nu})-4\xi_{ac}^{\nu}\xi_{ca}^{\nu}+5\xi_{aa}^{\nu}\xi_{cc}^{\nu}-4\xi_{bc}^{\nu}\xi_{cb}^{\nu}+5\xi_{bb}^{\nu}\xi_{cc}^{\nu})\right]\\
&+&A_8\left[\textstyle{\frac{1}{12}}(2(\xi_{ab}^{\nu5}\xi_{ba}^{\nu5}+\xi_{aa}^{\nu5}\xi_{bb}^{\nu5})-4\xi_{ac}^{\nu5}\xi_{ca}^{\nu5}+5\xi_{aa}^{\nu5}\xi_{cc}^{\nu5}-4\xi_{bc}^{\nu5}\xi_{cb}^{\nu5}+5\xi_{bb}^{\nu5}\xi_{cc}^{\nu5})\right],
\end{eqnarray*}
\begin{eqnarray*}
\bra{\Lambda_{abc}}\sigma^{FF(A)}\ket{\Lambda_{abc}}&=&A_1\left[-\textstyle{\frac{1}{4}}(\xi_{ac}^5\xi_{ca}^5+\xi_{aa}^5\xi_{cc}^5+\xi_{bc}^5\xi_{cb}^5+\xi_{bb}^5\xi_{cc}^5)\right]\\
&+&A_3\left[-\textstyle{\frac{1}{4}}(\xi_{ac}^{\mu\nu}\xi_{ca}^{\mu\nu}+\xi_{aa}^{\mu\nu}\xi_{cc}^{\mu\nu}+\xi_{bc}^{\mu\nu}\xi_{cb}^{\mu\nu}+\xi_{bb}^{\mu\nu}\xi_{cc}^{\mu\nu})\right]\\
&+&A_5\left[-\textstyle{\frac{1}{4}}(\xi_{ac}^{\nu}\xi_{ca}^{\nu}+\xi_{aa}^{\nu}\xi_{cc}^{\nu}+\xi_{bc}^{\nu}\xi_{cb}^{\nu}+\xi_{bb}^{\nu}\xi_{cc}^{\nu})\right]\\
&+&A_7\left[-\textstyle{\frac{1}{4}}(\xi_{ac}^{\nu5}\xi_{ca}^{\nu5}+\xi_{aa}^{\nu5}\xi_{cc}^{\nu5}+\xi_{bc}^{\nu5}\xi_{cb}^{\nu5}+\xi_{bb}^{\nu5}\xi_{cc}^{\nu5})\right]\\
&+&A_2\left[\textstyle{\frac{1}{4}}(2(\xi_{aa}^5\xi_{bb}^5-\xi_{ab}^5\xi_{ba}^5)+\xi_{aa}^5\xi_{cc}^5+\xi_{bb}^5\xi_{cc}^5)\right]\\
&+&A_4\left[\textstyle{\frac{1}{4}}(2(\xi_{aa}^{\mu\nu}\xi_{bb}^{\mu\nu}-\xi_{ab}^{\mu\nu}\xi_{ba}^{\mu\nu})+\xi_{aa}^{\mu\nu}\xi_{cc}^{\mu\nu}+\xi_{bb}^{\mu\nu}\xi_{cc}^{\mu\nu})\right]\\
&+&A_6\left[\textstyle{\frac{1}{4}}(2(\xi_{aa}^{\nu}\xi_{bb}^{\nu}-\xi_{ab}^{\nu}\xi_{ba}^{\nu})+\xi_{aa}^{\nu}\xi_{cc}^{\nu}+\xi_{bb}^{\nu}\xi_{cc}^{\nu})\right]\\
&+&A_8\left[\textstyle{\frac{1}{4}}(2(\xi_{aa}^{\nu5}\xi_{bb}^{\nu5}-\xi_{ab}^{\nu5}\xi_{ba}^{\nu5})+\xi_{aa}^{\nu5}\xi_{cc}^{\nu5}+\xi_{bb}^{\nu5}\xi_{cc}^{\nu5})\right],
\end{eqnarray*}
while the nontrivial off-diagonal terms are
\begin{eqnarray*}
\bra{N_{aab}}\sigma^{FF(A)}\ket{\Lambda_{dfg}}&=&A_1\bigl[\textstyle{\frac{1}{2{\sqrt 6}}}(\delta_{af}(\xi_{ad}^5\xi_{bg}^5+\xi_{ag}^5\xi_{bd}^5)+2\delta_{bd}\xi_{af}^5\xi_{ag}^5\\
&&\phantom{A_1}-\delta_{ad}(\xi_{af}^5\xi_{bg}^5+\xi_{ag}^5\xi_{bf}^5)-2\delta_{bf}\xi_{ad}^5\xi_{ag}^5)\bigr]\\
&+&A_3\bigl[\textstyle{\frac{1}{2{\sqrt 6}}}(\delta_{af}(\xi_{ad}^{\mu\nu}\xi_{bg}^{\mu\nu}+\xi_{ag}^{\mu\nu}\xi_{bd}^{\mu\nu})+2\delta_{bd}\xi_{af}^{\mu\nu}\xi_{ag}^{\mu\nu}\\
&&\phantom{A_3}-\delta_{ad}(\xi_{af}^{\mu\nu}\xi_{bg}^{\mu\nu}+\xi_{ag}^{\mu\nu}\xi_{bf}^{\mu\nu})-2\delta_{bf}\xi_{ad}^{\mu\nu}\xi_{ag}^{\mu\nu})\bigr]\\
&+&A_5\bigl[\textstyle{\frac{1}{2{\sqrt 6}}}(\delta_{af}(\xi_{ad}^{\nu}\xi_{bg}^{\nu}+\xi_{ag}^{\nu}\xi_{bd}^{\nu})+2\delta_{bd}\xi_{af}^{\nu}\xi_{ag}^{\nu}\\
&&\phantom{A_5}-\delta_{ad}(\xi_{af}^{\nu}\xi_{bg}^{\nu}+\xi_{ag}^{\nu}\xi_{bf}^{\nu})-2\delta_{bf}\xi_{ad}^{\nu}\xi_{ag}^{\nu})\bigr]\\
&+&A_7\bigl[\textstyle{\frac{1}{2{\sqrt 6}}}(\delta_{af}(\xi_{ad}^{\nu5}\xi_{bg}^{\nu5}+\xi_{ag}^{\nu5}\xi_{bd}^{\nu5})+2\delta_{bd}\xi_{af}^{\nu5}\xi_{ag}^{\nu5}\\
&&\phantom{A_7}-\delta_{ad}(\xi_{af}^{\nu5}\xi_{bg}^{\nu5}+\xi_{ag}^{\nu5}\xi_{bf}^{\nu5})-2\delta_{bf}\xi_{ad}^{\nu5}\xi_{ag}^{\nu5})\bigr]\\
&+&A_2\bigl[\textstyle{\frac{1}{2{\sqrt 6}}}(\delta_{af}(\xi_{ad}^5\xi_{bg}^5-2\xi_{ag}^5\xi_{bd}^5)+\delta_{ad}(2\xi_{bf}^5\xi_{ag}^5-\xi_{af}^5\xi_{bg}^5)\\
&&\phantom{A_2}+\delta_{bf}\xi_{ad}^5\xi_{ag}^5-\delta_{bd}\xi_{af}^5\xi_{ag}^5+3\delta_{ag}(\xi_{ad}^5\xi_{bf}^5-\xi_{af}^5\xi_{bd}^5))\bigr]\\
&+&A_4\bigl[\textstyle{\frac{1}{2{\sqrt 6}}}(\delta_{af}(\xi_{ad}^{\mu\nu}\xi_{bg}^{\mu\nu}-2\xi_{ag}^{\mu\nu}\xi_{bd}^{\mu\nu})+\delta_{ad}(2\xi_{bf}^{\mu\nu}\xi_{ag}^{\mu\nu}-\xi_{af}^{\mu\nu}\xi_{bg}^{\mu\nu})\\
&&\phantom{A_4}+\delta_{bf}\xi_{ad}^{\mu\nu}\xi_{ag}^{\mu\nu}-\delta_{bd}\xi_{af}^{\mu\nu}\xi_{ag}^{\mu\nu}+3\delta_{ag}(\xi_{ad}^{\mu\nu}\xi_{bf}^{\mu\nu}-\xi_{af}^{\mu\nu}\xi_{bd}^{\mu\nu}))\bigr]\\
&+&A_6\bigl[\textstyle{\frac{1}{2{\sqrt 6}}}(\delta_{af}(\xi_{ad}^{\nu}\xi_{bg}^{\nu}-2\xi_{ag}^{\nu}\xi_{bd}^{\nu})+\delta_{ad}(2\xi_{bf}^{\nu}\xi_{ag}^{\nu}-\xi_{af}^{\nu}\xi_{bg}^{\nu})\\
&&\phantom{A_6}+\delta_{bf}\xi_{ad}^{\nu}\xi_{ag}^{\nu}-\delta_{bd}\xi_{af}^{\nu}\xi_{ag}^{\nu}+3\delta_{ag}(\xi_{ad}^{\nu}\xi_{bf}^{\nu}-\xi_{af}^{\nu}\xi_{bd}^{\nu}))\bigr]\\
&+&A_8\bigl[\textstyle{\frac{1}{2{\sqrt 6}}}(\delta_{af}(\xi_{ad}^{\nu5}\xi_{bg}^{\nu5}-2\xi_{ag}^{\nu5}\xi_{bd}^{\nu5})+\delta_{ad}(2\xi_{bf}^{\nu5}\xi_{ag}^{\nu5}-\xi_{af}^{\nu5}\xi_{bg}^{\nu5})\\
&&\phantom{A_8}+\delta_{bf}\xi_{ad}^{\nu5}\xi_{ag}^{\nu5}-\delta_{bd}\xi_{af}^{\nu5}\xi_{ag}^{\nu5}+3\delta_{ag}(\xi_{ad}^{\nu5}\xi_{bf}^{\nu5}-\xi_{af}^{\nu5}\xi_{bd}^{\nu5}))\bigr].
\end{eqnarray*}
Repeated meson taste indices $\mu\nu$, $\nu$, and $\nu5$ are summed as in (\ref{FFA}); as for the quark-flow coefficients, the matrix is symmetric:  $\bra{\Lambda_{dfg}}\sigma^{FF(A)}\ket{N_{aab}}=\bra{N_{aab}}\sigma^{FF(A)}\ket{\Lambda_{dfg}}$.  In the Weyl representation, the matrix elements of $\sigma^{FF(A)}$ are consistent with the degeneracies and mixings parameterized in (\ref{rtmat}).  For this representation the distinct, nontrivial matrix elements are listed in Table~\ref{sigFFA}.

For the diagonal $\sigma^{FF(B)}$ contributions, first consider
\begin{eqnarray*}
\bra{N_{aab}}\sigma^{FF(B)}\ket{N_{aab}}&=&B_1\left[-\textstyle{\frac{1}{6}}(2\xi_{aa}^{4\nu}\xi_{aa}^{4\nu}+\xi_{ab}^{4\nu}\xi_{ba}^{4\nu}+\xi_{aa}^{4\nu}\xi_{bb}^{4\nu})\right]\\
&+&B_3\left[-\textstyle{\frac{1}{6}}(2\xi_{aa}^{4}\xi_{aa}^{4}+\xi_{ab}^{4}\xi_{ba}^{4}+\xi_{aa}^{4}\xi_{bb}^{4})\right]\\
&+&B_5\left[-\textstyle{\frac{1}{6}}(2\xi_{aa}^{45}\xi_{aa}^{45}+\xi_{ab}^{45}\xi_{ba}^{45}+\xi_{aa}^{45}\xi_{bb}^{45})\right]\\
&+&B_2\left[\textstyle{\frac{1}{6}}(\xi_{aa}^{4\nu}\xi_{aa}^{4\nu}-4\xi_{ab}^{4\nu}\xi_{ba}^{4\nu}+5\xi_{aa}^{4\nu}\xi_{bb}^{4\nu})\right]\\
&+&B_4\left[\textstyle{\frac{1}{6}}(\xi_{aa}^{4}\xi_{aa}^{4}-4\xi_{ab}^{4}\xi_{ba}^{4}+5\xi_{aa}^{4}\xi_{bb}^{4})\right]\\
&+&B_6\left[\textstyle{\frac{1}{6}}(\xi_{aa}^{45}\xi_{aa}^{45}-4\xi_{ab}^{45}\xi_{ba}^{45}+5\xi_{aa}^{45}\xi_{bb}^{45})\right],
\end{eqnarray*}
\begin{eqnarray*}
\bra{\Sigma_{abc}}\sigma^{FF(B)}\ket{\Sigma_{abc}}&=&B_1\left[-\textstyle{\frac{1}{12}}(4(\xi_{ab}^{4\nu}\xi_{ba}^{4\nu}+\xi_{aa}^{4\nu}\xi_{bb}^{4\nu})+\xi_{ac}^{4\nu}\xi_{ca}^{4\nu}+\xi_{aa}^{4\nu}\xi_{cc}^{4\nu}+\xi_{bc}^{4\nu}\xi_{cb}^{4\nu}+\xi_{bb}^{4\nu}\xi_{cc}^{4\nu})\right]\\
&+&B_3\left[-\textstyle{\frac{1}{12}}(4(\xi_{ab}^{4}\xi_{ba}^{4}+\xi_{aa}^{4}\xi_{bb}^{4})+\xi_{ac}^{4}\xi_{ca}^{4}+\xi_{aa}^{4}\xi_{cc}^{4}+\xi_{bc}^{4}\xi_{cb}^{4}+\xi_{bb}^{4}\xi_{cc}^{4})\right]\\
&+&B_5\left[-\textstyle{\frac{1}{12}}(4(\xi_{ab}^{45}\xi_{ba}^{45}+\xi_{aa}^{45}\xi_{bb}^{45})+\xi_{ac}^{45}\xi_{ca}^{45}+\xi_{aa}^{45}\xi_{cc}^{45}+\xi_{bc}^{45}\xi_{cb}^{45}+\xi_{bb}^{45}\xi_{cc}^{45})\right]\\
&+&B_2\left[\textstyle{\frac{1}{12}}(2(\xi_{ab}^{4\nu}\xi_{ba}^{4\nu}+\xi_{aa}^{4\nu}\xi_{bb}^{4\nu})-4\xi_{ac}^{4\nu}\xi_{ca}^{4\nu}+5\xi_{aa}^{4\nu}\xi_{cc}^{4\nu}-4\xi_{bc}^{4\nu}\xi_{cb}^{4\nu}+5\xi_{bb}^{4\nu}\xi_{cc}^{4\nu})\right]\\
&+&B_4\left[\textstyle{\frac{1}{12}}(2(\xi_{ab}^{4}\xi_{ba}^{4}+\xi_{aa}^{4}\xi_{bb}^{4})-4\xi_{ac}^{4}\xi_{ca}^{4}+5\xi_{aa}^{4}\xi_{cc}^{4}-4\xi_{bc}^{4}\xi_{cb}^{4}+5\xi_{bb}^{4}\xi_{cc}^{4})\right]\\
&+&B_6\left[\textstyle{\frac{1}{12}}(2(\xi_{ab}^{45}\xi_{ba}^{45}+\xi_{aa}^{45}\xi_{bb}^{45})-4\xi_{ac}^{45}\xi_{ca}^{45}+5\xi_{aa}^{45}\xi_{cc}^{45}-4\xi_{bc}^{45}\xi_{cb}^{45}+5\xi_{bb}^{45}\xi_{cc}^{45})\right],
\end{eqnarray*}
\begin{eqnarray*}
\bra{\Lambda_{abc}}\sigma^{FF(B)}\ket{\Lambda_{abc}}&=&B_1\left[-\textstyle{\frac{1}{4}}(\xi_{ac}^{4\nu}\xi_{ca}^{4\nu}+\xi_{aa}^{4\nu}\xi_{cc}^{4\nu}+\xi_{bc}^{4\nu}\xi_{cb}^{4\nu}+\xi_{bb}^{4\nu}\xi_{cc}^{4\nu})\right]\\
&+&B_3\left[-\textstyle{\frac{1}{4}}(\xi_{ac}^{4}\xi_{ca}^{4}+\xi_{aa}^{4}\xi_{cc}^{4}+\xi_{bc}^{4}\xi_{cb}^{4}+\xi_{bb}^{4}\xi_{cc}^{4})\right]\\
&+&B_5\left[-\textstyle{\frac{1}{4}}(\xi_{ac}^{45}\xi_{ca}^{45}+\xi_{aa}^{45}\xi_{cc}^{45}+\xi_{bc}^{45}\xi_{cb}^{45}+\xi_{bb}^{45}\xi_{cc}^{45})\right]\\
&+&B_2\left[\textstyle{\frac{1}{4}}(2(\xi_{aa}^{4\nu}\xi_{bb}^{4\nu}-\xi_{ab}^{4\nu}\xi_{ba}^{4\nu})+\xi_{aa}^{4\nu}\xi_{cc}^{4\nu}+\xi_{bb}^{4\nu}\xi_{cc}^{4\nu})\right]\\
&+&B_4\left[\textstyle{\frac{1}{4}}(2(\xi_{aa}^{4}\xi_{bb}^{4}-\xi_{ab}^{4}\xi_{ba}^{4})+\xi_{aa}^{4}\xi_{cc}^{4}+\xi_{bb}^{4}\xi_{cc}^{4})\right]\\
&+&B_6\left[\textstyle{\frac{1}{4}}(2(\xi_{aa}^{45}\xi_{bb}^{45}-\xi_{ab}^{45}\xi_{ba}^{45})+\xi_{aa}^{45}\xi_{cc}^{45}+\xi_{bb}^{45}\xi_{cc}^{45})\right].
\end{eqnarray*}
Repeated meson taste indices are summed as in (\ref{FFB}).  The $N_{aab}$ and $\Lambda_{abc}$ terms above include the diagonal $\sigma^{FF(B)}$ contributions.  However, the basis of \rrt\ eigenstates does not include the $\Sigma_{abc}$.  Instead, we consider the states of Table~\ref{RTeigen2}, $\rtwoorth N_{aab}-\oneorth\Sigma_{acd}$ and $\oneorth N_{aab}+\rtwoorth\Sigma_{acd}$.  In terms of the \rt\ basis, the corresponding diagonal contributions to $\sigma^{FF(B)}$ are
\begin{eqnarray*}
\bra{\rtwoorth N_{aab}-\oneorth\Sigma_{acd}}\sigma^{FF(B)}\ket{\rtwoorth N_{aab}-\oneorth\Sigma_{acd}}&=&2\bra{N_{aab}}\sigma^{FF(B)}\ket{N_{aab}}-\bra{\Sigma_{acd}}\sigma^{FF(B)}\ket{\Sigma_{acd}}\\
\bra{\oneorth N_{aab}+\rtwoorth\Sigma_{acd}}\sigma^{FF(B)}\ket{\oneorth N_{aab}+\rtwoorth\Sigma_{acd}}&=&2\bra{\Sigma_{acd}}\sigma^{FF(B)}\ket{\Sigma_{acd}}-\bra{N_{aab}}\sigma^{FF(B)}\ket{N_{aab}},
\end{eqnarray*}
where $abcd\in\{1423,\ 2314,\ 4132,\ 3241\}$, and I used $\ket{\Sigma_{abc}}=\ket{\Sigma_{bac}}$ (cf.~(\ref{siggy})).  The off-diagonal terms of $\sigma^{FF(B)}$ are
\begin{eqnarray*}
\bra{N_{aab}}\sigma^{FF(B)}\ket{\Lambda_{dfg}}&=&B_1\bigl[\textstyle{\frac{1}{2{\sqrt 6}}}(\delta_{af}(\xi_{ad}^{4\nu}\xi_{bg}^{4\nu}+\xi_{ag}^{4\nu}\xi_{bd}^{4\nu})+2\delta_{bd}\xi_{af}^{4\nu}\xi_{ag}^{4\nu}\\
&&\phantom{B_1}-\delta_{ad}(\xi_{af}^{4\nu}\xi_{bg}^{4\nu}+\xi_{ag}^{4\nu}\xi_{bf}^{4\nu})-2\delta_{bf}\xi_{ad}^{4\nu}\xi_{ag}^{4\nu})\bigr]\\
&+&B_3\bigl[\textstyle{\frac{1}{2{\sqrt 6}}}(\delta_{af}(\xi_{ad}^{4}\xi_{bg}^{4}+\xi_{ag}^{4}\xi_{bd}^{4})+2\delta_{bd}\xi_{af}^{4}\xi_{ag}^{4}\\
&&\phantom{B_3}-\delta_{ad}(\xi_{af}^{4}\xi_{bg}^{4}+\xi_{ag}^{4}\xi_{bf}^{4})-2\delta_{bf}\xi_{ad}^{4}\xi_{ag}^{4})\bigr]\\
&+&B_5\bigl[\textstyle{\frac{1}{2{\sqrt 6}}}(\delta_{af}(\xi_{ad}^{45}\xi_{bg}^{45}+\xi_{ag}^{45}\xi_{bd}^{45})+2\delta_{bd}\xi_{af}^{45}\xi_{ag}^{45}\\
&&\phantom{B_5}-\delta_{ad}(\xi_{af}^{45}\xi_{bg}^{45}+\xi_{ag}^{45}\xi_{bf}^{45})-2\delta_{bf}\xi_{ad}^{45}\xi_{ag}^{45})\bigr]\\
&+&B_2\bigl[\textstyle{\frac{1}{2{\sqrt 6}}}(\delta_{af}(\xi_{ad}^{4\nu}\xi_{bg}^{4\nu}-2\xi_{ag}^{4\nu}\xi_{bd}^{4\nu})+\delta_{ad}(2\xi_{bf}^{4\nu}\xi_{ag}^{4\nu}-\xi_{af}^{4\nu}\xi_{bg}^{4\nu})\\
&&\phantom{B_2}+\delta_{bf}\xi_{ad}^{4\nu}\xi_{ag}^{4\nu}-\delta_{bd}\xi_{af}^{4\nu}\xi_{ag}^{4\nu}+3\delta_{ag}(\xi_{ad}^{4\nu}\xi_{bf}^{4\nu}-\xi_{af}^{4\nu}\xi_{bd}^{4\nu}))\bigr]\\
&+&B_4\bigl[\textstyle{\frac{1}{2{\sqrt 6}}}(\delta_{af}(\xi_{ad}^{4}\xi_{bg}^{4}-2\xi_{ag}^{4}\xi_{bd}^{4})+\delta_{ad}(2\xi_{bf}^{4}\xi_{ag}^{4}-\xi_{af}^{4}\xi_{bg}^{4})\\
&&\phantom{B_4}+\delta_{bf}\xi_{ad}^{4}\xi_{ag}^{4}-\delta_{bd}\xi_{af}^{4}\xi_{ag}^{4}+3\delta_{ag}(\xi_{ad}^{4}\xi_{bf}^{4}-\xi_{af}^{4}\xi_{bd}^{4}))\bigr]\\
&+&B_6\bigl[\textstyle{\frac{1}{2{\sqrt 6}}}(\delta_{af}(\xi_{ad}^{45}\xi_{bg}^{45}-2\xi_{ag}^{45}\xi_{bd}^{45})+\delta_{ad}(2\xi_{bf}^{45}\xi_{ag}^{45}-\xi_{af}^{45}\xi_{bg}^{45})\\
&&\phantom{B_6}+\delta_{bf}\xi_{ad}^{45}\xi_{ag}^{45}-\delta_{bd}\xi_{af}^{45}\xi_{ag}^{45}+3\delta_{ag}(\xi_{ad}^{45}\xi_{bf}^{45}-\xi_{af}^{45}\xi_{bd}^{45}))\bigr],
\end{eqnarray*}
\begin{eqnarray*}
\bra{N_{aab}}\sigma^{FF(B)}\ket{\rtwoorth N_{ccd}-\oneorth\Sigma_{cfg}}&=&B_1\bigl[\textstyle{\frac{1}{6{\sqrt 6}}}(2\delta_{bc}(2\xi_{ac}^{4\nu}\xi_{ad}^{4\nu}-\xi_{af}^{4\nu}\xi_{ag}^{4\nu})+4(\delta_{ad}\xi_{ac}^{4\nu}\xi_{bc}^{4\nu}-\delta_{bd}\xi_{ac}^{4\nu}\xi_{ac}^{4\nu})\\
&&\phantom{B_1}+\delta_{ac}(\xi_{af}^{4\nu}\xi_{bg}^{4\nu}+\xi_{ag}^{4\nu}\xi_{bf}^{4\nu}-2\xi_{ac}^{4\nu}\xi_{bd}^{4\nu}-2\xi_{ad}^{4\nu}\xi_{bc}^{4\nu})\\
&&\phantom{B_1}+4\delta_{bg}\xi_{ac}^{4\nu}\xi_{af}^{4\nu}-2(\delta_{bf}\xi_{ag}^{4\nu}\xi_{ac}^{4\nu}+\delta_{ag}(\xi_{ac}^{4\nu}\xi_{bf}^{4\nu}+\xi_{af}^{4\nu}\xi_{bc}^{4\nu}))\\
&&\phantom{B_1}+\delta_{af}(\xi_{ag}^{4\nu}\xi_{bc}^{4\nu}+\xi_{ac}^{4\nu}\xi_{bg}^{4\nu}))\bigr]\\
&+&B_3\bigl[\textstyle{\frac{1}{6{\sqrt 6}}}(2\delta_{bc}(2\xi_{ac}^{4}\xi_{ad}^{4}-\xi_{af}^{4}\xi_{ag}^{4})+4(\delta_{ad}\xi_{ac}^{4}\xi_{bc}^{4}-\delta_{bd}\xi_{ac}^{4}\xi_{ac}^{4})\\
&&\phantom{B_3}+\delta_{ac}(\xi_{af}^{4}\xi_{bg}^{4}+\xi_{ag}^{4}\xi_{bf}^{4}-2\xi_{ac}^{4}\xi_{bd}^{4}-2\xi_{ad}^{4}\xi_{bc}^{4})\\
&&\phantom{B_3}+4\delta_{bg}\xi_{ac}^{4}\xi_{af}^{4}-2(\delta_{bf}\xi_{ag}^{4}\xi_{ac}^{4}+\delta_{ag}(\xi_{ac}^{4}\xi_{bf}^{4}+\xi_{af}^{4}\xi_{bc}^{4}))\\
&&\phantom{B_3}+\delta_{af}(\xi_{ag}^{4}\xi_{bc}^{4}+\xi_{ac}^{4}\xi_{bg}^{4}))\bigr]\\
&+&B_5\bigl[\textstyle{\frac{1}{6{\sqrt 6}}}(2\delta_{bc}(2\xi_{ac}^{45}\xi_{ad}^{45}-\xi_{af}^{45}\xi_{ag}^{45})+4(\delta_{ad}\xi_{ac}^{45}\xi_{bc}^{45}-\delta_{bd}\xi_{ac}^{45}\xi_{ac}^{45})\\
&&\phantom{B_5}+\delta_{ac}(\xi_{af}^{45}\xi_{bg}^{45}+\xi_{ag}^{45}\xi_{bf}^{45}-2\xi_{ac}^{45}\xi_{bd}^{45}-2\xi_{ad}^{45}\xi_{bc}^{45})\\
&&\phantom{B_5}+4\delta_{bg}\xi_{ac}^{45}\xi_{af}^{45}-2(\delta_{bf}\xi_{ag}^{45}\xi_{ac}^{45}+\delta_{ag}(\xi_{ac}^{45}\xi_{bf}^{45}+\xi_{af}^{45}\xi_{bc}^{45}))\\
&&\phantom{B_5}+\delta_{af}(\xi_{ag}^{45}\xi_{bc}^{45}+\xi_{ac}^{45}\xi_{bg}^{45}))\bigr]\\
&+&B_2\bigl[\textstyle{\frac{1}{6{\sqrt 6}}}(2(\delta_{bd}\xi_{ac}^{4\nu}\xi_{ac}^{4\nu}-\delta_{ad}\xi_{ac}^{4\nu}\xi_{bc}^{4\nu})+\delta_{bc}(\xi_{af}^{4\nu}\xi_{ag}^{4\nu}-2\xi_{ac}^{4\nu}\xi_{ad}^{4\nu})\\
&&\phantom{B_2}+\delta_{ac}(4\xi_{ag}^{4\nu}\xi_{bf}^{4\nu}+10\xi_{ac}^{4\nu}\xi_{bd}^{4\nu}-8\xi_{bc}^{4\nu}\xi_{ad}^{4\nu}-5\xi_{af}^{4\nu}\xi_{bg}^{4\nu})\\
&&\phantom{B_2}-2\delta_{bg}\xi_{ac}^{4\nu}\xi_{af}^{4\nu}+\delta_{bf}\xi_{ac}^{4\nu}\xi_{ag}^{4\nu}+\delta_{ag}(\xi_{ac}^{4\nu}\xi_{bf}^{4\nu}+\xi_{af}^{4\nu}\xi_{bc}^{4\nu})\\
&&\phantom{B_2}+\delta_{af}(4\xi_{ag}^{4\nu}\xi_{bc}^{4\nu}-5\xi_{ac}^{4\nu}\xi_{bg}^{4\nu}))\bigr]\\
&+&B_4\bigl[\textstyle{\frac{1}{6{\sqrt 6}}}(2(\delta_{bd}\xi_{ac}^{4}\xi_{ac}^{4}-\delta_{ad}\xi_{ac}^{4}\xi_{bc}^{4})+\delta_{bc}(\xi_{af}^{4}\xi_{ag}^{4}-2\xi_{ac}^{4}\xi_{ad}^{4})\\
&&\phantom{B_4}+\delta_{ac}(4\xi_{ag}^{4}\xi_{bf}^{4}+10\xi_{ac}^{4}\xi_{bd}^{4}-8\xi_{bc}^{4}\xi_{ad}^{4}-5\xi_{af}^{4}\xi_{bg}^{4})\\
&&\phantom{B_4}-2\delta_{bg}\xi_{ac}^{4}\xi_{af}^{4}+\delta_{bf}\xi_{ac}^{4}\xi_{ag}^{4}+\delta_{ag}(\xi_{ac}^{4}\xi_{bf}^{4}+\xi_{af}^{4}\xi_{bc}^{4})\\
&&\phantom{B_4}+\delta_{af}(4\xi_{ag}^{4}\xi_{bc}^{4}-5\xi_{ac}^{4}\xi_{bg}^{4}))\bigr]\\
&+&B_6\bigl[\textstyle{\frac{1}{6{\sqrt 6}}}(2(\delta_{bd}\xi_{ac}^{45}\xi_{ac}^{45}-\delta_{ad}\xi_{ac}^{45}\xi_{bc}^{45})+\delta_{bc}(\xi_{af}^{45}\xi_{ag}^{45}-2\xi_{ac}^{45}\xi_{ad}^{45})\\
&&\phantom{B_6}+\delta_{ac}(4\xi_{ag}^{45}\xi_{bf}^{45}+10\xi_{ac}^{45}\xi_{bd}^{45}-8\xi_{bc}^{45}\xi_{ad}^{45}-5\xi_{af}^{45}\xi_{bg}^{45})\\
&&\phantom{B_6}-2\delta_{bg}\xi_{ac}^{45}\xi_{af}^{45}+\delta_{bf}\xi_{ac}^{45}\xi_{ag}^{45}+\delta_{ag}(\xi_{ac}^{45}\xi_{bf}^{45}+\xi_{af}^{45}\xi_{bc}^{45})\\
&&\phantom{B_6}+\delta_{af}(4\xi_{ag}^{45}\xi_{bc}^{45}-5\xi_{ac}^{45}\xi_{bg}^{45}))\bigr],
\end{eqnarray*}
\begin{eqnarray*}
\bra{\Lambda_{abh}}\sigma^{FF(B)}\ket{\rtwoorth N_{ccd}-\oneorth\Sigma_{cfg}}&=&\rtwoorth\bra{N_{ccd}}\sigma^{FF(B)}\ket{\Lambda_{abh}}.
\end{eqnarray*}
In the Weyl representation, the matrix elements of $\sigma^{FF(B)}$ are consistent with the degeneracies and mixings parameterized in (\ref{rrtmat}).  The distinct, nontrivial matrix elements are listed in Table~\ref{sigFFB}.

\bibliography{bib}

\end{document}